\documentclass[12pt]{article}
\pdfoutput=1

\usepackage{epsfig}
\usepackage{psfrag}
\usepackage{latexsym}
\usepackage{indentfirst}
\usepackage{fancyhdr}
\usepackage{dsfont}
\usepackage{amssymb}
\usepackage{amsmath}
\usepackage{amsfonts}
\usepackage{pifont}
\usepackage{cite}
\usepackage{bbold}
\usepackage{color}
\usepackage{colordvi}
\usepackage{fancybox}
\usepackage[footnotesize]{caption2}
\usepackage{graphicx}
\usepackage[center,footnotesize,hang]{subfigure}
\usepackage{bbm}
\usepackage{url}
\usepackage{multirow}
\usepackage{array}
\usepackage{enumitem}
\newcommand{\PreserveBackslash}[1]{\let\temp=\\#1\let\\=\temp}
\newcolumntype{C}[1]{>{\PreserveBackslash\centering}p{#1}}
\newcolumntype{R}[1]{>{\PreserveBackslash\raggedleft}p{#1}}
\newcolumntype{L}[1]{>{\PreserveBackslash\raggedright}p{#1}}
\allowdisplaybreaks

\newcommand{\cleqn}{\setcounter{equation}{0}}

\newcommand{\bq}{\begin{eqnarray}}
\newcommand{\nq}{\end{eqnarray}}

\def\bvec#1{\raise1.5ex\hbox{$\rightarrow$}\mkern-16.5mu #1}

\newcommand{\cmark}{\ding{51}}
\newcommand{\xmark}{\ding{55}}
\makeatletter
\@addtoreset{equation}{section}
\makeatother

\begin{document}

\title{\hfill ~\\[0mm]
        \textbf{Lepton Mixing Predictions from Infinite Group Series $D^{(1)}_{9n, 3n}$ with Generalized CP}}

\date{}

\author{\\[1mm]Cai-Chang Li\footnote{E-mail: {\tt lcc0915@mail.ustc.edu.cn}}~,~~Chang-Yuan Yao\footnote{E-mail: {\tt phyman@mail.ustc.edu.cn}}~,~~Gui-Jun Ding\footnote{E-mail: {\tt dinggj@ustc.edu.cn}}\\ \\
\it{\small Department of Modern Physics, University of Science and
    Technology of China,}\\
  \it{\small Hefei, Anhui 230026, China}\\[4mm] }
\maketitle

\begin{abstract}

We have performed a comprehensive analysis of the type D group $D^{(1)}_{9n, 3n}$ as flavor symmetry and the generalized CP symmetry. All possible residual symmetries and their consequences for the prediction of the mixing parameters are studied.
We find that only one type of mixing pattern is able to accommodate the measured values of the mixing angles in both ``direct'' and ``variant of semidirect'' approaches, and four types of mixing patterns are phenomenologically viable in the ``semidirect'' approach. The admissible values of the mixing angles as well as CP violating phases are studied in detail for each case. It is remarkable that the first two smallest $D^{(1)}_{9n,3n}$ groups with $n=1, 2$ can fits the experimental data very well. The phenomenological predictions for neutrinoless double beta decay are discussed.

\end{abstract}
\thispagestyle{empty}
\vfill

\newpage
\setcounter{page}{1}

\section{\label{sec:Int}Introduction}

The precise measurement of the reactor mixing angle $\theta_{13}$~\cite{Abe:2011sj,Adamson:2011qu,Abe:2011fz,An:2012eh,Ahn:2012nd}
encourages the pursuit of the still missing results on leptonic CP violation and neutrino mass ordering as well as the characteristic neutrino nature.
Some low-significance hints for a maximally CP-violating value of the Dirac phase $\delta_{CP}\simeq3\pi/2$ have been observed~\cite{Abe:2015awa}. The global fits to lepton mixing parameters~\cite{Capozzi:2013csa,Forero:2014bxa,Gonzalez-Garcia:2014bfa} also provide weak evidence for the existence of Dirac type CP violation in neutrino oscillation. In the case that neutrinos are Majorana particles, two more Majorana CP phases $\alpha_{21}$ and $\alpha_{31}$ would be present, and they are crucial to the neutrinoless double beta decay process. However, the present experimental data don't impose any constraint on the values of the Majorana phases.

Finite discrete non-abelian flavor symmetries have been widely used
to make predictions for lepton flavor mixing.
Assuming the original flavor symmetry group is spontaneously broken to distinct abelian residual symmetries in the neutrino and charged lepton sectors  at a low energy scale, one can then determine mixing patterns from the residual symmetries and the structure of discrete flavor symmetry groups. Please see Refs.~\cite{Altarelli:2010gt,Ishimori:2010au,King:2013eh} for review on discrete flavor symmetries and the application in model building. For Majorana neutrinos,
if the residual symmetries of the charged lepton and neutrino mass matrices originate from a finite flavor group,
the lepton mixing matrix would be fully determined by residual symmetries up to independent row and column permutations. It turns out that the possible forms of the PMNS matrix are strongly constrained in this scenario such that the mixing patterns compatible with the data are of trimaximal form, and the Dirac CP phase is predicted to be $0$ or $\pi$~\cite{Fonseca:2014koa}. The same conclusion is reached for neutrinos being Dirac particles~\cite{Yao:2015dwa}. We note that the neutrino masses are not constrained in this approach and consequently the both Majorana phases $\alpha_{21}$ and $\alpha_{31}$ are undetermined. Their values can be fixed by considering a specific model. If the residual flavor symmetries of the neutrino and charged lepton mass matrix are partially contained in the underlying flavor group, the PMNS matrix would contains at least two free continuous parameters. As a result, the predictivity of the model would be lessened to a certain extent.

Besides the extensively discussed residual flavor symmetries, the neutrino and charged lepton mass matrices also admit residual CP transformations, and the residual CP symmetries can be generated by performing two residual CP transformations~\cite{Chen:2014wxa,Chen:2015nha,Everett:2015oka}. Analogous to residual flavor symmetries, the residual CP transformations can also constraint the lepton flavor mixing in particular the CP violating phases~\cite{Chen:2014wxa}. The simplest nontrivial CP transformation is known as $\mu-\tau$ reflection which gives rise to maximal atmospheric
mixing and maximal Dirac phase~\cite{Harrison:2002kp,Grimus:2003yn,Farzan:2006vj}. The deviation from maximal atmospheric mixing and non-maximal Dirac CP violation can be naturally obtained from the so-called generalized $\mu-\tau$ reflection~\cite{Chen:2015siy}.

Recently the flavor symmetry has been extended to combine with the generalized CP symmetry~\cite{Feruglio:2012cw,Holthausen:2012dk}.
This can lead to rather predictive scenario where both mixing angles and CP phases determined by a small number of (frequently only one) input parameters~\cite{Feruglio:2012cw}. In this case, the CP transformation matrix is generally non-diagonal and it is also called generalized CP. The generalized CP symmetry and the corresponding constraints on quark mass matrices have been exploited about thirty year ago~\cite{Ecker:1981wv,Grimus:1995zi}.
In this case the interplay between CP and flavor symmetries has to be carefully treated in order to make the theory consistent~\cite{Feruglio:2012cw,Holthausen:2012dk,Chen:2014tpa}.
There have been some models and model independent analysis of CP and flavor symmetries, such as  $A_{4}$~\cite{Ding:2013bpa}, $S_{4}$~\cite{Feruglio:2012cw,Ding:2013hpa,Li:2014eia,Feruglio:2013hia, Luhn:2013vna, Li:2013jya},
$\Delta(27)$~\cite{Branco:2015gna}, $\Delta(48)$~\cite{Ding:2013nsa}, $A_{5}$~\cite{Li:2015jxa,DiIura:2015kfa,Ballett:2015wia}, $\Delta(96)$~\cite{Ding:2014ssa}, and the group series  $\Delta(3n^{2})$~\cite{Hagedorn:2014wha,Ding:2015rwa} and $\Delta(6n^{2})$~\cite{King:2014rwa,Hagedorn:2014wha,Ding:2014ora} for general integer $n$. It is notable that smaller group for instance $A_4$~\cite{Ding:2013bpa}, $S_4$~\cite{Feruglio:2012cw,Ding:2013hpa,Li:2014eia,Feruglio:2013hia, Luhn:2013vna, Li:2013jya} and $A_5$~\cite{Li:2015jxa,DiIura:2015kfa,Ballett:2015wia} can already describe the experimentally measured values of the mixing angles, and the Dirac CP phase is predicted to be conserved or maximal while the Majorana phases are trivial. On the other hand, all the three CP violating phases generally depend on the free real parameter $\theta$ for $\Delta(3n^{2})$~\cite{Hagedorn:2014wha,Ding:2015rwa} and $\Delta(6n^{2})$~\cite{King:2014rwa,Hagedorn:2014wha,Ding:2014ora} flavor symmetries.

In the present work, we shall thoroughly analyze the lepton mixing patterns which can be obtained from the breaking of $D^{(1)}_{9n, 3n}$ flavor symmetry and generalized CP. All possible residual symmetries in the ``direct'', ``semidirect'' and ``variant of semidirect'' approaches and their consequences for the prediction of the mixing parameters are studied. We shall perform a detailed numerical analysis for all the possible mixing patterns. The admissible values of the mixing parameters for each $n$ and the possible values of the effective mass $|m_{ee}|$ will be explored.

The outline of this paper is as follows. In section~\ref{sec:GCP} we find the class-inverting automorphism of the $D^{(1)}_{9n,3n}$ group and the corresponding physically well-defined generalized CP transformations are determined by solving the consistency condition. In section~\ref{sec:framework} we review the approach to determining the lepton flavor mixing from residual flavor and CP symmetries of the neutrino and the charged lepton sectors. All possible residual symmetries and the consequences for the prediction of the flavor mixing are studied in the method of the direct approach in section~\ref{sec:direct_approach}. The PMNS matrix is determined to be of the trimaximal form, both Dirac phase $\delta_{CP}$ and the Majorana phase $\alpha_{31}$ are conserved, and the values of $\alpha_{21}$ are integer multiple of $2\pi/(3n)$. We investigate the possible mixing patterns which can be derived from the semidirect approach and variant of semidirect approach in section~\ref{sec:Z2xCP_neutrino} and section~\ref{sec:Z2xCP_charged_lepton}. The analytical expressions of the PMNS matrices, mixing angles and CP invariants are presented, the admissible values of the mixing angles and CP violation phases are analyzed numerically in detail, and phenomenological predictions for neutrinoless double beta decay are studied.
For the lowest order $D^{(1)}_{9n ,3n}$ group with $n=1, 2$, we find all the mixing patterns that can describe the experimentally measured values of the mixing angles, and a $\chi^2$ analysis is performed. Finally we summarize and present our conclusions in section~\ref{sec:Conclusion}. The group theory of $D^{(1)}_{9n,3n}$ is presented in Appendix~\ref{app:group_theory} including the conjugacy classes, the irreducible representations, the character table, the Kronecker products and the Clebsch-Gordan coefficients.

\section{\label{sec:GCP}Generalized CP consistent with $D^{(1)}_{9n,3n}$ family symmetry}

The finite subgroups of $SU(3)$ have been systematically classified by mathematicians~\cite{su(3)_subgroups_book} (see Refs.~\cite{Grimus:2010ak,Grimus:2011fk,Grimus:2013apa} for recent work). It is well-established that all discrete subgroups of $SU(3)$ can be divided into five categories: type A, type B, type C, type D, and type E~\cite{Grimus:2011fk,Grimus:2013apa}.  The type D group turns out to be particularly significant in flavor symmetry theory~\cite{Fonseca:2014koa,King:2013vna}.
Type D group is isomorphic to $(Z_{m}\times Z_{n})\rtimes S_3$, and it can be generated by four generators $a$, $b$, $c$ and $d$ subject to the following rules~\cite{Grimus:2013apa}:
\begin{eqnarray}
\nonumber&&a^3=b^2=(ab)^2=c^m=d^n=1,\quad cd=dc,\\
\nonumber&&\quad~ aca^{-1}=c^kd,\quad ada^{-1}=c^{-m/n}d^{-(k+1)},\\ \quad \label{eq:group_relation}&&\qquad~~ bcb^{-1}=cd,\quad bdb^{-1}=d^{-1}\,.
\end{eqnarray}
It is found that the type D group exists only for~\cite{Grimus:2013apa}
\begin{equation}
k=0, m=n ~~~~~ \text{or}~~~~~  k=1, m=3n\,.
\end{equation}
In the case of $k=0$, $m=n$, the corresponding group denoted as $D^{(0)}_{n, n}$ is exactly the well-known $\Delta(6n^2)$ group~\cite{Escobar:2008vc}. For another case of $k=1$, $m=3n$, the corresponding type D group denoted as $D^{1)}_{3n, n}$ is isomorphic to $Z_3\times\Delta(6n^2)$ if $n$ is not divisible by 3~\cite{Grimus:2013apa}. Therefore the representation of $D^{(1)}_{3n, n}$ can be obtained by multiplying the representation matrices of $\Delta(6n^2)$ with 1, $e^{2\pi i/3}$ and $e^{4\pi i/4}$ for $3 \nmid n$. As a consequence, the $D^{(1)}_{3n, n}$ group for $3 \nmid n$ would give rise to the same set of lepton flavor mixing as $\Delta(6n^2)$~group no matter whether the generalized CP symmetry is considered or not. The $\Delta(6n^2)$ as flavor symmetry group has been comprehensively explored in the literature~\cite{King:2013vna,Ding:2014ora,Hagedorn:2014wha}, we shall focus on the second independent type D infinite series of groups $D^{(1)}_{9n,3n}$ where $n$ is any positive integer. It is remarkable that $D^{(1)}_{9n,3n}$ can generate experimentally viable lepton and quark mixing simultaneously~\cite{Yao:2015dwa}. In the present work, we shall include the generalized CP symmetry compatible with $D^{(1)}_{9n,3n}$ and investigate its predictions for lepton mixing angles and CP violating phases. The group theory of $D^{(1)}_{9n,3n}$, its irreducible representations and the Clebsch-Gordan coefficients are presented in Appendix~\ref{app:group_theory}.

\begin{table}[t!]
\begin{center}
\item \begin{tabular}{|c|c|c|c|c|c|}
\hline\hline

$n$ &  {\tt $G_f$} &  \texttt{GAP-Id}  &  \texttt{Inn($G_f$)}  &  \texttt{Out($G_f$)} \\  \hline

1   &  $D^{(1)}_{9,3}$   &  [162,14]    &   $\left(\left(Z_3\times Z_3\right)\rtimes Z_3\right)\rtimes Z_2$ &    $Z_6$     \\ \hline

2   &  $D^{(1)}_{18, 6}$   & [648,259]   &   $\left(\left(Z_6\times Z_6\right)\rtimes Z_3\right)\rtimes Z_2$    &    $Z_6$     \\ \hline

3  &  $D^{(1)}_{27, 9}$  & [1458,659]   &   $\left(\left(Z_9\times Z_9\right)\rtimes Z_3\right)\rtimes Z_2$    &    $Z_{18}$     \\
\hline\hline

\end{tabular}
\caption{\label{tab:automorphism}The automorphism groups of the $D^{(1)}_{9n, 3n}$ group with $n=1, 2, 3$, where \texttt{Inn($G_f$)} and  \texttt{Out($G_f$)} denote inner automorphism group and outer automorphism group of \texttt{$G_f$} respectively. Note that each of these three groups has a unique class-inverting outer automorphism.}
\end{center}
\end{table}

It is highly nontrivial to introduce the generalized CP symmetry in the presence of a discrete flavor symmetry $G_{f}$. In order to consistently combine the generalized CP symmetry with flavor symmetry, the following consistency condition has to be fulfilled~\cite{Feruglio:2012cw,Holthausen:2012dk,Chen:2014tpa},
\begin{equation}
\label{eq:consistency_condition}X_{\mathbf{r}}\rho^{*}_{\mathbf{r}}(g)X^{\dagger}_{\mathbf{r}}=\rho_{\mathbf{r}}(g^{\prime}),\quad
g,g^{\prime}\in G_{f}\,,
\end{equation}
where $\rho_{\mathbf{r}}(g)$ is the representation matrix of the element $g$ in the irreducible representation $\mathbf{r}$ of $G_{f}$, and $X_{\mathbf{r}}$ is the generalized CP transformation. Obviously the CP transformation $X_{\mathbf{r}}$ maps $g$ into another group element $g^{\prime}$. Therefore the generalized CP symmetry corresponds the automorphism group of $G_{f}$. Moreover, it was shown that the physically well-defined CP transformations should be given by class-inverting automorphism of $G_f$~\cite{Chen:2014tpa}. We have exploited the computer algebra system \texttt{GAP}~\cite{GAP4.4} to calculate the automorphism group of the first three $D^{(1)}_{9n,3n}$ groups with $n=1, 2, 3$, the results are listed in table~\ref{tab:automorphism}. Notice that larger $D^{(1)}_{9n,3n}$ group for $n\geq4$ is not stored in \texttt{GAP} at present. We see that the automorphism group of $D^{(1)}_{9n,3n}$ is quite complex but each one of $D^{(1)}_{9, 3}$, $D^{(1)}_{18, 6}$ and $D^{(1)}_{27, 9}$ has a unique class-inverting outer automorphism. Furthermore, we find a generic class-inverting automorphism $u$ of the $D^{(1)}_{9n,3n}$ group, and its actions on the generators $a$, $b$, $c$, $d$ are as follows
\begin{equation}
\label{eq:physical_aut}a\stackrel{u}{\longmapsto}a,\quad
b\stackrel{u}{\longmapsto}b,\quad c\stackrel{u}{\longmapsto}c^{-1},\quad
d\stackrel{u}{\longmapsto}d^{-1}\,.
\end{equation}
It is easy to check that $u$ indeed maps each element into the class of its inverse element for any value of the parameter $n$. We denote the physical CP transformation corresponding to the automorphism $u$ as $X_{\mathbf{r}}(u)$, and its explicit form is determined by the following consistency equations:
\begin{eqnarray}
\nonumber&&X_{\mathbf{r}}\left(u\right)\rho^{*}_{\mathbf{r}}(a)X^{\dagger}_{\mathbf{r}}\left(u\right)=\rho_{\mathbf{r}}\left(u\left(a\right)\right)=\rho_{\mathbf{r}}(a)\,,\\
\nonumber&&X_{\mathbf{r}}\left(u\right)\rho^{*}_{\mathbf{r}}(b)X^{\dagger}_{\mathbf{r}}\left(u\right)=\rho_{\mathbf{r}}\left(u\left(b\right)\right)=\rho_{\mathbf{r}}\left(b\right)\,,\\
\nonumber&&X_{\mathbf{r}}\left(u\right)\rho^{*}_{\mathbf{r}}(c)X^{\dagger}_{\mathbf{r}}\left(u\right)=\rho_{\mathbf{r}}\left(u\left(c\right)\right)=\rho_{\mathbf{r}}\left(c^{-1}\right)\,,\\
\label{eq:conmsistency_equations}&&X_{\mathbf{r}}\left(u\right)\rho^{*}_{\mathbf{r}}(d)X^{\dagger}_{\mathbf{r}}\left(u\right)=\rho_{\mathbf{r}}\left(u\left(d\right)\right)=\rho_{\mathbf{r}}\left(d^{-1}\right)\,.
\end{eqnarray}
In our working basis shown in Appendix~\ref{app:group_theory}, the representation matrices of $a$ and $b$ are real while the representation matrices of $c$ and $d$ are complex and diagonal for any irreducible representations of $D^{(1)}_{9n, 3n}$. Therefore the CP transformation $X_{\mathbf{r}}(u)$ is a unit matrix, i.e.
\begin{equation}\label{eq:gcp_trans}
X_{\mathbf{r}}\left(u\right)=\mathbb{1}_{\bf{r}}\,.
\end{equation}
Given this CP transformation $X_{\mathbf{r}}(u)$, the matrix $\rho_{\mathbf{r}}(g)X_{\mathbf{r}}(u)=\rho_{\mathbf{r}}(g)$ is also an admissible CP transformation for any $g\in D^{(1)}_{9n, 3n}$. It corresponds to performing a conventional CP transformation followed by a group transformation $\rho_{\mathbf{r}}(g)$. As a consequence, we conclude that the generalized CP transformation compatible with the $D^{(1)}_{9n, 3n}$ family symmetry is of the same form as the flavor symmetry transformation in our basis, i.e.
\begin{equation}
\label{eq:GCP_full}
X_{\mathbf{r}}=\rho_{\mathbf{r}}(g),\quad g\in D^{(1)}_{9n,3n}\,.
\end{equation}
Note that other possible CP transformations can also be defined if a model contains only a subset of irreducible representations. Lepton mixing can be derived from the remnant symmetries in the
charged lepton and neutrino mass matrices, while the mechanism of symmetry breaking is irrelevant. The basic procedure and the resulting master formulae are given in Refs.~\cite{Ding:2013bpa,Ding:2013hpa,Ding:2014ora,Chen:2014wxa,Chen:2015nha}. In the following, we shall consider all possible remnant symmetries of the neutrino and charged lepton sectors and discuss the predictions for the PMNS matrix and the lepton mixing parameters.

\section{\label{sec:framework}Framework}

In the present work, the family symmetry is taken to be $D^{(1)}_{9n, 3n}$, and the generalized CP symmetry is considered in order to predict the lepton mixing parameters including the CP violating phases. Without loss of generality,
we assume that the three left-handed leptons transform as a triplet $\mathbf{3}_{1, 0}$ under $D^{(1)}_{9n, 3n}$. For brevity we shall denote the faithful irreducible representation $\mathbf{3}_{1, 0}$ as $\mathbf{3}$. The representation matrices of the generators $a$, $b$, $c$ and $d$ in $\mathbf{3}_{1, 0}$ are given in Eq.~\eqref{eq:3dimrep1}. The light neutrinos are assumed to be Majorana particles. From the bottom-up perspective, the most general symmetry of a generic charged lepton mass matrices is $U(1)\times U(1)\times U(1)$, which has finite subgroups isomorphic to a cyclic group $Z_m$ for any integer $m$ or a direct product of several cyclic groups~\cite{Chen:2014wxa,Chen:2015nha,Yao:2015dwa}. On the other hand, the largest possible symmetry of the neutrino mass matrix is $Z_2\times Z_2$~\cite{Chen:2014wxa,Chen:2015nha,Yao:2015dwa,Lam:2007qc}. Moreover the neutrino and charged lepton mass matrices are invariant under a set of CP transformations, and both the $U(1)\times U(1)\times U(1)$ symmetry group of the charged-lepton mass term and the $Z_2\times Z_2$ symmetry of the neutrino mass terms can be generated by performing two CP symmetry transformations~\cite{Chen:2014wxa,Chen:2015nha}. Conversely, the lepton mass matrices are strongly constrained by the postulated remnant symmetry such that the lepton mixing matrix can be derived from the remnant symmetries in the
charged lepton and neutrino sectors, while the mechanism of dynamically realizing the assumed remnant symmetries is irrelevant~\cite{Chen:2014wxa,Chen:2015nha}. From the view of the top-down method, the remnant flavor and CP symmetries of the neutrino and charged lepton mass matrices may originate from certain symmetry group implemented at high energy scales. In the present work, both flavor symmetry $D^{(1)}_{9n, 3n}$ and the generalized CP are imposed, i.e., the parent symmetry is $D^{(1)}_{9n, 3n}\rtimes H_{CP}$, where $H_{CP}$ denotes the generalized CP transformations consistent with $D^{(1)}_{9n, 3n}$ and it is given by Eq.~\eqref{eq:GCP_full}. $D^{(1)}_{9n, 3n}\rtimes H_{CP}$ is assumed to be broken down into $G_{l}\rtimes H^{l}_{CP}$ and $G_{\nu}\times H^{\nu}_{CP}$ in the charged lepton and neutrino sectors respectively. The allowed forms of the neutrino and charged lepton mass matrices are constrained by the remnant symmetries, and subsequently we can diagonalize them to get the PMNS matrix.

The requirement that a subgroup $G_{l}\rtimes H^{l}_{CP}$ is preserved at low energies entails that the combination $m^{\dagger}_{l}m_{l}$ has to fulfill
\begin{eqnarray}
\nonumber&&\rho^{\dagger}_{\mathbf{3}}(g_{l})m^{\dagger}_{l}m_{l}\rho_{\mathbf{3}}(g_{l})=m^{\dagger}_{l}m_{l},\quad g_{l}\in G_{l},\\
\label{eq:constr_clep}&&X^{\dagger}_{l\mathbf{3}}m^{\dagger}_{l}m_{l}X_{l\mathbf{3}}=\left(m^{\dagger}_{l}m_{l}\right)^{\ast},\quad X_{l\mathbf{3}}\in H^{l}_{CP}\,,
\end{eqnarray}
where the charged lepton mass matrix $m_l$ is given in the convention $l^{c}m_{l}l$. The hermitian combination $m^{\dagger}_{l}m_{l}$ is diagonalized by the unitary transformation $U_{l}$ with $U^{\dagger}_{l}m^{\dagger}_{l}m_{l}U_{l}=\text{diag}(m^2_{e}, m^2_{\mu}, m^2_{\tau})$. The three charged leptons have distinct
masses $m_{e}\neq m_{\mu}\neq m_{\tau}$. From Eq.~\eqref{eq:constr_clep}, it is straightforward to derive that the remnant symmetry $G_{l}\rtimes H^{l}_{CP}$ leads to the following constraints on $U_{l}$
\begin{eqnarray}
\nonumber&&U^{\dagger}_{l}\rho_{\mathbf{3}}(g_{l})U_{l}=\rho^{diag}_{\mathbf{3}}(g_{l}),\quad g_{l}\in G_{l},\\
\label{eq:constr_Ul}&&U^{\dagger}_{l}X_{l\mathbf{3}}U^{\ast}_{l}=X^{diag}_{l\mathbf{3}},\quad X_{l\mathbf{3}}\in H^{l}_{CP}\,,
\end{eqnarray}
where both $\rho^{diag}_{\mathbf{3}}(g_{l})$ and $X^{diag}_{l\mathbf{3}}$ are diagonal phase matrices. As a consequence, we see that $U_{l}$ also diagonalizes the residual flavor symmetry transformation matrix $\rho_{\mathbf{3}}(g_{l})$, the residual CP transformation $X_{l\mathbf{3}}$ is a symmetric matrix, and the following restricted consistency condition should be satisfied~\cite{Li:2014eia},
\begin{equation}
\label{eq:res_cons_cond_clep}X_{l\mathbf{r}}\rho^{\ast}_{\mathbf{r}}(g_{l})X^{-1}_{l\mathbf{r}}=\rho_{\mathbf{r}}(g^{-1}_{l}),\qquad
g_{l}\in G_{l},~X_{l\mathbf{r}}\in H^{l}_{CP} \,.
\end{equation}
In the same fashion, the neutrino mass matrix is invariant under the action of the elements of the residual subgroup $G_{\nu}\times H^{\nu}_{CP}$:
\begin{eqnarray}
\nonumber&&\rho^{T}_{\mathbf{3}}(g_{\nu})m_{\nu}\rho_{\mathbf{3}}(g_{\nu})=m_{\nu},\qquad g_{\nu}\in G_{\nu},\\
\label{eq:constr_nu}&&X^{T}_{\nu\mathbf{3}}m_{\nu}X_{\nu\mathbf{3}}=m^{*}_{\nu},\qquad
\quad X_{\nu\mathbf{3}}\in H^{\nu}_{CP}\,.
\end{eqnarray}
We denote the unitary diagonalization matrix of $m_{\nu}$ as $U_{\nu}$ fulfilling $U^{T}_{\nu}m_{\nu}U_{\nu}=\text{diag}\left(m_1, m_2, m_3\right)$. Then $U_{\nu}$ would be subject to the following constraints from the postulated residual symmetry~\cite{Chen:2014wxa,Chen:2015nha,Everett:2015oka}:
\begin{eqnarray}
\nonumber&&U^{\dagger}_{\nu}\rho_{\mathbf{3}}(g_{\nu})U_{\nu}=\textrm{diag}(\pm1, \pm1, \pm1),\\
\label{eq:constr_Unu}&&U^{\dagger}_{\nu}X_{\nu\mathbf{3}}U^{*}_{\nu}=\text{diag}(\pm1, \pm1, \pm1)\,,
\end{eqnarray}
where the ``$\pm$'' signs can be chosen independently. Therefore the residual CP transformation $X_{\nu\mathbf{3}}$ is a symmetric unitary matrix as well, and the restricted consistency condition on the neutrino sector takes the form~\cite{Chen:2014wxa,Chen:2015nha,Everett:2015oka,Feruglio:2012cw}:
\begin{equation}
\label{eq:res_cons_cond_nu}X_{\nu\mathbf{r}}\rho^{*}_{\mathbf{r}}(g_{\nu})X^{-1}_{\nu\mathbf{r}}=\rho_{\mathbf{r}}(g_{\nu}),\qquad
g_{\nu}\in G_{\nu},~X_{\nu\mathbf{r}}\in H^{\nu}_{CP}\,.
\end{equation}
Obviously $X_{\nu\mathbf{r}}$ maps any element $g_{\nu}$ of the neutrino residual flavor symmetry $G_{\nu}$ into itself. Hence the mathematical structure of the remnant subgroup comprising $G_{\nu}$ and $H^{\nu}_{CP}$ is generally a direct product instead of a semidirect product. Given a pair of well-defined remnant symmetries $G_{l}\rtimes H^{l}_{CP}$ and $G_{\nu}\times H^{\nu}_{CP}$ for which the consistency equations in Eqs.~(\ref{eq:res_cons_cond_clep}, \ref{eq:res_cons_cond_nu}) are fulfilled, the allowed forms of the mass matrices $m^{\dagger}_{l}m_{l}$ and $m_{\nu}$ can be determined from Eqs.~(\ref{eq:constr_clep}, \ref{eq:constr_nu}), and subsequently the prediction for the PMNS matrix $U_{\text{PMNS}}=U^{\dagger}_{l}U_{\nu}$ can be obtained by diagonalizing $m^{\dagger}_{l}m_{l}$ and $m_{\nu}$.

For two pair of remnant symmetry subgroups $\left\{G_{l}\rtimes H^{l}_{CP}, G_{\nu}\times H^{\nu}_{CP}\right\}$ and  $\big\{G^{\prime}_{l}\rtimes H^{l^{\prime}}_{CP}$, $G^{\prime}_{\nu}\times H^{\nu^{\prime}}_{CP}\big\}$, if $G_{l}$, $G_{\nu}$ and $G^{\prime}_{l}$, $G^{\prime}_{\nu}$ are related by a similarity transformation, for example if they are conjugate,
\begin{equation}
G^{\prime}_{l}=hG_{l}h^{-1},\qquad G^{\prime}_{\nu}=hG_{\nu}h^{-1},\qquad h\in D^{(1)}_{9n, 3n}\,.
\end{equation}
The remnant CP would also be related by
\begin{equation}
\label{eq:CP_conju}H^{l^{\prime}}_{CP}=\rho_{\mathbf{r}}(h)H^{l}_{CP}\rho^{T}_{\mathbf{r}}(h)
,\qquad H^{\nu^{\prime}}_{CP}=\rho_{\mathbf{r}}(h)H^{\nu}_{CP}\rho^{T}_{\mathbf{r}}(h)
\end{equation}
in order to fulfill the consistency conditions in Eqs.~(\ref{eq:res_cons_cond_clep}, \ref{eq:res_cons_cond_nu}). That is to say the elements of $H^{l^{\prime}}_{CP}$ and $H^{\nu^{\prime}}_{CP}$ are given by $\rho_{\mathbf{r}}(h)X_{l\mathbf{r}}\rho^{T}_{\mathbf{r}}(h)$ and $\rho_{\mathbf{r}}(h)X_{\nu\mathbf{r}}\rho^{T}_{\mathbf{r}}(h)$ respectively, where $X_{\nu\mathbf{r}}\in H^{\nu}_{CP}$ and $X_{l\mathbf{r}}\in H^{l}_{CP}$. Notice that all the possible remnant CP transformations compatible with the remnant flavor symmetry have been considered in this work. Hence if $G_{l}\rtimes H^{l}_{CP}$ and  $G_{\nu}\times H^{\nu}_{CP}$ fix the charged lepton and neutrino mass matrices to be $m^{\dagger}_{l}m_{l}$ and $m_{\nu}$, then $m^{\prime\dagger}_{l}m^{\prime}_{l}\equiv\rho_{\mathbf{3}}(h)m^{\dagger}_{l}m_{l}\rho^\dagger_{\mathbf{3}}(h)$ and $m^{\prime}_{\nu}\equiv\rho^{\ast}_{\mathbf{3}}(h)m_{\nu}\rho^{\dagger}_{\mathbf{3}}(h)$ would be invariant under the remnant symmetries $G^{\prime}_{l}\rtimes H^{l^{\prime}}_{CP}$ and $G^{\prime}_{\nu}\times H^{\nu^{\prime}}_{CP}$ respectively. As a result, two pair of remnant symmetries $\left\{G_{l}\rtimes H^{l}_{CP}, G_{\nu}\times H^{\nu}_{CP}\right\}$ and  $\big\{G^{\prime}_{l}\rtimes H^{l^{\prime}}_{CP}$, $G^{\prime}_{\nu}\times H^{\nu^{\prime}}_{CP}\big\}$ would yield the same results for the PMNS matrix $U_{\text{PMNS}}$. In this work, we shall perform a comprehensive analysis of the mixing patterns which can be derived from the group $D^{(1)}_{9n, 3n}\rtimes H_{CP}$.
It is sufficient to only analyze a few representative remnant symmetries which give rise to different results for $U_{\text{PMNS}}$ and lepton mixing parameters, as other possible choices for the remnant symmetry groups are related to the representative ones by similarity transformation and consequently no new results are obtained.

\section{\label{sec:direct_approach}Lepton mixing from direct approach}

In the direct approach, the residual flavor symmetry $G_{\nu}$ is a Klein four subgroup, and the residual flavor symmetry $G_{l}$ is a cyclic group $Z_{m}$ with index $m\geq3$ or a product of cyclic groups. We assume that the residual flavor symmetry group $G_{l}$ can distinguish the three generations of charged lepton. In other words, the restricted representation of the triplet representation $\mathbf{3}$ on $G_{l}$ should decompose into three inequivalent 1-dimensional representations of $G_{l}$. From Eq.~\eqref{eq:constr_clep} and Eq.~\eqref{eq:constr_Ul}, we see that $U_{l}$ not only diagonalizes the mass matrix $m^{\dagger}_{l}m_{l}$ but also the residual flavor symmetry transformation matrix $\rho_{\mathbf{3}}(g_{l})$ with $g_{l}\in G_{l}$. As a result, the requirement that $U^{\dagger}_{l}\rho_{\mathbf{3}}(g_{l})U_{l}=\rho^{diag}_{\mathbf{3}}(g_{l})$ is diagonal allows us to determine $U_{l}$ and no knowledge of $m^{\dagger}_{l}m_{l}$ is necessary. Notice that the remnant CP invariant condition in Eq.~\eqref{eq:constr_clep} is automatically satisfied, the reason is that the residual CP transformation $X_{l\mathbf{3}}$ has to be compatible with residual flavor symmetry and its allowed form is strongly constrained by the restricted consistency condition of Eq.~\eqref{eq:res_cons_cond_clep}.

As shown in the Appendix~\ref{app:group_theory}, the group structure of the $D^{(1)}_{9n, 3n}$ has been studied in detail. The residual subgroup $G_{l}$ is an abelian subgroup, and it can be generated by the generator $c^{s}d^{t}$, $bc^{s}d^{t}$, $ac^{s}d^{t}$,
$a^2c^{s}d^{t}$, $abc^{s}d^{t}$ or $a^2bc^{s}d^{t}$ with $s=0,1,\dots,9n-1,t=0,1,\dots,3n-1$.
The diagonalization of $\rho_{\mathbf{3}}(g_{l})$ determines the unitary transformation $U_{l}$ up to permutations and phases of the column vectors if $\rho_{\mathbf{3}}(g_{l})$  has non-degenerate eigenvalues, where $g_{l}$ can be taken to be the generator of $G_{l}$. The explicit form of $U_{l}$ for different $G_{l}$ and the corresponding remnant CP transformations compatible with $G_{l}$ are summarized in table~\ref{tab:cle_diagonal_matrix}. If the eigenvalues of $\rho_{\mathbf{3}}(g_{l})$ are degenerate so
that its diagonalization matrix $U_{l}$ can not be determined uniquely, we would extend $G_{l}$ from a single cyclic subgroup to a product of cyclic groups, for example $G_{l}=G_{1}\times G_{2}$ where the generators of $G_1$ and $G_2$ should be commutable with each other. If $G_1$ (or $G_2$) is sufficient to distinguish among the generations such that its eigenvalues are not degenerate, then another subgroup $G_2$ (or $G_1$) would not impose any new constraint on the lepton mixing. On the other hand, if the three eigenvalues of the generator of either $G_1$ or $G_2$ are completely degenerate, e.g. $G_1 (~\text{or}~G_2)=\langle c^{3n}\rangle$, its three-dimensional representation matrix would be proportional to a unit matrix. As a result, we shall concentrate on the case that the representation matrices of both $G_1$ and $G_2$ have two degenerate eigenvalues, therefore either $G_1$ or $G_2$ alone fixes only a column of $U_{l}$ and the third column can be determined by unitary condition.
The possible extension of remnant flavor symmetry group $G_{l}$, the corresponding remnant CP transformations and the unitary transformations $U_{l}$ are collected in table~\ref{tab:extension_Gch}. We see that the diagonalization matrix $U_{l}$ can only take five distinct forms $U_l^{(1)}$, $U_l^{(2)}$, $U_l^{(3)}$, $U_l^{(4)}$ or $U_l^{(5)}$ such that the constraints on $s$ and $t$ shown in table~\ref{tab:cle_diagonal_matrix} are relaxed.

\begin{table}[t!]
\renewcommand{\tabcolsep}{0.5mm}
\centering
\small
\begin{tabular}{|c|c|c|c|}
\hline \hline
 &  &  &   \\ [-0.16in]
 $G_{l}$ &  $U_{l}$  &  \texttt{Constraints} & $H^{l}_{CP}$  \\

  &   & &     \\ [-0.16in]\hline
 &   &  &     \\ [-0.16in]

 &   &  $t\neq0$ &  \\[-0.16in]

$\langle c^{s}d^{t}\rangle$  &   $U^{(1)}_l=\begin{pmatrix}
 1  &  0   &  0  \\
 0  &  1   &  0  \\
 0  &  0   &  1
\end{pmatrix}$  &  $s-t\neq0$ $\text{mod}(3n)$ & $\{c^{\gamma}d^{\delta}\}$ \\[-0.16in]

  &    &   $s-2t\neq0$ $\text{mod}(3n)$ &   \\

&   &   &    \\ [-0.16in]\hline
 &   &  &     \\ [-0.10in]

 &   &   &  $\{c^{2t-s+2\delta+3n\tau}d^{\delta},$  \\ [-0.25in]

$\langle bc^{s}d^{t}\rangle$  &   $U^{(2)}_l=\frac{1}{\sqrt{2}}
\begin{pmatrix}
 \sqrt{2} &~ 0 ~& 0 \\
 0 &~ -e^{\frac{i\pi(2t-s)}{3 n}} ~& e^{\frac{i\pi(2t-s)}{3 n}} \\
 0 &~ 1 ~& 1 \\
\end{pmatrix}$  &  $s\neq0,3n,6n$ &   \\[-0.25in]

&    &   &  $bc^{2\delta+3n\tau}d^{\delta}\}$ \\

&   & &      \\ [-0.10in]\hline
 &   & &      \\ [-0.16in]

  &   &   & $\{bc^{-2t+3n\tau}d^{-t},$ \\[-0.16in]

$\langle ac^{s}d^{t}\rangle$  &   $U^{(3)}_l=\frac{1}{\sqrt{3}}
\begin{pmatrix}
 e^{-\frac{2 i \pi  s}{9 n}} &~ \omega^2 e^{-\frac{2 i \pi  s}{9 n}} ~& \omega e^{-\frac{2 i \pi  s}{9 n}} \\
 e^{\frac{2 i \pi  (3t-2 s)}{9 n}} &~ \omega e^{\frac{2 i \pi  (3t-2 s)}{9 n}} ~& \omega^2 e^{\frac{2 i \pi  (3t-2 s)}{9 n}} \\
 1 &~ 1 ~& 1 \\
\end{pmatrix}$  &  --- & $abc^{s-2t+3n\tau}d^{s-2t},$ \\[-0.16in]

 &    &    & $a^{2}bc^{t-s+3n\tau}\}$  \\

&   & &     \\ [-0.16in]\hline
 &   &  &     \\ [-0.16in]

  &   &   & $\{bc^{2(t-s)+3n\tau}d^{t-s},$ \\[-0.16in]

$\langle a^2c^{s}d^{t}\rangle$  &   $U^{(3')}_l=\frac{1}{\sqrt{3}}\left(
\begin{array}{ccc}
 e^{\frac{2 i \pi  (3t-2s)}{9 n}} &~ \omega e^{\frac{2 i \pi  (3t-2s)}{9 n}} ~& \omega^2 e^{\frac{2 i \pi  (3t-2s)}{9 n}} \\
 e^{\frac{2 i \pi  (3t-s)}{9 n}} &~ \omega^2 e^{\frac{2 i \pi  (3t-s)}{9 n}} ~& \omega e^{\frac{2 i \pi  (3t-s)}{9 n}} \\
 1 &~ 1 ~& 1 \\
\end{array}
\right)$  &  --- & $abc^{-t+3n\tau}d^{-t},$  \\[-0.16in]

 &    &    & $a^{2}bc^{2t-s+3n\tau}\}$  \\

&   &   &   \\ [-0.16in]\hline
 &   &   &    \\ [-0.16in]

$\langle abc^{s}d^{t}\rangle$  &   $U^{(4)}_l=\frac{1}{\sqrt{2}}
\begin{pmatrix}
 e^{\frac{i \pi  (t-s)}{3 n}} & 0 & -e^{\frac{i \pi  (t-s)}{3 n}} \\
 0 & \sqrt{2} & 0 \\
 1 & 0 & 1 \\
\end{pmatrix}$  &  $s-3t\neq0,3n,6n$  & $\{c^{\gamma}d^{\gamma+s-t},abc^{\gamma}d^{\gamma}\}$ \\

&   &   &   \\ [-0.16in]\hline
 &   &   &    \\ [-0.16in]

$\langle a^2bc^{s}d^{t}\rangle$  &   $U^{(5)}_l=\frac{1}{\sqrt{2}}
\begin{pmatrix}
 -e^{-\frac{i \pi  t}{3 n}} &~ e^{-\frac{i \pi  t}{3 n}} ~& 0 \\
 1 &~ 1 ~& 0 \\
 0 &~ 0 ~& \sqrt{2} \\
\end{pmatrix}$  &  $2s-3t\neq0,3n,6n$ & $\{c^{\gamma}d^{-t},a^{2}bc^{\gamma}\}$  \\
&   &  &    \\ [-0.16in]
\hline\hline
\end{tabular}
\caption{\label{tab:cle_diagonal_matrix}  The form of $U_{l}$ for different residual subgroup $G_{l}$ generated by a single element $g$, and here we denote $G_{l}=\langle g\rangle$. $H^{l}_{CP}$ is the residual CP transformations consistent with $G_l$. The allowed values of the parameters $s$, $t$, $\gamma$, $\delta$ and $\tau$ are   $t,\delta=0,1,...3n-1$, $s,\gamma=0,1,...,9n-1$ and $\tau=0,1,2$. The parameter $\omega$ is the cube root of unit with $\omega=e^{2\pi i/3}$. Note that because $\left(ac^{2s-3t}d^{s-t}\right)^2=a^2c^sd^t$ holds, the $U_{l}$ for $G_{l}=\langle a^2c^{s}d^{t}\rangle$ can be obtained from the that corresponding to $G_{l}=\langle ac^{s}d^{t}\rangle$ by the replacement $s\rightarrow 2s-3t$ and $t\rightarrow s-t$. The constraints on the parameters $s$ and $t$ is to remove the degeneracy among the eigenvalues. }
\end{table}

\begin{table}[t!]
\begin{center}
\renewcommand\arraystretch{1.3}
\begin{tabular}{|c|c|c|c|c|}
\hline\hline
$\mathcal{G}_1$ & $\mathcal{G}_2$ & \texttt{Constraints on group parameters}  & \texttt{Form of} $U_{l}$  & $H^{l}_{CP}$ \\\hline
$\langle c^sd^t\rangle$   & $\langle c^{s^{\prime}}d^{t^{\prime}}\rangle$                   & $\begin{array}{l}
\left\{\begin{array}{c}
s-2t=0\pmod{3n}\\
s^{\prime}-t^{\prime}=0\pmod{3n}
\end{array}\right.\\[0.15in]
\hskip-0.16in\text{or} \left\{\begin{array}{c}
s-2t=0\pmod{3n}\\
t^{\prime}=0\pmod{3n}
\end{array}\right.\\[0.15in]
\hskip-0.16in\text{or}
\left\{\begin{array}{c}
s-t=0\pmod{3n}\\
t^{\prime}=0\pmod{3n}
\end{array}\right.\\[0.1in]
\qquad\quad(s\leftrightarrow s^{\prime},~~ t\leftrightarrow t^{\prime})
\end{array}$  &   $U^{(1)}_{l}$  & $\{c^\gamma d^\delta  \}$ \\\hline

\multirow{2}{*}[-12pt]{$\langle bc^sd^t\rangle$}  &   $\langle c^{s^{\prime}}d^{t^{\prime}}\rangle$                  & $\begin{array}{c}
s^{\prime}-2t^{\prime}=0\pmod{3n}\\
s=0\pmod{3n}
\end{array}$  &  \multirow{2}{*}[-12pt]{  $U^{(2)}_{l}$} &
\multirow{2}{*}[-12pt]{$\begin{array}{l} \{c^{2t+2\delta+3n\tau}d^\delta,\\ bc^{2\delta+3n\tau}d^\delta\} \end{array}$} \\ \cline{2-3}

   &   $\langle bc^{s^{\prime}}d^{t^{\prime}}\rangle$                  &  $\begin{array}{c}
(s-s^{\prime})-2(t-t^{\prime})=3l_1n\pmod{6n}\\
s=3l_2n\pmod{6n},\quad s^{\prime}=3l_3n\pmod{6n}
\end{array}$ & &  \\\hline

\multirow{2}{*}[-22pt]{$\langle abc^sd^t\rangle$}   & $\langle c^{s^{\prime}}d^{t^{\prime}}\rangle$                & $\begin{array}{c}
s^{\prime}-t^{\prime}=0\pmod{3n}\\
3t-s=0\pmod{3n}
\end{array}$ & \multirow{2}{*}[-22pt]{  $U^{(4)}_{l}$}
& \multirow{2}{*}[-12pt]{$\begin{array}{l} \{c^{\gamma}d^{\gamma+2t},\\ abc^{\gamma}d^\gamma\} \end{array}$}  \\\cline{2-3}

  &   $\langle abc^{s^{\prime}}d^{t^{\prime}}\rangle$  & $\begin{array}{c}
(s-s^{\prime})-(t-t^{\prime})=3l_1n\pmod{6n}\\
3t-s=3l_2n\pmod{6n}\\
3t^{\prime}-s^{\prime}=3l_3n\pmod{6n}
\end{array}$  & & \\\hline

\multirow{2}{*}[-22pt]{$\langle a^2bc^sd^t\rangle$}   & $\langle c^{s^{\prime}}d^{t^{\prime}}\rangle$     & $\begin{array}{c}
t^{\prime}=0\pmod{3n}\\
2s-3t=0\pmod{3n}
\end{array}$ &  \multirow{2}{*}[-22pt]{ $U^{(5)}_{l}$}
& \multirow{2}{*}[-12pt]{$\begin{array}{l} \{c^{\gamma}d^{-t},\\ a^2bc^{\gamma}\} \end{array}$}  \\\cline{2-3}

   &   $\langle a^2bc^{s^{\prime}}d^{t^{\prime}}\rangle$               & $\begin{array}{c}
t-t^{\prime}=3l_1n\pmod{6n}\\
2s-3t=3l_2n\pmod{6n}\\
2s^{\prime}-3t^{\prime}=3l_3n\pmod{6n}
\end{array}$ & &  \\\hline\hline
\end{tabular}
\end{center}
\caption{\label{tab:extension_Gch}The product extension of the remnant flavor symmetry $G_{l}=G_1\times G_2$, the remnant CP transformation compatible with $G_{l}$, and the corresponding unitary transformation $U_{l}$. We require the column vectors fixed by $\mathcal{G}_1$ and $\mathcal{G}_2$ be different. Consequently we have the parameters $l_{1,2,3}=0,1$ and $l_1+l_2+l_3=1,3$. The values of parameters $s$, $t$, $s^\prime$, $t^\prime$, $\gamma$, $\delta$ and $\tau$ are $s, s^{\prime},\gamma=0,1,\cdots,9n-1$, $t,t^\prime,\delta=0,1,\cdots,3n-1$ and $\tau=0, 1, 2$.}
\end{table}

\begin{table}[t!]
\renewcommand{\tabcolsep}{2.0mm}
\centering
\begin{tabular}{|c|c|}
\hline \hline
    &    \\ [-0.16in]
$G_{\nu}$ &  $X_{\nu}$   \\

  &       \\ [-0.16in]\hline
 &       \\ [-0.16in]

$K^{(c^{9n/2},d^{3n/2})}_4$  &  $\rho_{\mathbf{r}}(c^{\gamma}d^{\delta})$  \\
  &        \\ [-0.16in]\hline
  &       \\ [-0.16in]

$K^{(d^{3n/2},bd^{x})}_4$   & $\rho_{\mathbf{r}}(c^{2\delta+2x+3n\tau}d^{\delta}),
\rho_{\mathbf{r}}(bc^{2\delta+3n\tau}d^{\delta})$  \\

  &        \\ [-0.16in]\hline
  &       \\ [-0.16in]

$K^{(c^{9n/2}d^{3n/2},abc^{3y}d^{y})}_4$  & $\rho_{\mathbf{r}}(c^{\delta-2y-3n\tau}d^{\delta}),\rho_{\mathbf{r}}(abc^{\delta-3n\tau}d^{\delta})$  \\

  &        \\ [-0.16in]\hline
  &       \\ [-0.16in]

$K^{(c^{9n/2},a^2bc^{3z}d^{2z})}_4$   & $\rho_{\mathbf{r}}(c^{\gamma}d^{-2z}),\rho_{\mathbf{r}}(a^2bc^{\gamma})$  \\ \hline\hline

\end{tabular}
\caption{\label{tab:K4_RCP} The $K_{4}$ subgroups of the $D^{(1)}_{9n, 3n}$ group and eligible remnant CP transformations, where the superscript of the $K_4$ subgroup denotes its generators. The allowed values of the parameters are $\gamma=0, 1, \ldots, 9n-1$, $x, y, z, \delta=0, 1, \ldots, 3n-1$, and $\tau=0, 1, 2$.}
\end{table}

In the direct approach, the flavor symmetry group $D^{(1)}_{9n, 3n}$ group is broken down to a Klein four subgroup in the neutrino sector. From Appendix~\ref{app:group_theory}, we see that $D^{(1)}_{9n, 3n}$ for even $n$ has only four Klein four subgroups:
\begin{eqnarray}
\nonumber&&K^{(c^{9n/2},d^{3n/2})}_4\equiv\left\{1, c^{9n/2}, d^{3n/2}, c^{9n/2}d^{3n/2}\right\},\quad K^{(d^{3n/2},bd^{x})}_4\equiv\left\{1, d^{3n/2}, bd^{x}, bd^{x+3n/2} \right\},\\
\nonumber&&K^{(c^{9n/2}d^{3n/2}, abc^{3y}d^{y})}_4\equiv\left\{1, c^{9n/2}d^{3n/2}, abc^{3y}d^{y}, abc^{3y+9n/2}d^{y+3n/2}\right\},\\
\label{eq:K4_conjugate}&&K^{(c^{9n/2}, a^2bc^{3z}d^{2z})}_4\equiv\left\{1, c^{9n/2}, a^2bc^{3z}d^{2z}, a^2bc^{3z+9n/2}d^{2z}\right\}\,,
\end{eqnarray}
where $x, y, z=0, 1, \ldots, 3n-1$. We note that $K^{(c^{9n/2},d^{3n/2})}_4$ is a normal subgroup of $D^{(1)}_{9n, 3n}$, and the remaining three $K_4$ subgroups are conjugate:
\begin{eqnarray}
\nonumber&&(a^{2}c^{y-x+2\delta}d^{\delta})K^{(d^{3n/2},bd^{x})}_4(a^{2}c^{y-x+2\delta}d^{\delta})^{-1}=K^{(c^{9n/2}d^{3n/2}, abc^{3y}d^{y})}_4,\\
&&(ac^{-z-x+2\delta}d^{\delta})K^{(d^{3n/2},bd^{x})}_4(ac^{-z-x+2\delta}d^{\delta})^{-1}=K^{(c^{9n/2}, a^2bc^{3z}d^{2z})}_4\,,
\end{eqnarray}
with $\delta=0, 1, \ldots, 3n-1$. Furthermore, the residual CP symmetry $H^{\nu}_{CP}$ in the neutrino sector has to be compatible with the remnant $K_4$ symmetry, and the following restricted consistency condition must be fulfilled,
\begin{equation}
X_{\nu\mathbf{r}}\rho^{\ast}_{\mathbf{r}}(g)X^{-1}_{\nu\mathbf{r}}=\rho_{\mathbf{r}}(g),\qquad
g\in K_4\,.
\end{equation}
Solving this equation, we can straightforwardly find the eligible remnant CP transformations for different $K_4$ subgroups. The results are collected in table~\ref{tab:K4_RCP}. Then we proceed to determine the neutrino mass matrix $m_{\nu}$ invariant under the action of both remnant CP and remnant flavor symmetry for each case, i.e., $m_{\nu}$ is subject to the constraints in Eq.~\eqref{eq:constr_nu}.
\begin{itemize}[labelindent=-0.7em, leftmargin=1.2em]

\item{$G_{\nu}=K^{(c^{9n/2}, d^{3n/2})}_4$, $X_{\nu\mathbf{r}}=\rho_{\mathbf{r}}(c^{\gamma}d^{\delta})$}

In our working basis, the representation matrices for both $a$ and $c$ are diagonal with
\begin{equation}
\rho_{\mathbf{3}}(c^{9n/2})=\left(\begin{array}{ccc}
-1  &   0  &  0 \\
0   &  -1  &  0 \\
0   &  0   &  1
\end{array}\right),\quad \rho_{\mathbf{3}}(d^{3n/2})=\left(\begin{array}{ccc}
1  &   0  &  0 \\
0   &  -1  &  0 \\
0   &  0   &  -1
\end{array}\right)\,.
\end{equation}
Consequently the residual flavor symmetry enforces the neutrino mass matrix to be diagonal as well. Taking into account the remnant CP symmetry further, we find
\begin{equation}
m_{\nu}=\begin{pmatrix}
m_{11} e^{-2i\pi\frac{\gamma}{9n}} & 0 & 0 \\
 0 & m_{22}e^{-2i\pi\frac{\gamma-3\delta}{9n}} & 0 \\
 0 & 0 & m_{33}e^{2i\pi\frac{2\gamma-3\delta}{9n}}
\end{pmatrix}\,,
\end{equation}
where $m_{11}$, $m_{22}$ and $m_{33}$ are real parameters. We can read out the neutrino diagonalization matrix $U_{\nu}$ as
\begin{equation}\label{eq:unu_K41}
U_{\nu}=\text{diag}\left(e^{i\pi\frac{\gamma}{9n}},e^{i\pi\frac{\gamma-3\delta}{9n}}, e^{-i\pi\frac{2\gamma-3\delta}{9 n}}\right)Q_{\nu}\,,
\end{equation}
where $Q_{\nu}$ is a diagonal phase matrix with entry being $\pm1$ or $\pm i$, and it encodes the CP parity of the neutrino states.
The light neutrino mass eigenvalues are
\begin{equation}
m_1=\left|m_{11}\right|,\qquad m_2=\left|m_{22}\right|,\qquad
m_3=\left|m_{33}\right|\,.
\end{equation}
Obviously the light neutrino masses depend on only three real parameters, and the order of the light neutrino masses can not be fixed by remnant symmetries. Therefore the unitary transformation $U_{\nu}$ is determined up to independent row and column permutations in the present framework, and the neutrino mass spectrum can be wither normal ordering (NO) or inverted ordering (IO).

\item{$G_{\nu}=K^{(d^{3n/2},bd^{x})}_4$, $X_{\nu\mathbf{r}}=\left\{\rho_{\mathbf{r}}(c^{2\delta+2x+3n\tau}d^{\delta}),\rho_{\mathbf{r}}(bc^{2\delta+3n\tau}d^{\delta})\right\}$}

In the same fashion as previous case, we find that the light neutrino mass matrix takes the following form:
\begin{equation}
m_{\nu}=\begin{pmatrix}
m_{11}e^{-2i\pi\frac{2x+2\delta+3n\tau}{9n}}   &  0   &~   0 \\
 0  &   m_{22}e^{-2i\pi\frac{2x-\delta+3n\tau}{9 n}}   &~ m_{23}e^{i\pi\frac{2x+2\delta+3n\tau}{9n}} \\
 0 & m_{23}e^{i\pi\frac{2x+2\delta+3n\tau}{9n}}   &~   m_{22}e^{2i\pi\frac{4x+\delta-3n\tau}{9 n}} \\
\end{pmatrix}\,,
\end{equation}
where $m_{11}$, $m_{22}$ and $m_{23}$ are real. It is diagonalized by the
unitary transformation $U_{\nu}$ with
\begin{equation}\label{eq:unu_abcy_2}
U_{\nu}=\frac{1}{\sqrt{2}}
\begin{pmatrix}
\sqrt{2}e^{i\pi\frac{2x+2\delta+3n\tau}{9n}}   &   0   &~  0 \\
 0    & e^{i\pi\frac{2x-\delta+3n\tau}{9n}}   &~   -e^{i\pi\frac{2x-\delta+3n\tau}{9n}} \\
 0    &    e^{-i\pi\frac{4x+\delta-3n\tau}{9n}}   &~   e^{-i\pi\frac{4x+\delta-3n\tau}{9n}}
\end{pmatrix}\,,
\end{equation}
where the matrix $Q_{\nu}$ is omitted for simplicity and we will also not explicitly write out this factor hereafter. The light neutrino masses are
\begin{equation}
m_1=\left|m_{11}\right|,\qquad
m_2=\left|m_{22}+(-1)^{\tau}m_{23}\right|,\qquad m_3=\left|m_{22}-(-1)^{\tau}m_{23}\right|\,.
\end{equation}

\item{$G_{\nu}=K^{(c^{9n/2}d^{3n/2},abc^{3y}d^{y})}_4$, $X_{\nu\mathbf{r}}=\left\{\rho_{\mathbf{r}}(c^{\delta-2y-3n\tau}d^{\delta}),\rho_{\mathbf{r}}(abc^{\delta-3n\tau}d^{\delta})\right\}$}

In this case, we find that the light neutrino mass matrix takes the form
\begin{equation}
m_{\nu}=\begin{pmatrix}
m_{11}e^{2i\pi\frac{2y-\delta+3n\tau}{9n}}  & 0 &   m_{13}e^{-i\pi\frac{2y+2\delta+3n\tau}{9n}} \\
 0   & m_{22}e^{2i\pi\frac{2y+2\delta+3n\tau}{9n}} & 0 \\
m_{13}e^{-i\pi\frac{2y+2\delta+3n\tau}{9n}}   &   0   &  m_{11}e^{-2i\pi\frac{4y+\delta+6n\tau}{9n}}
\end{pmatrix}\,,
\end{equation}
where $m_{11}$, $m_{13}$ and $m_{22}$ are real. Consequently the unitary transformation $U_{\nu}$ is
\begin{equation}
U_{\nu}=\frac{1}{\sqrt{2}}\left(
\begin{array}{ccc}
 -e^{i\pi\frac{-2y+\delta+6 n \tau }{9 n}} & 0 & e^{i \pi \frac{-2 y+\delta +6 n \tau }{9 n}} \\
 0 &   \sqrt{2} e^{-i\pi\frac{2y+2\delta+3n\tau}{9 n}} & 0 \\
 e^{i\pi\frac{4y+\delta+6n\tau}{9n}} & 0 & e^{i \pi \frac{4y+\delta+6n\tau}{9n}}
\end{array}
\right)\,.
\end{equation}
The light neutrino masses are
\begin{equation}
m_1=\left|m_{11}-(-1)^{\tau}m_{13}\right|,\qquad  m_2=\left|m_{22}\right|,\qquad
m_3=\left|m_{11}+(-1)^{\tau}m_{13}\right|\,.
\end{equation}

\item{$G_{\nu}=K^{(c^{9n/2},a^2 b c^{3z}d^{2z})}_4$, $X_{\nu\mathbf{r}}=\left\{\rho_{\mathbf{r}}(c^{\gamma}d^{-2z}),\rho_{\mathbf{r}}(a^2bc^{\gamma})\right\}$}

The light neutrino mass matrix $m_{\nu}$ is constrained by the remnant symmetry to be of the form
\begin{equation}
\label{eq:mnu_k4_4}m_{\nu}=\left(
\begin{array}{ccc}
 m_{11} e^{-2i \pi \frac{ \gamma }{9 n}}   &~ m_{12} e^{-2i \pi\frac{3z+\gamma}{9n}} & 0 \\
 m_{12} e^{-2i\pi\frac{3z+\gamma}{9n}} &~  m_{11} e^{-2i\pi\frac{6z+\gamma}{9n}} & 0 \\
 0 &~ 0 & m_{33}e^{4i\pi\frac{3z+\gamma}{9n}}
\end{array}
\right)\,,
\end{equation}
where $m_{11}$, $m_{12}$ and $m_{33}$ are real. The unitary matrix $U_{\nu}$ diagonalizing the above neutrino mass matrix is determined to be
\begin{equation}
\label{eq:unu_k4_4}
U_{\nu}=\frac{1}{\sqrt{2}}
\left(\begin{array}{ccc}
 -e^{i\pi\frac{\gamma}{9n}} &~   e^{i\pi\frac{\gamma}{9 n}} & 0 \\
 e^{i\pi\frac{6z+\gamma}{9n}} &~ e^{i\pi\frac{6z+\gamma}{9n}} & 0 \\
 0 & 0 & \sqrt{2} e^{-2 i \pi \frac{ 3 z+\gamma }{9 n}}
\end{array}
\right)\,.
\end{equation}
The neutrino masses are given by
\begin{equation}
m_1=\left|m_{11}-m_{12}\right|,\qquad  m_2=\left|m_{11}+m_{22}\right|,\qquad
m_3=\left|m_{33}\right|\,.
\end{equation}
\end{itemize}

Then we proceed to discuss the possible mixing patterns achievable in direct approach by combining the different remnant symmetries of the charged lepton sector with those of the neutrino sector. As shown in section~\ref{sec:framework}, two pairs of subgroups $\left\{G_{l}, G_{\nu}\right\}$ and $\left\{G^{\prime}_{l}, G^{\prime}_{\nu}\right\}$ would yield the same results for the PMNS matrix after considering all the eligible residual CP transformations, if these two pairs of groups are conjugate. Notice the conjugate relations between distinct $K_{4}$ subgroups in Eq.~\eqref{eq:K4_conjugate} and the identities $(bc^{2\epsilon}d^{2\epsilon})(abc^sd^t)(bc^{2\epsilon}d^{2\epsilon})^{-1}=a^2bc^sd^{s-t}$, $(bc^{2\epsilon}d^{2\epsilon})K^{(c^{9n/2}, d^{3n/2})}_{4}(bc^{2\epsilon}d^{2\epsilon})^{-1}=K^{(c^{9n/2}, d^{3n/2})}_{4}$ and $(bc^{2\epsilon}d^{2\epsilon})K^{(d^{3n/2},bd^x)}_{4}(bc^{2\epsilon}d^{2\epsilon})^{-1}=K^{(d^{3n/2},bd^x)}_{4}$ for any integer $\epsilon$, we find it is sufficient to only consider eight kinds of remnant symmetries with $G_{l}=\langle c^{s}d^{t}\rangle$, $\langle bc^{s}d^{t}\rangle$, $\langle ac^{s}d^{t}\rangle$, $\langle abc^{s}d^{t}\rangle$ and $G_{\nu}=K^{(c^{9n/2},d^{3n/2})}_4$, $K^{(d^{3n/2},bd^x)}_4$. In this scenarios, all mixing parameters including Majorana phases are completely fixed by remnant symmetries.

\begin{description}[labelindent=-0.8em, leftmargin=0.3em]

\item[~~(\romannumeral1)]{$G_{l}=\langle c^{s}d^{t}\rangle$, $G_{\nu}=K^{(c^{9n/2},d^{3n/2})}_4$, $X_{\nu\bf{r}}=\{\rho_{\bf{r}}(c^{\gamma}d^{\delta})\}$}  \\
In this case, the unitary transformation $U_{l}$ is a unit matrix, as shown in table~\ref{tab:cle_diagonal_matrix}. $U_{\nu}$ is a diagonal phase matrix and it is given by Eq.~\eqref{eq:unu_K41}. As a result, the PMNS matrix is also a diagonal matrix up to row and column permutations, and obviously it doesn't agree with the present neutrino oscillation data~\cite{Capozzi:2013csa,Forero:2014bxa,Gonzalez-Garcia:2014bfa}.

\item[~~(\romannumeral2)]{$G_{l}=\langle bc^{s}d^{t}\rangle$, $G_{\nu}=K^{(c^{9n/2},d^{3n/2})}_4$, $X_{\nu\bf{r}}=\{\rho_{\bf{r}}(c^{\gamma}d^{\delta})\}$}  \\
In this case, the postulated residual subgroups lead to the mixing pattern    \begin{equation}
U_{\text{PMNS}}=\frac{1}{\sqrt{2}}
\begin{pmatrix}
 \sqrt{2} &~ 0 ~& 0 \\
 0 &~ 1 ~& -e^{i \varphi_{1}} \\
 0 &~ 1 ~& e^{i \varphi_{1}}
\end{pmatrix}\,,
\end{equation}
with
\begin{equation}
\varphi_{1}=-\frac{\pi(\gamma-2\delta+s-2t)}{3n}\,.
\end{equation}
The lepton mixing angles are $\theta_{13}=\theta_{12}=0$, $\theta_{23}=45^\circ$, and therefore large corrections to both $\theta_{12}$ and $\theta_{13}$ are necessary in order to be compatible with experimental data.

\item[~~(\romannumeral3)]{$G_{l}=\langle ac^{s}d^{t}\rangle$, $G_{\nu}=K^{(c^{9n/2},d^{3n/2})}_4$, $X_{\nu\bf{r}}=\{\rho_{\bf{r}}(c^{\gamma}d^{\delta})\}$}  \\
This residual symmetry allows us to pin down the lepton mixing matrix as:
\begin{equation}
U_{\text{PMNS}}=\frac{1}{\sqrt{3}}
\begin{pmatrix}
 e^{i\varphi_1} &~ 1 ~& e^{i\varphi_2} \\
 \omega e^{i\varphi_1}   &~ 1 ~& \omega^2e^{i\varphi_2}  \\
 \omega^2e^{i\varphi_1}  &~ 1 ~& \omega e^{i\varphi_2}
\end{pmatrix}\,,
\end{equation}
where
\begin{equation}
\varphi_{1}=\frac{\pi(3\gamma-3\delta+2 s)}{9n}, \qquad \varphi_{2}=\frac{\pi  (3 \gamma -6 \delta +4 s-6 t)}{9 n}\,.
\end{equation}
This pattern leads to $\sin^2\theta_{12}=\sin^2\theta_{23}=1/2$, $\sin^2\theta_{13}=1/3$ and a maximal Dirac CP phase $|\delta_{CP}|=\pi/2$. The solar as well as the reactor mixing angle have to acquire appropriate corrections in order to be in accordance with experimental data.

\item[~~(\romannumeral4)]{$G_{l}=\langle abc^{s}d^{t}\rangle$, $G_{\nu}=K^{(c^{9n/2},d^{3n/2})}_4$, $X_{\nu\bf{r}}=\{\rho_{\bf{r}}(c^{\gamma}d^{\delta})\}$}  \\
In this case we find the lepton mixing matrix is
\begin{equation}
U_{\text{PMNS}}=\frac{1}{\sqrt{2}}
\begin{pmatrix}
 \sqrt{2} &~ 0 ~& 0 \\
 0 &~ 1 ~& e^{i\varphi_1} \\
 0 &~ -1 ~& e^{i\varphi_1}
\end{pmatrix}\,, \quad \mathrm{with} \quad \varphi_{1}=-\frac{\pi(\gamma-\delta+s-t)}{3n}\,,
\end{equation}
which leads to $\theta_{12}=\theta_{13}=0$, $\theta_{23}=45^\circ$. Large corrections to $\theta_{12}$ and $\theta_{13}$ are needed to be compatible with experimental data.

\item[~~(\romannumeral5)]{$G_{l}=\langle c^{s}d^{t}\rangle$, $G_{\nu}=K^{(d^{3n/2},bd^x)}_4$, $X_{\nu\bf{r}}=\{\rho_{\bf{r}}(c^{2\delta+2x+3n\tau}d^{\delta}),\rho_{\bf{r}}(bc^{2\delta+3n\tau}d^{\delta})\}$}\\
The unitary transformation $U_{\nu}$ is fixed by residual subgroup to be Eq.~\eqref{eq:unu_abcy_2}, and the PMNS matrix takes the form
\begin{equation}
U_{\text{PMNS}}=
\begin{pmatrix}
 1 &~ 0 ~& 0 \\
 0 &~ \frac{1}{\sqrt{2}} ~& -\frac{1}{\sqrt{2}} \\
 0 &~ \frac{1}{\sqrt{2}} ~& \frac{1}{\sqrt{2}}
\end{pmatrix}\,,
\end{equation}
which lead to $\theta_{12}=\theta_{13}=0$, $\theta_{23}=45^\circ$. Again $\theta_{12}$ and $\theta_{13}$ require large corrections in order to be in the experimentally preferred range.

\item[~~(\romannumeral6)]{$G_{l}=\langle bc^{s}d^{t}\rangle$, $G_{\nu}=K^{(d^{3n/2},bd^x)}_4$, $X_{\nu\bf{r}}=\{\rho_{\bf{r}}(c^{2\delta+2x+3n\tau}d^{\delta}),\rho_{\bf{r}}(bc^{2\delta+3n\tau}d^{\delta})\}$}\\
Using these residual symmetries, we can derive the lepton mixing matrix
\begin{equation}
U_{\text{PMNS}}=
\begin{pmatrix}
 1 &~ 0 ~& 0 \\
 0 &~ -\sin \varphi_1 ~& \cos \varphi_1 \\
 0 &~ \cos \varphi_1 ~& \sin\varphi_1
\end{pmatrix}\,, \quad \mathrm{with} \quad \varphi_1=\frac{\pi(s-2t+2x)}{6n}\,.
\end{equation}
The mixing angles are $\theta_{12}=\theta_{13}=0$, $\sin^2\theta_{23}=\cos^2\varphi_1$ which is strongly disfavored by the experimental data~\cite{Capozzi:2013csa,Forero:2014bxa,Gonzalez-Garcia:2014bfa}.

\item[~~(\romannumeral7)]{$G_{l}=\langle ac^{s}d^{t}\rangle$, $G_{\nu}=K^{(d^{3n/2},bd^x)}_4$, $X_{\nu\bf{r}}=\{\rho_{\bf{r}}(c^{2\delta+2x+3n\tau}d^{\delta}),\rho_{\bf{r}}(bc^{2\delta+3n\tau}d^{\delta})\}$}\\
In this case, the lepton mixing matrix is determined to be of the trimaximal form, i.e., the second column of the PMNS matrix is $(1, 1, 1)^{T}/\sqrt{3}$ with
\begin{equation}
\label{eq:acsdt_K4}
U_{\text{PMNS}}=\frac{1}{\sqrt{3}}
\begin{pmatrix}
 -\sqrt{2} e^{i\varphi_2}\cos\varphi_1 &~ 1 ~& \sqrt{2}e^{i\varphi_2}\sin\varphi_1 \\
 \sqrt{2} e^{i\varphi_2} \sin \left(\frac{\pi }{6}+\varphi_1\right) &~ 1 ~& \sqrt{2} e^{i\varphi_2} \cos \left(\frac{\pi }{6}+\varphi_1\right) \\
 \sqrt{2} e^{i\varphi_2} \sin \left(\frac{\pi }{6}-\varphi_1\right) &~ 1 ~& -\sqrt{2} e^{i\varphi_2} \cos \left(\frac{\pi }{6}-\varphi_1\right) \\
\end{pmatrix}\,,
\end{equation}
where
\begin{equation}
\label{eq:varphi_1_2_values}\varphi_1=\frac{2s-3t+3x}{9n}\pi,\qquad
\varphi_2=-\frac{\delta+t+x}{3n}\pi\,.
\end{equation}
These two parameters $\varphi_1$ and $\varphi_2$ are independent from each other, and they can take the following discrete values
\begin{eqnarray}
\nonumber&&\varphi_1~(\textrm{mod}~2\pi)=0, \frac{1}{9n}\pi, \frac{2}{9n}\pi, \ldots, \frac{18n-1}{9n}\pi\,,\\
&&\varphi_2~(\textrm{mod}~2\pi)=0, \frac{1}{3n}\pi, \frac{2}{3n}\pi, \ldots, \frac{6n-1}{3n}\pi\,.
\end{eqnarray}
We can read out the mixing angles as
\begin{eqnarray}
\nonumber &&  \sin^2\theta_{13}=\frac{2}{3}\sin^2\varphi_1, \quad
\sin^2\theta_{12}=\frac{1}{2+\cos2\varphi_{1}},\quad
 \sin^2\theta_{23}=\frac{1+\cos(\frac{\pi}{6}+2\varphi_1)}{2+\cos2\varphi_1}\,.
\end{eqnarray}
All possible predictions of $\sin\theta_{13}$ for each $D^{(1)}_{9n, 3n}$ of even $n$ are displayed in figure~\ref{fig:theta13_alpha21}. It is remarkable that viable reactor mixing angle $\theta_{13}$ can always be achieved for each $n$. Moreover, the three mixing angles are closely related as follows
\begin{equation}
3\sin^2\theta_{12}\cos^2\theta_{13}=1,\quad \sin^2\theta_{23}=\frac{1}{2}\pm\frac{1}{2}\tan\theta_{13}\sqrt{2-\tan^2\theta_{13}}\,.
\end{equation}
Inputting the experimentally preferred $3\sigma$ range $0.0176\leq\sin^2\theta_{13}\leq0.0295$~\cite{Capozzi:2013csa}, we obtain predictions for solar as well as atmospheric mixing angles:
\begin{equation}
0.339\leq\sin^2\theta_{12}\leq0.343,\qquad 0.378\leq\sin^2\theta_{23}\leq0.406,~~\mathrm{or}~~0.594\leq\sin^2\theta_{23}\leq0.622\,.
\end{equation}
From the PMNS matrix of Eq.~\eqref{eq:acsdt_K4}, we can also extract the CP violating phases
\begin{eqnarray}
\sin\delta_{CP}=\sin\alpha_{31}=0, \quad \tan\alpha_{21}=-\tan2\varphi_2\,,
\end{eqnarray}
where the contribution of the CP parity matrix $Q_{\nu}$ is considered. We see that both Dirac phase $\delta_{CP}$ and the Majorana phase $\alpha_{31}$ are trivial, and another Majorana phase $\alpha_{21}$ is
\begin{equation}
\alpha_{21}=-2\varphi_{2}~~\mathrm{or}~~ \alpha_{21}=\pi-2\varphi_{2}\,.
\end{equation}
The admissible values of $\alpha_{21}$ are
\begin{equation}
\alpha_{21}=0, \frac{2}{3n}\pi, \frac{4}{3n}\pi, \ldots, \frac{6n-2}{3n}\pi\,,
\end{equation}
which are plotted in figure~\ref{fig:theta13_alpha21}. Note that here the predictions for the CP phases are consistent with the general results of Ref.~\cite{Chen:2015nha}.
\begin{figure}[t!]
\begin{center}
\subfigure{\includegraphics[width=0.9\linewidth]{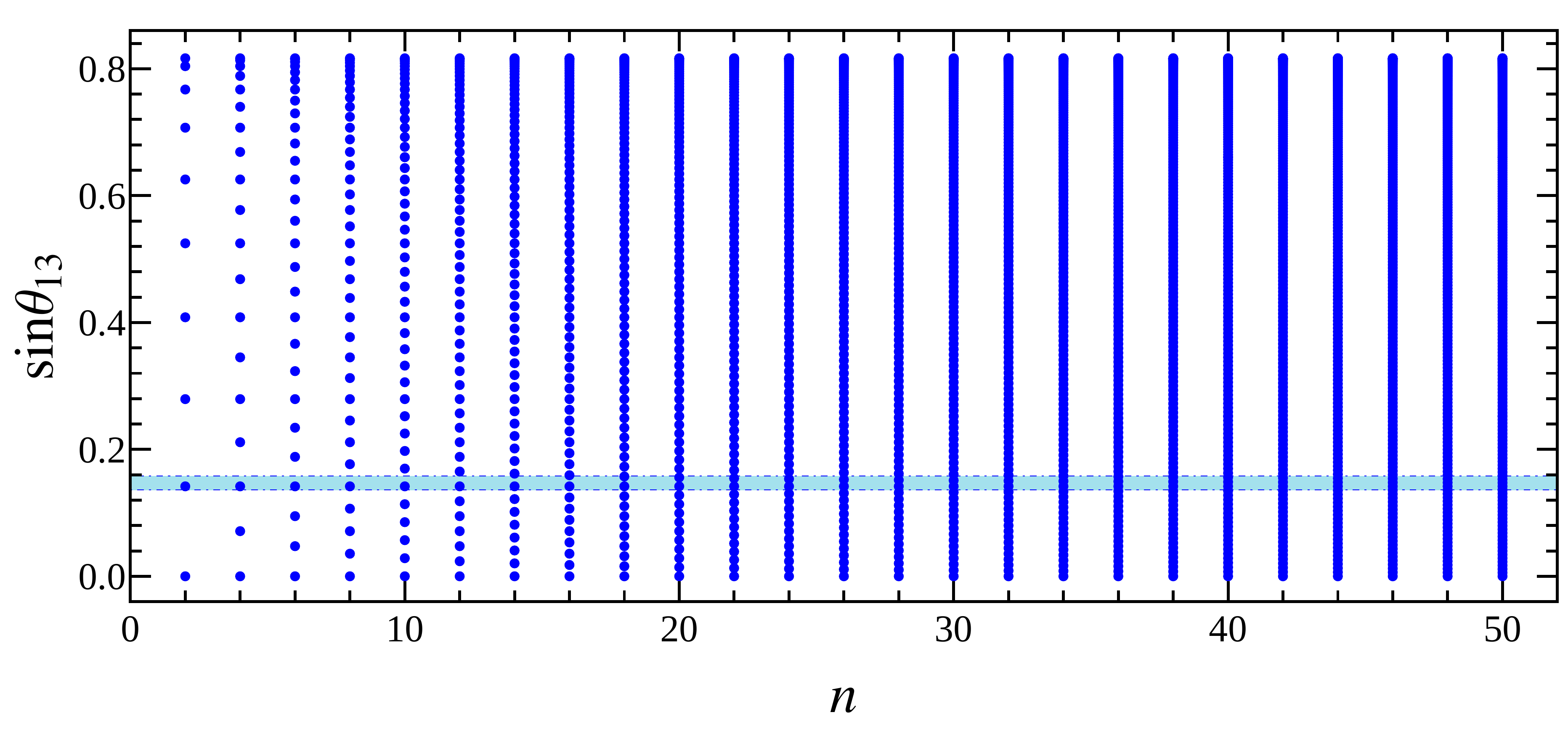}}
\subfigure{\includegraphics[width=0.9\linewidth]{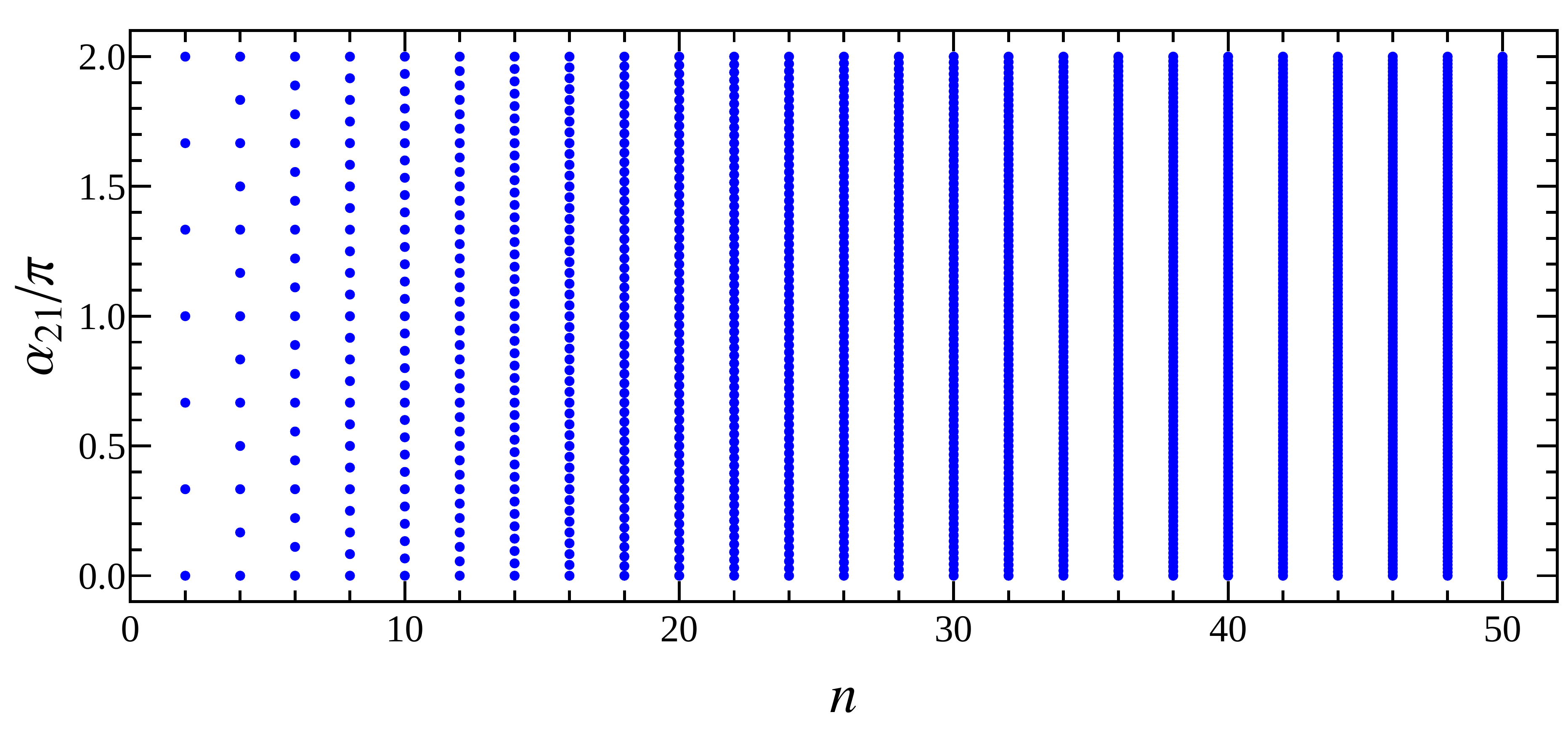}}
\caption{\label{fig:theta13_alpha21} The possible values of the mixing angle $\sin\theta_{13}$ and the Majorana phase $\alpha_{21}$ for each $D^{(1)}_{9n, 3n}$ group with even $n$ when the remnant symmetries are
$G_{l}=\langle ac^{s}d^{t}\rangle$, $G_{\nu}=K^{(d^{3n/2},bd^x)}_4$, $X_{\nu\bf{r}}=\{\rho_{\bf{r}}(c^{2\delta+2x+3n\tau}d^{\delta}),\rho_{\bf{r}}(bc^{2\delta+3n\tau}d^{\delta})\}$. The light blue region denotes the $3\sigma$ bound of $\sin\theta_{13}$, which is taken from Ref.~\cite{Capozzi:2013csa}.
}
\end{center}
\end{figure}

\item[~~(\romannumeral8)]{$G_{l}=\langle abc^{s}d^{t}\rangle$, $G_{\nu}=K^{(d^{3n/2},bd^x)}_4$, $X_{\nu\bf{r}}=\{\rho_{\bf{r}}(c^{2\delta+2x+3n\tau}d^{\delta}),\rho_{\bf{r}}(bc^{2\delta+3n\tau}d^{\delta})\}$}\\
In this case, we find the lepton mixing matrix is the well-known bimaximal pattern
\begin{equation}
U_{\text{PMNS}}=\frac{1}{2}
\begin{pmatrix}
 -\sqrt{2} &~ \sqrt{2} ~& 0 \\
 1 &~ 1 ~& \sqrt{2} e^{i\varphi_1} \\
 1 &~ 1 ~& -\sqrt{2} e^{i\varphi_1}
\end{pmatrix}\,,
\quad \mathrm{with} \quad \varphi_1=\frac{\pi  (s-2 t+2 x)}{6 n}\,.
\end{equation}
The bimaximal mixing can be a valid first approximation in a model where corrections of order of the Cabibbo angle can naturally arise~\cite{Li:2014eia,Altarelli:2009gn}.

\end{description}

\section{\label{sec:Z2xCP_neutrino}Lepton mixing from semidirect approach}

In the semidirect approach, the original symmetry $D^{(1)}_{9n, 3n}\rtimes H_{CP}$ is broken at low energies into $G_{l}\rtimes H^{l}_{CP}$ in the charged lepton sector and to $Z_2\times H^{\nu}_{CP}$ in the neutrino sector. The PMNS matrix turns out to depend on only a single real parameter in this scenario. It is generally assumed that the residual flavor symmetry $G_{l}$ is able to distinguish the three generations of charged leptons such that the unitary matrix $U_{l}$ can be determined from the requirement that all the generators of $G_{l}$ should be simultaneously diagonalized by $U_{l}$.
The possible candidates for the subgroup $G_{l}$, the remnant CP transformations compatible with $G_{l}$ and the corresponding unitary transformation $U_{l}$ are summarized in table~\ref{tab:cle_diagonal_matrix} and table~\ref{tab:extension_Gch}. Then we turn to the neutrino sector. From the multiplication rules given in Eq.~\eqref{eq:multiplication_rules}, we see that the order 2 elements of the $D^{(1)}_{9n, 3n}$ group are
\begin{equation}
\label{eq:z2_1} bd^{x},\quad abc^{3y}d^{y}, \quad a^{2}bc^{3z}d^{2z},\quad x, y, z=0,1\ldots 3n-1\,,
\end{equation}
and additionally
\begin{equation}
\label{eq:z2_2}c^{9n/2},\quad d^{3n/2},\quad c^{9n/2}d^{3n/2}\,,
\end{equation}
for even $n$. The residual CP transformation $X_{\nu\mathbf{r}}$ is a symmetric unitary matrix, and it should map the element of the neutrino residual flavor symmetry to itself,
\begin{equation}\label{eq:consistency_neutrino}
X_{\nu\mathbf{r}}\rho^{*}_{\mathbf{r}}(g_{\nu})X^{-1}_{\nu\mathbf{r}}=\rho_{\mathbf{r}}(g_{\nu}),\quad
g_{\nu}\in G_{\nu}\,.
\end{equation}
The eligible residual CP transformations for different $Z_2$ subgroups are collected in table~\ref{tab:Z2_RCP}. Furthermore, we notice that all the $Z_2$ elements in Eq.~\eqref{eq:z2_1} are conjugate:
\begin{equation}\label{eq:neutrino_conjugate_1}
\begin{array}{ll}
\left(c^{\gamma}d^{\delta}\right)bd^{x}\left(c^{\gamma}d^{\delta}\right)^{-1}=bd^{x^{\prime}},
~&~ \left(bc^{\gamma}d^{\delta}\right)bd^{x}\left(bc^{\gamma}d^{\delta}\right)^{-1}=bd^{-x^{\prime}}\,,\\
\left(ac^{\gamma}d^{\delta}\right)bd^{x}\left(ac^{\gamma}d^{\delta}\right)^{-1}=a^2bc^{-3 x^{\prime}}d^{-2 x^{\prime}},
~&~ \left(a^2c^{\gamma}d^{\delta}\right)bd^{x}\left(a^2c^{\gamma}d^{\delta}\right)^{-1}=abc^{3 x^{\prime}}d^{x^{\prime}}\,, \\
\left(abc^{\gamma}d^{\delta}\right)bd^{x}\left(abc^{\gamma}d^{\delta}\right)^{-1}=a^2bc^{3 x^{\prime}}d^{2 x^{\prime}},
~&~ \left(a^2bc^{\gamma}d^{\delta}\right)bd^{x}\left(a^2bc^{\gamma}d^{\delta}\right)^{-1}=abc^{-3 x^{\prime}}d^{-x^{\prime}}\,,
\end{array}
\end{equation}
where $x^{\prime}=x+\gamma-2\delta$. Similarly the three elements in Eq.~\eqref{eq:z2_2} are also conjugate to each other:
\begin{equation}\label{eq:neutrino_conjugate_4}
\begin{array}{ll}
\left(c^{\gamma}d^{\delta}\right)c^{9n/2}\left(c^{\gamma}d^{\delta}\right)^{-1}=c^{9n/2},
~&~\left(bc^{\gamma}d^{\delta}\right)c^{9n/2}\left(bc^{\gamma}d^{\delta}\right)^{-1}=c^{9n/2}d^{3n/2}\,,\\
\left(ac^{\gamma}d^{\delta}\right)c^{9n/2}\left(ac^{\gamma}d^{\delta}\right)^{-1}=c^{9n/2}d^{3n/2},
~&~\left(a^2c^{\gamma}d^{\delta}\right)c^{9n/2}\left(a^2c^{\gamma}d^{\delta}\right)^{-1}=d^{3n/2}\,,\\
\left(abc^{\gamma}d^{\delta}\right)c^{n/2}\left(abc^{\gamma}d^{\delta}\right)^{-1}=d^{3n/2},
~&~\left(a^2bc^{\gamma}d^{\delta}\right)c^{9n/2}\left(a^2bc^{\gamma}d^{\delta}\right)^{-1}=c^{9n/2}\,.
\end{array}
\end{equation}
\begin{table}[t!]
\renewcommand{\tabcolsep}{2.0mm}
\centering
\begin{tabular}{|c||c|}
  \hline \hline
    &    \\ [-0.16in]
$G_{\nu}$ &  $X_{\nu}$   \\
 &       \\ [-0.16in]\hline
 &       \\ [-0.16in]
$Z^{bd^x}_2$ & $\rho_{\mathbf{r}}(c^{2\delta+2x+3n\tau}d^{\delta})$, $\rho_{\mathbf{r}}(bc^{2\delta+3n\tau}d^{\delta})$\\
 &       \\ [-0.16in]\hline
 &       \\ [-0.16in]
$Z^{abc^{3y}d^{y}}_2$ & $\rho_{\mathbf{r}}(c^{\delta-2y-3n\tau}d^{\delta})$, $\rho_{\mathbf{r}}(abc^{\delta-3n\tau}d^{\delta})$\\
 &       \\ [-0.16in]\hline
 &       \\ [-0.16in]
$Z^{a^2bc^{3z}d^{2z}}_2$ & $\rho_{\mathbf{r}}(c^{\gamma}d^{-2z})$, $\rho_{\mathbf{r}}(a^2bc^{\gamma})$\\
 &       \\ [-0.16in]\hline
 &       \\ [-0.16in]
$Z^{c^{9n/2}}_2$ & $\rho_{\mathbf{r}}(c^{\gamma}d^{\delta})$, $\rho_{\mathbf{r}}(a^2bc^{\gamma})$\\
 &       \\ [-0.16in]\hline
 &       \\ [-0.16in]
$Z^{d^{3n/2}}_2$ & $\rho_{\mathbf{r}}(c^{\gamma}d^{\delta})$, $\rho_{\mathbf{r}}(bc^{2\delta+3n\tau}d^{\delta})$\\
 &       \\ [-0.16in]\hline
 &       \\ [-0.16in]
$Z^{c^{9n/2}d^{3n/2}}_2$ & $\rho_{\mathbf{r}}(c^{\gamma}d^{\delta})$, $\rho_{\mathbf{r}}(abc^{\delta+3n\tau}d^{\delta})$\\ \hline\hline
\end{tabular}
\caption{\label{tab:Z2_RCP} Different types of remnant $Z_2$ subgroup $G_{\nu}$ and viable remnant CP transformations, where the superscript of the $Z_2$ subgroup denotes its generators. The allowed values of the parameters are $\gamma=0, 1, \ldots, 9n-1$, $x, y, z, \delta=0, 1, \ldots, 3n-1$, and $\tau=0, 1, 2$.}
\end{table}
As a result, it is sufficient to consider the representative residual symmetry $G_{\nu}=Z^{bd^x}_2$, $Z^{c^{9n/2}}_2$ and $G_{l}=\left\langle c^{s}d^{t}\right\rangle$, $\left\langle
bc^{s}d^{t}\right\rangle$, $\left\langle ac^{s}d^{t}\right\rangle$,
$\left\langle abc^{s}d^{t}\right\rangle$ and $\left\langle
a^2bc^{s}d^{t}\right\rangle$. Since only a $Z_2$ subgroup instead of a full Klein subgroup is preserved by the neutrino mass matrix, the postulated remnant flavor symmetries can only fix one column of the PMNS matrix. We list the explicit form of the determined columns for different remnant flavor symmetry in table~\ref{tab:PMNS_column_Z2}. Global analysis of the neutrino oscillation data gives the $3\sigma$ ranges on the absolute values of the elements of the PMNS matrix~\cite{Capozzi:2013csa}:
\begin{equation}
\label{eq:3sigma_ranges}||U_{\text{PMNS}}||=\left(
\begin{array}{ccc}
0.789\sim0.853 ~&~ 0.501\sim0.594 ~&~
0.133\sim0.172\\
0.194\sim0.558 ~&~ 0.408\sim0.735 ~&~
0.602\sim0.784 \\
0.194\sim0.558 ~&~ 0.408\sim0.735 ~&~
0.602\sim0.784 \\
\end{array}
\right)\,,
\end{equation}
It is obvious that none entry of the PMNS matrix is vanishing~\cite{Capozzi:2013csa,Forero:2014bxa,Gonzalez-Garcia:2014bfa}.
Therefore if one element of the fixed column is predicted to be zero, it would be excluded by the experimental data.  From table~\ref{tab:PMNS_column_Z2} we see that only three independent cases are viable with the residual flavor symmetries $(G_{\nu}, G_{l})=(Z^{bd^{x}}_{2}, \langle ac^{s}d^{t}\rangle)$, $(Z^{c^{9n/2}}_{2}, \langle ac^{s}d^{t}\rangle)$ and $(Z^{bd^{x}}_{2}, \langle abc^{s}d^{t}\rangle)$. In the following, the contribution of all admissible remnant CP transformations will be included further. We shall find the neutrino mass matrix invariant under the residual flavor and CP symmetries, and then the unitary transformation $U_{\nu}$ as well as the PMNS matrix $U_{\text{PMNS}}$ will be presented for each case.

\begin{table}[t!]
\renewcommand{\tabcolsep}{2.0mm}
\centering
\begin{tabular}{|c||c|c|}
\hline \hline
 &  &     \\ [-0.15in]
 &  $G_{\nu}=Z^{bd^x}_2$  &  $G_{\nu}=Z^{c^{9n/2}}_2$  \\

  &   &      \\ [-0.15in]\hline
 &   &       \\ [-0.15in]

$G_l=\langle c^{s}d^{t}\rangle$  &   $\frac{1}{\sqrt{2}}\left(\begin{array}{c}
0 \\
-1\\
1
\end{array}\right)$\xmark  &  $\left(\begin{array}{c}
0\\
0\\
1
\end{array}\right)$ \xmark  \\
 &   &         \\ [-0.15in]\hline
 &   &        \\ [-0.15in]

$G_l=\langle bc^{s}d^{t}\rangle$ &   $\left(\begin{array}{c}
0\\
\cos\left(\frac{s-2t+2x}{6n}\pi\right)\\
\sin\left(\frac{s-2t+2x}{6n}\pi\right)
\end{array}
\right)$ \xmark   &  $\frac{1}{\sqrt{2}}\left(\begin{array}{c}
0\\
-1\\
1
\end{array}
\right)$\xmark \\
 &   &         \\ [-0.15in]\hline
 &   &        \\ [-0.15in]

$G_l=\langle ac^{s}d^{t}\rangle$  &  $\sqrt{\frac{2}{3}}\left(
\begin{array}{c}
\sin\left(\frac{2s-3t+3x}{9n}\pi\right) \\
\cos\left(\frac{\pi}{6}+\frac{2s-3t+3x}{9n}\pi\right) \\
\cos\left(\frac{\pi}{6}-\frac{2s-3t+3x}{9n}\pi\right)
\end{array}
\right)$ \cmark  &  $\frac{1}{\sqrt{3}}\left(\begin{array}{c}
1\\
1\\
1
\end{array}
\right)$ \cmark \\
 &   &         \\ [-0.15in]\hline
 &   &        \\ [-0.15in]

$G_l=\langle abc^{s}d^{t}\rangle$  &  $\frac{1}{2}\left(\begin{array}{c}
1\\
1\\
\sqrt{2}
\end{array}
\right)$ \cmark  & $\frac{1}{\sqrt{2}}\left(\begin{array}{c}
0\\
-1\\
1
\end{array}
\right)$ \xmark   \\
 &   &         \\ [-0.15in]\hline
 &   &        \\ [-0.15in]

$G_l=\langle a^2bc^{s}d^{t}\rangle$ &  $\frac{1}{2}\left(\begin{array}{c}
1\\
1\\
\sqrt{2}
\end{array}
\right)$ \cmark  & $\left(\begin{array}{c}
0\\
0\\
1
\end{array}
\right)$ \xmark   \\

 &   &      \\ [-0.15in]\hline\hline

\end{tabular}
\caption{\label{tab:PMNS_column_Z2} The column vector of the PMNS matrix determined by the residual flavor symmetries $G_{\nu}$ and $G_{l}$. If one (or two) element of the fixed column is vanishing, we would use the notation ``\xmark'' to indicate that it is disfavored by the present experimental data otherwise the notation ``\cmark'' is labelled to indicate that agreement with the experimental data could be achieved. Notice that two pair of subgroups $(G_{\nu}, G_{l})=(Z^{bd^x}_2, \langle abc^{s}d^{t}\rangle)$ and $(Z^{bd^x}_2, \langle a^2bc^{s}d^{s-t}\rangle)$ are conjugate under the element $bc^{2x}d^{2x}$.}
\end{table}
\begin{description}[labelindent=-0.8em, leftmargin=0.3em]
\item[~~(\uppercase\expandafter{\romannumeral1})]{
$G_{l}=\left\langle ac^{s}d^{t}\right\rangle$,
$G_{\nu}=Z^{bd^x}_2$,
$X_{\nu\mathbf{r}}=\left\{\rho_{\mathbf{r}}(c^{2\delta+2x+3n\tau}d^{\delta}),
\rho_{\mathbf{r}}(bc^{2\delta+2x+3n\tau}d^{\delta+x})\right\}$}  \\

The residual symmetry transformation $G_{\nu}\times H^{\nu}_{CP}$ of the neutrino fields leaves the neutrino mass term invariant. Therefore the neutrino mass matrix $m_{\nu}$ must satisfy
\begin{eqnarray}
\nonumber&&\rho^{T}_{\mathbf{3}}(g_{\nu})m_{\nu}\rho_{\mathbf{3}}(g_{\nu})=m_{\nu},\qquad g_{\nu}\in G_{\nu},\\
\label{eq:constr_nu_again}&&X^{T}_{\nu\mathbf{3}}m_{\nu}X_{\nu\mathbf{3}}=m^{*}_{\nu},\qquad
\quad X_{\nu\mathbf{3}}\in H^{\nu}_{CP}\,.
\end{eqnarray}
In our working basis, it is straightforward to find that the neutrino mass matrix is constrained to take the form
\begin{equation}
\label{eq:mnu_I}m_{\nu}=
\begin{pmatrix}
e^{-\frac{4i\pi(x+\delta)}{9 n}}m_{11}  &~ e^{-\frac{i\pi(4x+\delta)}{9n}}m_{12}  &~  e^{\frac{i\pi(2 x-\delta)}{9n}}m_{12}  \\
e^{-\frac{i\pi(4 x+\delta )}{9n}}m_{12}  &~ e^{-\frac{2i\pi(2 x-\delta)}{9 n}}m_{22}  &~ e^{\frac{2i\pi(x+\delta)}{9n}}m_{23}  \\
e^{\frac{i\pi(2 x-\delta)}{9n}}m_{12}  &~ e^{\frac{2i\pi(x+\delta)}{9 n}} m_{23}  &~ e^{\frac{2i\pi(4x+\delta)}{9n}}m_{22}
\end{pmatrix}\,,
\end{equation}
where $m_{11}$, $m_{12}$, $m_{22}$ and $m_{23}$ are real. It follows that the neutrino mass matrix $m_{\nu}$ can be diagonalized by
\begin{equation}
U^{T}_{\nu}m_{\nu}U_{\nu}=\text{diag}\left(m_1,m_2,m_3\right)\,,
\end{equation}
with the unitary transformation
\begin{equation}\label{eq:unu_bdx}
U_{\nu}=\frac{1}{\sqrt{2}}
\begin{pmatrix}
 \sqrt{2} e^{\frac{2 i \pi  (x+\delta )}{9 n}} \cos \theta  &~ 0 ~& \sqrt{2} e^{\frac{2 i \pi  (x+\delta )}{9 n}} \sin\theta \\
 -e^{\frac{i \pi  (2 x-\delta )}{9 n}} \sin\theta &~ -e^{\frac{i \pi  (2 x-\delta )}{9 n}} ~& e^{\frac{i \pi  (2 x-\delta )}{9 n}} \cos\theta \\
 -e^{-\frac{i \pi  (4 x+\delta )}{9 n}} \sin\theta &~ e^{-\frac{i \pi  (4 x+\delta )}{9 n}} ~& e^{-\frac{i \pi  (4 x+\delta )}{9 n}} \cos\theta \\
\end{pmatrix}Q_{\nu}\,,
\end{equation}
where the angle $\theta$ is
\begin{equation}
\tan2\theta=\frac{2\sqrt{2}m_{12}}{m_{22}+m_{23}-m_{11}}\,.
\end{equation}
The factor $Q_{\nu}$ is a diagonal phase matrix with elements equal to $\pm1$ and $\pm i$ and it is necessary to make the light neutrino masses positive definite.
The neutrino mass eigenvalues are given by
\begin{eqnarray}
\nonumber&&m_1=\frac{1}{2}\left|m_{11}+m_{22}+m_{23}-\frac{m_{22}+m_{23}-m_{11}}{\cos2\theta}\right|,\\
\nonumber&&m_2=\left|m_{22}-m_{23}\right|,\\
\nonumber&&m_3=\frac{1}{2}\left|m_{11}+m_{22}+m_{23}+\frac{m_{22}+m_{23}-m_{11}}{\cos2\theta}\right|\,.
\end{eqnarray}
We see that the neutrino masses depend on four parameters $m_{11}$, $m_{12}$, $m_{22}$ and $m_{23}$, the experimentally measured mass squared differences could be easily accommodated.
The order of the three neutrino masses $m_1$, $m_2$ and $m_3$ can not be pinned down in the present framework, hence the unitary matrix $U_{\nu}$ is determined up to permutations of the columns (the same holds true in the following cases), and the neutrino mass spectrum can be either normal ordering (NO) or inverted ordering (IO). Taking into account the corresponding charged lepton diagonalization matrix $U_{l}$ listed in table~\ref{tab:cle_diagonal_matrix} and table~\ref{tab:extension_Gch}, we find the PMNS matrix $U_{\text{PMNS}}\equiv U^{\dagger}_{l}U_{\nu}$ up to row and column permutations is
{\footnotesize \begin{eqnarray*}
U^{I}_{\text{PMNS}}=\frac{1}{\sqrt{3}}
\begin{pmatrix}
 e^{i \varphi_{2}} \cos\theta-\sqrt{2} \cos\varphi_{1}\sin\theta &~ \sqrt{2} \sin\varphi_{1} ~& e^{i \varphi_{2}} \sin\theta+\sqrt{2} \cos\theta \cos\varphi_{1}  \\
 -e^{i \varphi_{2}}\cos\theta-\sqrt{2}\sin\theta\sin\left(\frac{\pi }{6}-\varphi_{1}\right) &~ \sqrt{2} \cos \left(\frac{\pi }{6}-\varphi_{1}\right) ~& -e^{i\varphi_{2}}\sin\theta+\sqrt{2}\cos\theta\sin\left(\frac{\pi }{6}-\varphi_{1}\right) \\
 e^{i\varphi_{2}} \cos\theta+\sqrt{2} \sin\theta \sin \left(\frac{\pi }{6}+\varphi_{1}\right) &~ \sqrt{2} \cos \left(\frac{\pi }{6}+\varphi_{1}\right) ~& e^{i \varphi_{2}} \sin\theta-\sqrt{2} \cos\theta\sin \left(\frac{\pi }{6}+\varphi_{1}\right)
\end{pmatrix}Q_{\nu}\,,
\end{eqnarray*}}
with
\begin{equation}
\varphi_1=\frac{2s-3t+3x}{9n}\pi,\qquad
\varphi_2=\frac{\delta+t+x}{3n}\pi\,.
\end{equation}
Both $\varphi_1$ and $\varphi_2$ are determined by the postulated remnant symmetries, they are independent of each other, and their values can be multiple of $\frac{\pi}{9n}$ and $\frac{\pi}{3n}$ respectively
\begin{eqnarray}
\nonumber&&\varphi_1~(\mathrm{mod}~2\pi)=0, \frac{1}{9n}\pi, \frac{2}{9n}\pi, \ldots, \frac{18n-1}{9n}\pi\,,\\
\label{eq:para_values_caseI}&&\varphi_2~(\mathrm{mod}~2\pi)=0, \frac{1}{3n}\pi, \frac{2}{3n}\pi, \ldots, \frac{6n-1}{3n}\pi\,.
\end{eqnarray}
We see that one column of the PMNS matrix is determined to be
\begin{equation}
\sqrt{\frac{2}{3}}\left(\begin{array}{c}
\sin\varphi_1\\
\cos\left(\pi/6-\varphi_1\right)\\
\cos\left(\pi/6+\varphi_1\right)
\end{array}
\right)
\end{equation}
in this case. As the neutrino mass ordering isn't constrained in the present framework, this column vector can be any of the three column of the PMNS matrix. As a consequence, the PMNS matrix can take the following three possible forms:
{\footnotesize \begin{eqnarray}
\nonumber U^{I,1}_{\text{PMNS}}&=&\frac{1}{\sqrt{3}}
\begin{pmatrix}
 \sqrt{2} \sin\varphi_{1} &~ e^{i \varphi_{2}} \cos\theta-\sqrt{2} \cos\varphi_{1}\sin\theta &~ e^{i \varphi_{2}} \sin\theta+\sqrt{2} \cos\theta \cos\varphi_{1}  \\
\sqrt{2} \cos \left(\frac{\pi }{6}-\varphi_{1}\right) &  -e^{i \varphi_{2}}\cos\theta-\sqrt{2}\sin\theta\sin\left(\frac{\pi }{6}-\varphi_{1}\right) &~ -e^{i\varphi_{2}}\sin\theta+\sqrt{2}\cos\theta\sin\left(\frac{\pi }{6}-\varphi_{1}\right) \\
\sqrt{2} \cos \left(\frac{\pi }{6}+\varphi_{1}\right) & ~e^{i\varphi_{2}} \cos\theta+\sqrt{2} \sin\theta \sin \left(\frac{\pi }{6}+\varphi_{1}\right) &~  e^{i \varphi_{2}} \sin\theta-\sqrt{2} \cos\theta\sin \left(\frac{\pi }{6}+\varphi_{1}\right)
\end{pmatrix}\,,\\
\nonumber U^{I,2}_{\text{PMNS}}&=&\frac{1}{\sqrt{3}}
\begin{pmatrix}
 e^{i \varphi_{2}} \cos\theta-\sqrt{2} \cos\varphi_{1}\sin\theta &~ \sqrt{2} \sin\varphi_{1} ~& e^{i \varphi_{2}} \sin\theta+\sqrt{2} \cos\theta \cos\varphi_{1}  \\
 -e^{i \varphi_{2}}\cos\theta-\sqrt{2}\sin\theta\sin\left(\frac{\pi }{6}-\varphi_{1}\right) &~ \sqrt{2} \cos \left(\frac{\pi }{6}-\varphi_{1}\right) ~& -e^{i\varphi_{2}}\sin\theta+\sqrt{2}\cos\theta\sin\left(\frac{\pi }{6}-\varphi_{1}\right) \\
 e^{i\varphi_{2}} \cos\theta+\sqrt{2} \sin\theta \sin \left(\frac{\pi }{6}+\varphi_{1}\right) &~ \sqrt{2} \cos \left(\frac{\pi }{6}+\varphi_{1}\right) ~& e^{i \varphi_{2}} \sin\theta-\sqrt{2} \cos\theta\sin \left(\frac{\pi }{6}+\varphi_{1}\right)
\end{pmatrix}\,,\\
\nonumber U^{I,3}_{\text{PMNS}}&=&\frac{1}{\sqrt{3}}
\begin{pmatrix}
e^{i \varphi_{2}} \sin\theta+\sqrt{2} \cos\theta \cos\varphi_{1} &~ e^{i \varphi_{2}} \cos\theta-\sqrt{2} \cos\varphi_{1}\sin\theta &~ \sqrt{2} \sin\varphi_{1} \\
-e^{i\varphi_{2}}\sin\theta+\sqrt{2}\cos\theta\sin\left(\frac{\pi }{6}-\varphi_{1}\right)  &~ -e^{i \varphi_{2}}\cos\theta-\sqrt{2}\sin\theta\sin\left(\frac{\pi }{6}-\varphi_{1}\right) &~ \sqrt{2} \cos \left(\frac{\pi }{6}-\varphi_{1}\right) \\
e^{i \varphi_{2}} \sin\theta-\sqrt{2} \cos\theta\sin \left(\frac{\pi }{6}+\varphi_{1}\right) &~ e^{i\varphi_{2}} \cos\theta+\sqrt{2} \sin\theta \sin \left(\frac{\pi }{6}+\varphi_{1}\right) &~ \sqrt{2} \cos \left(\frac{\pi }{6}+\varphi_{1}\right)
\end{pmatrix}\,.\label{eq:PMNS_caseI}
\end{eqnarray}}
The effect of row permutation is equivalent to redefinitions of the parameters $\theta$, $\varphi_1$ and $\varphi_2$, and no new possible values of $\varphi_1$ and $\varphi_2$ beyond those in Eq.~\eqref{eq:para_values_caseI} are obtained. Then we proceed to discuss the phenomenological predictions of each mixing pattern. For $U^{I,1}_{\text{PMNS}}$ the three lepton mixing angles read as
\begin{eqnarray}
\nonumber && \sin^{2}\theta_{13}=\frac{1}{3}\left(1+\cos^2\theta\cos2\varphi_1+\sqrt{2}\sin2\theta\cos\varphi_1\cos\varphi_2\right),\\
\nonumber&&\sin^2\theta_{12}=\frac{1+\sin^2\theta\cos2\varphi_1-\sqrt{2}\sin2\theta\cos\varphi_1\cos\varphi_2}
{2-\cos^2\theta\cos2\varphi_1-\sqrt{2}\sin2\theta\cos\varphi_1\cos\varphi_2},\\
\label{eq:mixing_para_caseI_1st}&&\sin^2\theta_{23}=\frac{1-\cos^2\theta\sin\left(\pi/6+2\varphi_1\right)-\sqrt{2}\sin2\theta\cos\varphi_2\sin\left(\pi/6-\varphi_1\right)}
{2-\cos^2\theta\cos2\varphi_1-\sqrt{2}\sin2\theta\cos\varphi_1\cos\varphi_2}\,,
\end{eqnarray}
which yield the correlation
\begin{equation}
3\cos^2\theta_{12}\cos^2\theta_{13}=2\sin^2\varphi_{1}\,,
\end{equation}
In order to accommodate the experimentally favored $3\sigma$ ranges $0.259\leq\sin^2\theta_{12}\leq0.359$ and $0.0176\leq\sin^2\theta_{13}\leq0.0295$ from the global fit~\cite{Capozzi:2013csa}, we find the allowed region of the parameter $\varphi_1$ is
\begin{equation}
\varphi_1\in\left[0.417\pi, 0.583\pi\right]\cup\left[1.417\pi, 1.583\pi\right]\,.
\end{equation}
Obviously $\varphi_1$ should be around $\pi/2$ or $3\pi/2$. Furthermore, the three CP rephasing invariants $J_{CP}$, $I_1$ and $I_2$ are predicted to be
\begin{eqnarray}
\nonumber && \left|J_{CP}\right|=\frac{1}{6\sqrt{6}}\left|\sin2\theta\sin3\varphi_1\sin\varphi_2\right|  \,, \\
\nonumber && \left|I_1\right|=\frac{2\sqrt{2}}{9}\left|\sin ^2\varphi_1 \sin \varphi_2 \left(\sqrt{2}\cos^2 \theta \cos \varphi_2- \sin 2\theta  \cos \varphi _1\right)\right|  \,, \\
\label{eq:CP_para_caseI_1st} && \left|I_2\right|= \frac{2\sqrt{2}}{9}\left|\sin ^2\varphi_1 \sin \varphi_2 \left(\sqrt{2}\sin^2 \theta \cos \varphi_2+ \sin 2\theta  \cos \varphi _1\right)\right| \,,
\end{eqnarray}
where $J_{CP}$ is well-known Jarlskog invariant, and $I_1$ and $I_2$ are defined for the Majorana phases with
\begin{eqnarray}
\nonumber J_{CP}&=&\mathrm{Im}\left[\left(U_{\text{PMNS}}\right)_{11}\left(U_{\text{PMNS}}\right)_{33}\left(U^{\ast}_{\text{PMNS}}\right)_{13}\left(U^{\ast}_{\text{PMNS}}\right)_{31}\right]\\
\nonumber&=&\frac{1}{8}\sin2\theta_{12}\sin2\theta_{13}\sin2\theta_{23}\cos\theta_{13}\sin\delta_{CP}\,,\\
\nonumber I_1&=&\text{Im}\left[\left(U_{\text{PMNS}}\right)^{\ast2}_{11}\left(U_{\text{PMNS}}\right)^2_{12}\right]=\frac{1}{4}\sin^22\theta_{12}\cos^4\theta_{13}\sin\alpha_{21}\,,\\
I_2&=&\text{Im}\left[\left(U_{\text{PMNS}}\right)^{\ast2}_{11}\left(U_{\text{PMNS}}\right)^2_{13}\right]=\frac{1}{4}\sin^22\theta_{13}\cos^2\theta_{12}\sin\alpha^{\prime}_{31}\,,
\end{eqnarray}
where $\alpha^{\prime}_{31}\equiv\alpha_{31}-2\delta_{CP}$, $\delta_{CP}$ is the Dirac CP violating phase, $\alpha_{21}$ and $\alpha_{31}$ are the Majorana CP phases in the standard parameterization of the PMNS matrix~\cite{Agashe:2014kda}. We show the absolute values of $J_{CP}$, $I_1$ and $I_2$ in Eq.~\eqref{eq:CP_para_caseI_1st}, the reason is because the sign of the $J_{CP}$ depends on the ordering of rows and columns and the sign of $I_1$ and $I_2$ could be changed by the CP parity matrix $Q_{\nu}$. Moreover, if the lepton doublet fields are assigned to the triplet $\mathbf{3}_{9n-1, 0}$ instead of $\mathbf{3}_{1, 0}$, the prediction for $U_{\text{PMNS}}$ would be complex conjugated such that the signs of $J_{CP}$, $I_1$ and $I_2$ are all inversed.
We show the possible predictions for the mixing parameters $\sin^2\theta_{12}$, $\sin\theta_{13}$, $\sin^2\theta_{23}$ as well as $\left|\sin\delta_{CP}\right|$, $\left|\sin\alpha_{21}\right|$ and $\left|\sin\alpha_{31}\right|$ for each $D^{(1)}_{9n, 3n}$ group in figure~\ref{fig:caseI_1cl_mixing_para}, where all the admissible values of $\varphi_1$ and $\varphi_2$ shown in Eq.~\eqref{eq:para_values_caseI} are considered and all the three mixing angles are required to lie in the $3\sigma$ allowed regions adapted from~\cite{Capozzi:2013csa}. It is notable that the solar mixing angle is predicted to be within the narrow interval of $0.313\leq\sin^2\theta_{12}\leq0.344$. The near future medium-baseline reactor neutrino oscillation experiments, such as JUNO~\cite{An:2015jdp} and RENO-50~\cite{Kim:2014rfa} are expected to make very precise, sub-percent measurements of the solar mixing angle $\theta_{12}$. They provide the one of the most significant test of this mixing pattern. The allowed values of the CP violation phases increase with group index $n$ and they are strongly constrained for smaller $n$. From figure~\ref{fig:caseI_1cl_mixing_para} we can read $0 \leq\left|\sin\delta_{CP}\right|\leq 0.226$, $0.847 \leq\left|\sin\alpha_{21}\right|\leq 0.873$ and $0 \leq\left|\sin\alpha'_{31}\right|\leq 0.488$ in the case of $n=1$. However, almost any values of the CP phases can be achieved for sufficient large value of $n$.

\begin{figure}[t!]
\begin{center}
\includegraphics[width=0.99\textwidth]{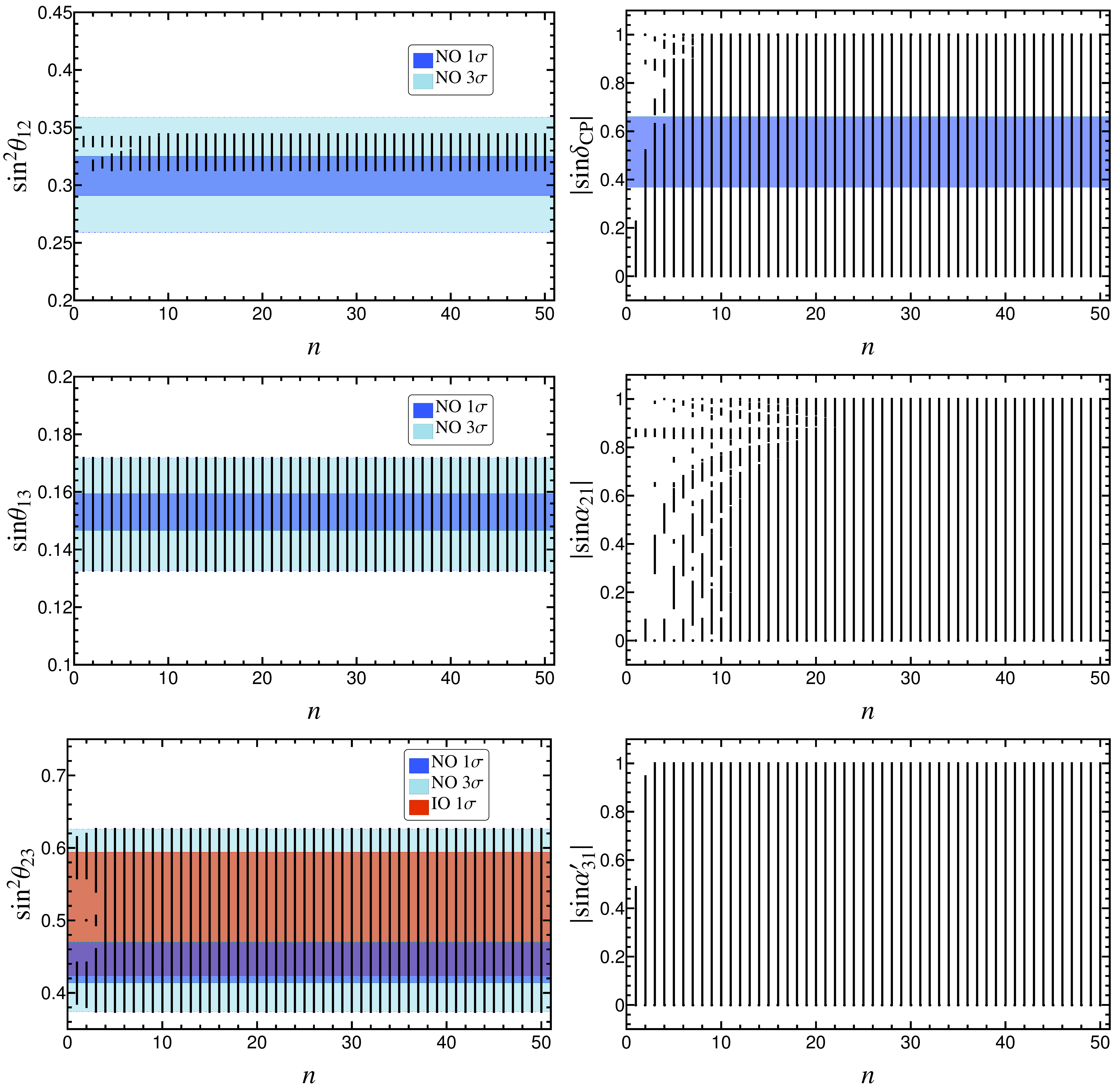}
\caption{\label{fig:caseI_1cl_mixing_para}The possible values of $\sin^2\theta_{12}$, $\sin\theta_{13}$, $\sin^2\theta_{23}$, $\left|\sin\delta_{CP}\right|$, $\left|\sin\alpha_{21}\right|$ and $\left|\sin\alpha^{\prime}_{31}\right|$ with respect to $n$ for the mixing pattern $U^{I,1}_{\text{PMNS}}$ in the case I, where the three lepton mixing angles are required to be within the experimentally preferred $3\sigma$ ranges. The $1\sigma$ and $3\sigma$ regions of the three neutrino mixing angles are adapted from global fit~\cite{Capozzi:2013csa}.}
\end{center}
\end{figure}

Then we turn to the second mixing pattern $U^{I,2}_{\text{PMNS}}$ in which
$\sqrt{\frac{2}{3}}\left(\sin\varphi_1, \cos\left(\frac{\pi}{6}-\varphi_1\right), \cos\left(\frac{\pi}{6}+\varphi_1\right)\right)^{T}$is the second column vector.
Its predictions for the mixing angles are
\begin{eqnarray}
\nonumber&&\sin^2\theta_{13}=\frac{1}{3}\left(1+\cos^2\theta\cos2\varphi_1+\sqrt{2}\sin2\theta\cos\varphi_1\cos\varphi_2\right),\\
\nonumber&&\sin^2\theta_{12}=\frac{2\sin^2\varphi_1}{2-\cos^2\theta\cos2\varphi_1-\sqrt{2}\sin2\theta\cos\varphi_1\cos\varphi_2}\,,\\
\label{eq:mixing_para_caseI_2nd}&&\sin^2\theta_{23}=\frac{1-\cos^2\theta\sin\left(\pi/6+2\varphi_1\right)-\sqrt{2}\sin2\theta\cos\varphi_2\sin\left(\pi/6-\varphi_1\right)}
{2-\cos^2\theta\cos2\varphi_1-\sqrt{2}\sin2\theta\cos\varphi_1\cos\varphi_2}\,.
\end{eqnarray}
We see that the solar and reactor mixing angles are correlated as
\begin{equation}
3\sin^2\theta_{12}\cos^2\theta_{13}=2\sin^2\varphi_1\,.
\end{equation}
In order to accommodate the experimental results on $\theta_{12}$ and $\theta_{13}$, $\varphi_1$ should vary in the interval:
\begin{equation}
\varphi_{1}\in\left[0.210\pi,0.259\pi\right]\cup\left[0.741\pi,0.790\pi\right]\cup \left[1.210\pi,1.259\pi\right]\cup \left[1.741\pi,1.790\pi\right]\,.
\end{equation}
Consequently we have
\begin{equation}
|\cos(\varphi_1-\frac{\pi}{6})|,|\cos(\varphi_1+\frac{\pi}{6})|\in[0.230,0.377]\cup[0.958,0.991]\,.
\end{equation}
We see that both $(22)$ and $(32)$ entries of $U^{I,2}_{\text{PMNS}}$ are not in  agreement with the experimental data given by Eq.~\eqref{eq:3sigma_ranges}. Hence this mixing pattern is phenomenologically disfavored.

\begin{figure}[t!]
\begin{center}
\includegraphics[width=0.90\textwidth]{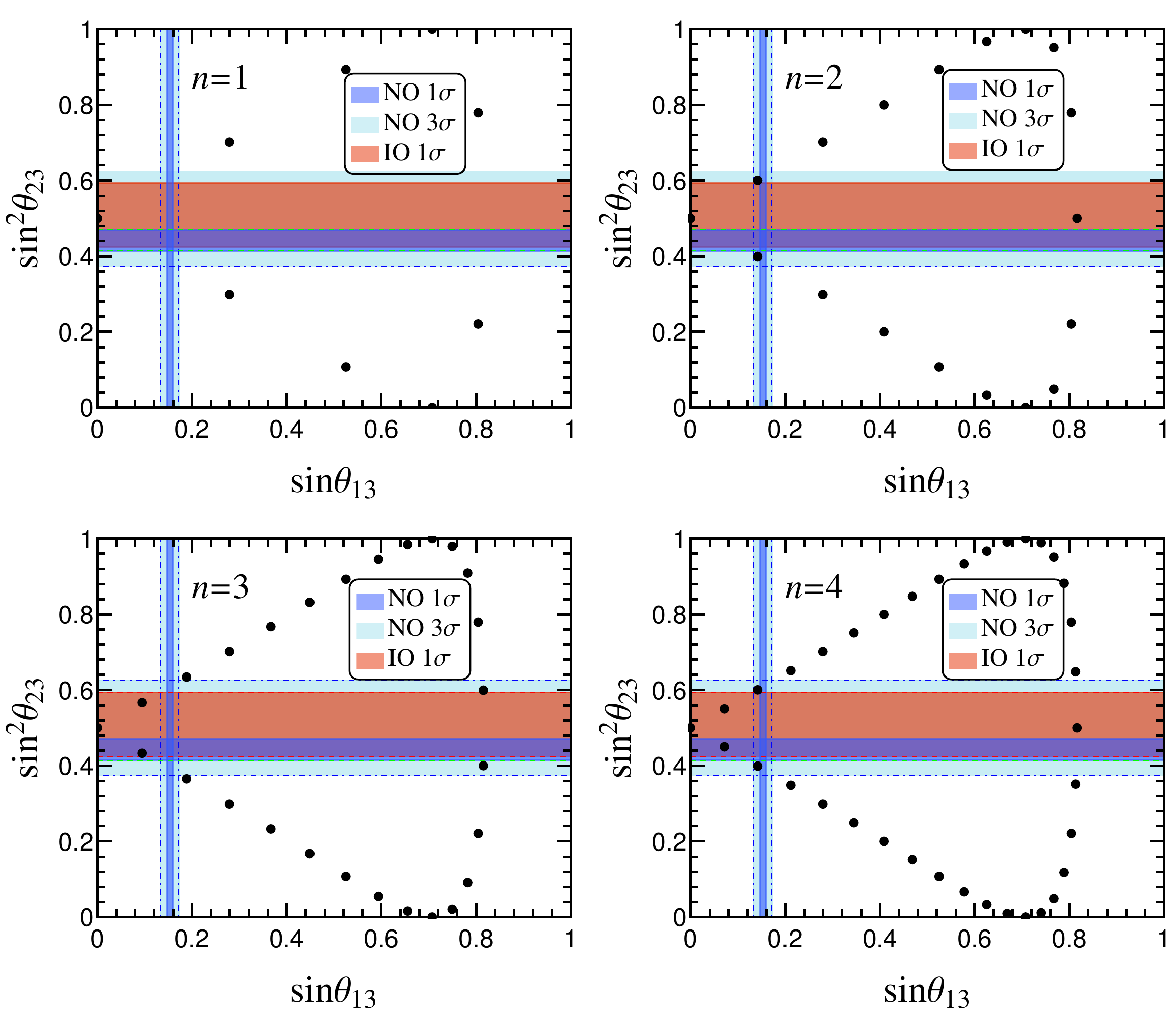}
\caption{\label{fig:caseI_2cl_mixing_para} The allowed values of $\sin^2\theta_{23}$ and $\sin\theta_{13}$ for the mixing pattern $U^{I,3}_{\text{PMNS}}$ in case I, where the first four smallest $D^{(1)}_{9n, 3n}$ group with $n=1, 2, 3, 4$ are considered. The $1\sigma$ and $3\sigma$ regions of the three neutrino mixing angles are adapted from global fit~\cite{Capozzi:2013csa}. }
\end{center}
\end{figure}

For the third possible arrangement of the rows and columns, the PMNS matrix is $U^{I, 3}_{\text{PMNS}}$. In this case, the third column of the PMNS matrix doesn't depend on the continuous parameter $\theta$ and it is completely fixed the remnant CP symmetry. It is straightforward to extract the mixing angles.
\begin{eqnarray}
\nonumber&&\sin^2\theta_{13}=\frac{2}{3}\sin^2\varphi_1,\qquad\quad \sin^2\theta_{23}=\frac{1+\sin\left(\pi/6+2\varphi_1\right)}{2+\cos2\varphi_1}\,,\\
\label{eq:mixing_para_caseI_3rd}&&\sin^2\theta_{12}=\frac{1+\sin^2\theta\cos2\varphi_1-\sqrt{2}\sin2\theta\cos\varphi_1\cos\varphi_2}{2+\cos2\varphi_1}\,.
\end{eqnarray}
The experimental data  $1.76\times10^{-2}\leq\sin^2\theta_{13}\leq2.95\times10^{-2}$ at $3\sigma$ level~\cite{Capozzi:2013csa} can be accommodated for the following values of the parameter $\varphi_1$:
\begin{equation}
\begin{split}
\varphi_1\in\left[0.0519\pi,0.0675\pi\right]\cup\left[0.933\pi,0.948\pi\right]\cup \left[1.0519\pi,1.0675\pi\right]\cup \left[1.933\pi,1.948\pi\right]\,.
\end{split}
\end{equation}
As both $\theta_{13}$ and $\theta_{23}$ depend on a single parameter $\varphi_1$, we can derive a sum rule between them,
\begin{equation}
2\sin^2\theta_{23}=1\pm\tan\theta_{13}\sqrt{2-\tan^2\theta_{13}}\;.
\end{equation}
Given the experimental best fitting value of the reactor mixing angle
$\sin^2\theta_{13}=2.34\times 10^{-2}$~\cite{Capozzi:2013csa}, we have
\begin{equation}
\sin^2\theta_{23}\simeq0.391,\quad\mathrm{or}\quad \sin^2\theta_{23}\simeq0.609\,,
\end{equation}
which is within the $3\sigma$ range although it is non-maximal. For a given $D^{(1)}_{9n, 3n}$ group, the atmospheric and reactor mixing angles can only take a set of discrete values. The possible values of $\sin^2\theta_{23}$ and $\sin\theta_{13}$ for the first four smallest $n=1, 2, 3, 4$ are displayed in figure~\ref{fig:caseI_2cl_mixing_para}. We see that the values $\varphi_1=\pm\pi/18, \pm17\pi/18$ in the case of $n=2, 4$ lead to $(\theta_{13}, \theta_{23})=(8.151^{\circ}, 50.813^{\circ})$ or $(8.151^{\circ}, 39.187^{\circ})$
which are compatible with the present experimental data~\cite{Capozzi:2013csa}.
The next generation of superbeam neutrino oscillation experiments would provide a high-precision determination of $\theta_{23}$. If no significant deviations from maximal mixing of $\theta_{23}$ will be detected, our present scheme will be excluded. Furthermore, we find that the CP invariants are
\begin{eqnarray}
\nonumber && \left|J_{CP}\right|=\frac{1}{6\sqrt{6}}\left|\sin2\theta\sin3\varphi_1\sin\varphi_2\right|  \,, \\
\nonumber && \left|I_1\right|=\frac{1}{9} \left|\cos\varphi_1\sin\varphi_2\left(4\cos2\theta\cos\varphi_1\cos \varphi_2-\sqrt{2}\sin2\theta\cos2\varphi_1\right)\right|\,, \\
\label{eq:CP_para_caseI_3rd}&&\left|I_2\right|=\frac{2\sqrt{2}}{9} \left|\sin^2\varphi_1 \sin \varphi_2 \left(\sqrt{2}\sin^2 \theta \cos \varphi_2+ \sin 2\theta  \cos \varphi _1\right)\right| \,.
\end{eqnarray}
Furthermore, we study the admissible values of mixing angles and CP phases for each $D^{(1)}_{9n, 3n}$ group. The numerical results are displayed in figure~\ref{fig:caseI_3cl_mixing_para}. We easily see that the atmospheric mixing angle $\theta_{23}$ is not maximal and it is around the $3\sigma$ upper or lower bounds. Similar to the $\Delta(6n^2)$ group~\cite{Ding:2014ora}, maximal value of the Majorana phase $\alpha^{\prime}_{31}$ can not be achieved in this case and it is found to be in the range of $\left|\sin\alpha^{\prime}_{31}\right|\leq 0.910$ while almost any values of $\delta_{CP}$ and $\alpha_{21}$ can be possible for large $n$.
\begin{figure}[t!]
\begin{center}
\includegraphics[width=0.99\textwidth]{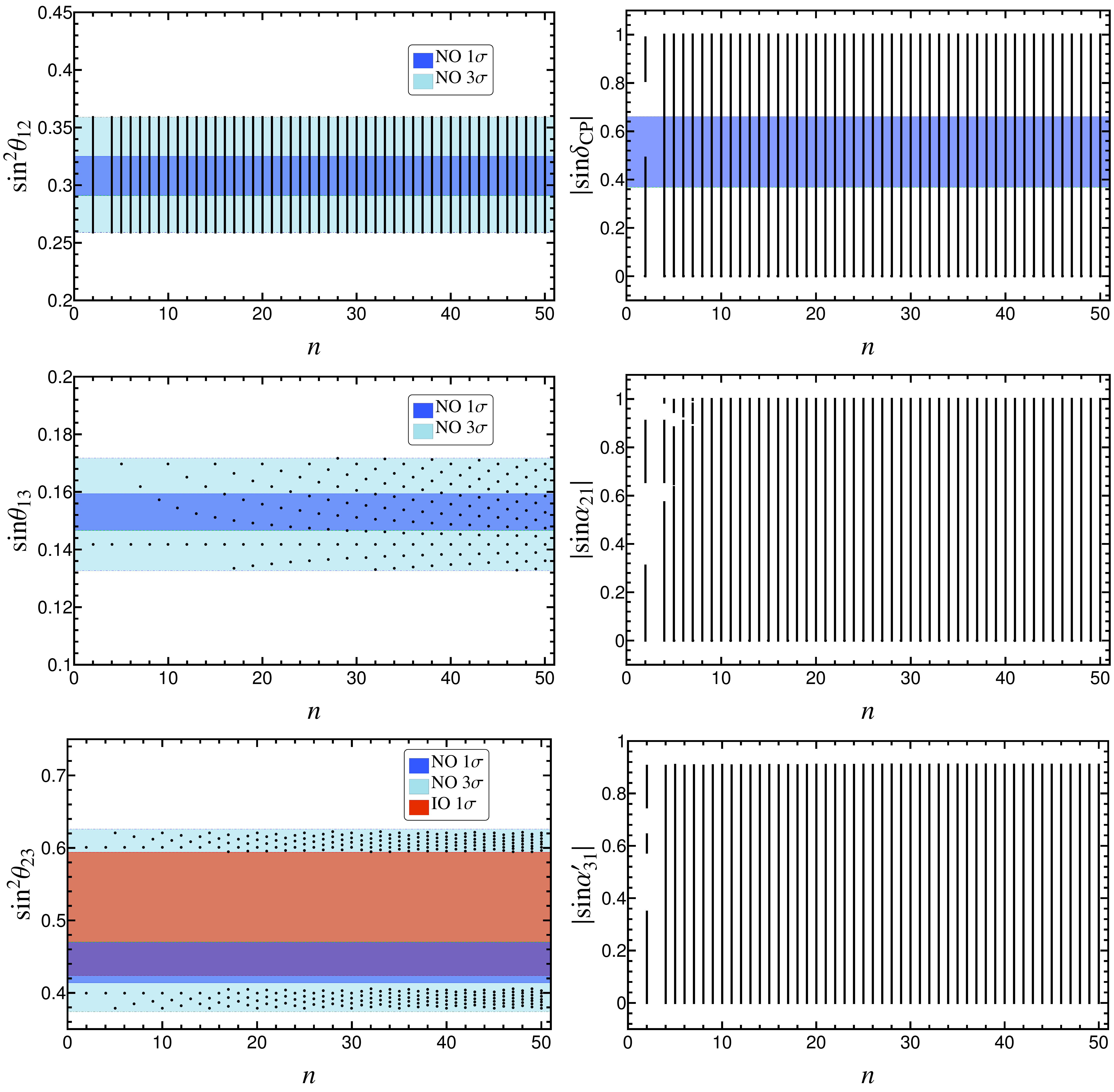}
\caption{\label{fig:caseI_3cl_mixing_para}The possible values of $\sin^2\theta_{12}$, $\sin\theta_{13}$, $\sin^2\theta_{23}$, $\left|\sin\delta_{CP}\right|$, $\left|\sin\alpha_{21}\right|$ and $\left|\sin\alpha^{\prime}_{31}\right|$ with respect to $n$ for the mixing pattern $U^{I,3}_{\text{PMNS}}$ in the case I, where the three lepton mixing angles are required to be within the experimentally preferred $3\sigma$ ranges. The $1\sigma$ and $3\sigma$ regions of the three neutrino mixing angles are adapted from global fit~\cite{Capozzi:2013csa}.}
\end{center}
\end{figure}

As a concrete example, we shall study the first two smallest $D^{(1)}_{9n, 3n}$ group with $n=1$ and $n=2$. From the expression of the PMNS matrix, we know that $U^{I, 1}_{\text{PMNS}}$ has the following symmetry properties:
\begin{eqnarray}
\nonumber&&U^{I, 1}_{\text{PMNS}}(\theta, \varphi_1, \pi+\varphi_2)=U^{I, 1}_{\text{PMNS}}(-\theta, \varphi_1, \varphi_2)\text{diag}(1, -1, 1),\\
\nonumber&&U^{I, 1}_{\text{PMNS}}(\theta, \varphi_1, \pi-\varphi_2)=[U^{I, 1}_{\text{PMNS}}(-\theta, \varphi_1, \varphi_2)]^{*}\text{diag}(1, -1, 1),\\
\nonumber&&U^{I, 1}_{\text{PMNS}}(\theta, \varphi_1, -\varphi_2)=[U^{I, 1}_{\text{PMNS}}(\theta, \varphi_1, \varphi_2)]^{*},\\
\label{eq:caseI_symmetry}&&U^{I, 1}_{\text{PMNS}}(\theta, \pi+\varphi_1, \varphi_2)=U^{I, 1}_{\text{PMNS}}(-\theta, \varphi_1, \varphi_2)\text{diag}(-1, 1, -1)\,,
\end{eqnarray}
where the diagonal matrix can be absorbed into the matrix $Q_{\nu}$. Similar relations are satisfied for the PMNS matrix $U^{I, 3}_{\text{PMNS}}$. Note that the PMNS matrix would become its complex conjugation if the three generations of leptons are assigned to the triplet $\mathbf{3}_{9n-1, 0}\cong\mathbf{3}^{*}_{1, 0}$. As a result, without loss of generality, we shall focus on the case of $0\leq\varphi_1\leq\pi$ and $0\leq\varphi_2\leq\pi/2$. A conventional $\chi^2$ analysis is performed. Notice that we don't include the information of the Dirac CP phase $\delta_{CP}$ into the $\chi^2$ function, since the evidence for a preferred value of $\delta_{CP}$ coming from both present experiments and the global fitting is rather weak.
The numerical results are reported in table~\ref{tab:caseI_n12}, where we exclude all patterns that can not accommodate the experimental data at the best fitting point $\theta=\theta_{bf}$ for which the $\chi^2$ function is minimized.
Since the global fit results of the mixing angles are slightly distinct for NO and IO neutrino mass spectrums~\cite{Capozzi:2013csa}, the $\chi^2$ function has been defined for NO and IO respectively. The values in the parentheses are the results for the IO case. Applying the symmetry transformations in Eq.~\eqref{eq:caseI_symmetry}, we can obtain other values of $\varphi_1$ and $\varphi_2$ which yield the same best fit values for the mixing angles such that the same $\chi^2_{min}$ is obtained. For both mixing patterns $U^{I, 1}_{\text{PMNS}}$ and $U^{I, 3}_{\text{PMNS}}$, we can check that the formulae in Eqs.~(\ref{eq:mixing_para_caseI_1st},\ref{eq:mixing_para_caseI_3rd}) for the mixing angles $\sin^2\theta_{12}$ and $\sin^2\theta_{13}$ are invariant while $\sin^2\theta_{23}$ turns into
$\cos^2\theta_{23}$ under the transformation $\varphi_1\rightarrow\pi-\varphi_1$, $\theta\rightarrow\pi-\theta$. As a result, the sum of the best fitting value $\theta_{bf}$ for $\varphi_1$ and $\pi-\varphi_1$ is approximately equal to $\pi$. It is remarkable that even the smallest $D^{(1)}_{9n, 3n}$ group with $n=1$ allows a reasonable fit to the experimental data, for instance, the mixing patterns with $(\varphi_1, \varphi_2)=(4\pi/9, 0)$, $(4\pi/9, \pi/3)$, $(5\pi/9, 0)$ and $(5\pi/9, \pi/3)$ can describe the experimentally measured values of the mixing angles, as can be seen from table~\ref{tab:caseI_n12}. In particular, the CP violating phases are neither conserved nor maximal in the case of $(\varphi_1, \varphi_2)=(4\pi/9, \pi/3)$ and $(5\pi/9, \pi/3)$. The PMNS matrix $U^{I, 1}_{\text{PMNS}}$ for $n=2$ as well as $(\varphi_1, \varphi_2)=(\pi/2, \pi/2)$ give rise to maximal atmospheric mixing and maximal Dirac phase. On the other hand, the group index $n$ should be equal or greater than 2 in order to obtain phenomenologically viable mixing pattern of the form $U^{I, 3}_{\text{PMNS}}$. Scrutinizing all the admissible cases listed in table~\ref{tab:caseI_n12}, we find that the predictions for $\theta_{13}$ are almost the same, nevertheless $\theta_{12}$, $\theta_{23}$ and $\delta_{CP}$ are predicted to be considerably different. The JUNO experiment will be capable of reducing the error of $\sin^2\theta_{12}$ to about $0.1^{\circ}$ or around $0.3\%$~\cite{An:2015jdp}. Future long baseline experiments such as DUNE~\cite{Adams:2013qkq}, LBNO~\cite{::2013kaa}, T2HK~\cite{Kearns:2013lea} and possibly ESS$\nu$SB~\cite{Baussan:2012cw} at the European Spallation Source can make very precise measurements of the oscillation parameters $\theta_{12}$, $\theta_{23}$ and $\delta_{CP}$.
Therefore future neutrino facilities have the potential to discriminate between the above possible cases, or to rule them out entirely. Furthermore, we expect that a more ambitious facility such as the neutrino factory~\cite{Geer:1997iz} could provide a more stringent tests of our approach.

\begin{table}[t!]
\centering
\footnotesize
\renewcommand{\tabcolsep}{1.8mm}
\begin{tabular}{|c|c|c|c|c|c|c|c|c|c|c|}
\hline \hline
\multicolumn{11}{|c|}{Case I}   \\ \hline
\multicolumn{11}{|c|}{$n=1$ and $n=2$}   \\ \hline
& $\varphi_1$ & $\varphi_2$  & $\theta_{bf}$ & $\chi^2_{min}$   & $\sin^2\theta_{13}$ & $\sin^2\theta_{12}$ & $\sin^2\theta_{23}$  & $|\sin\delta_{CP}|$ & $|\sin\alpha_{21}|$ & $|\sin\alpha^{\prime}_{31}|$  \\ \hline
\multirow{8}{*}{$U^{I, 1}_{\text{PMNS}}$}&\multirow{4}{*}{$\frac{4\pi}{9}$} & \multirow{2}{*}{$0$} & $0.0245$ & $3.789$  & $0.0243$ & $0.337$ & $0.419$   & \multirow{2}{*}{$0$ ($0$)} & \multirow{2}{*}{$0$ ($0$)} & \multirow{2}{*}{$0$ ($0$)}  \\
&&&($0.0274$) & ($4.267$)  & ($0.0248$) & ($0.337$) & ($0.421$) &&& \\ \cline{3-11}

&& \multirow{2}{*}{$\frac{\pi}{3}$} & $0.0435$ & $3.928$  & $0.0242$ & $0.337$ & $0.417$  &$0.125$ & $0.857$ & $0.276$ \\
&&& ($0.0480$) & ($4.438$)  & ($0.0247$) & ($0.337$) & ($0.419$)  &($0.137$) & ($0.856$) & ($0.302$)    \\  \cline{2-11}

& \multirow{4}{*}{$\frac{5\pi}{9}$} & \multirow{2}{*}{$0$} & $3.108$ & $21.499$ & $0.0259$ & $0.336$ & $0.574$   & \multirow{2}{*}{$0$ ($0$)} & \multirow{2}{*}{$0$ ($0$)} & \multirow{2}{*}{$0$ ($0$)}\\
 &&& ($3.117$) & ($3.822$)  & ($0.0244$) & ($0.337$) & ($0.581$)  &&&\\ \cline{3-11}

&& \multirow{2}{*}{$\frac{\pi}{3}$} & $3.087$ & $22.307$  & $0.0255$ & $0.337$ & $0.578$  &$0.154$ & $0.854$ & $0.338$   \\
&&  &  ($3.097$) & ($3.849$)  & ($0.0243$) & ($0.337$) & ($0.583$)  & ($0.127$) & ($0.856$) & ($0.281$) \\ \hline\hline

\multicolumn{11}{|c|}{$n=2$}   \\ \hline\hline

\multirow{14}{*}{$U^{I, 1}_{\text{PMNS}}$}&\multirow{4}{*}{$\frac{4\pi}{9}$} &  \multirow{2}{*}{$\frac{\pi}{6}$} & $0.0278$ & $3.807$  & $0.0243$ & $0.337$ & $0.419$ &$0.0462$ & $0.869$ & $0.103$  \\
&&& ($0.0311$) & ($4.289$)  & ($0.0248$) & ($0.337$) & ($0.421$)  &($0.0510$) & ($0.870$) & ($0.114$)   \\  \cline{3-11}

&& \multirow{2}{*}{$\frac{\pi}{2}$} & $0.108$ & $5.666$  & $0.0237$ & $0.338$ & $0.400$  &$0.362$ & $0.0532$ & $0.739$  \\
&&& ($0.116$) & ($6.131$)  & ($0.0243$) & ($0.337$) & ($0.400$)  &($0.384$) & ($0.0572$) & ($0.774$)   \\  \cline{2-11}

&\multirow{4}{*}{$\frac{5\pi}{9}$} &  \multirow{2}{*}{$\frac{\pi}{6}$} &$3.104$ & $21.616$  & $0.0258$ & $0.336$ & $0.575$  &$0.0602$ & $0.871$ & $0.134$   \\
&&&  ($3.113$) & ($3.826$)  & ($0.0244$) & ($0.337$) & ($0.581$)  &($0.0468$) & ($0.869$) & ($0.104$)   \\  \cline{3-11}

&& \multirow{2}{*}{$\frac{\pi}{2}$} &   $3.033$ & $27.468$  & $0.0238$ & $0.337$ & $0.600$  &$0.365$ & $0.0537$ & $0.744$  \\
&&& ($3.026$) & ($4.087$)  & ($0.0243$) & ($0.337$) & ($0.600$)  &($0.383$) & ($0.0569$) & ($0.772$)   \\  \cline{2-11}

&\multirow{6}{*}{$\frac{\pi}{2}$} & \multirow{2}{*}{$\frac{\pi}{3}$} &$0.261$ & $26.399$  & $0.0222$ & $0.318$ & $0.604$  &$0.885$ & $0.866$ & $0.866$   \\
&&  & ($0.272$) & ($1.490$)  & ($0.0240$) & ($0.317$) & ($0.608$)  &($0.887$) & ($0.866$) & ($0.866$)  \\  \cline{3-11}

 && \multirow{2}{*}{$\frac{\pi}{3}$} &$2.877$ & $3.838$  & $0.0228$ & $0.318$ & $0.394$  &$0.886$ & $0.866$ & $0.866$   \\
&&  & ($2.873$) & ($4.352$)  & ($0.0234$) & ($0.317$) & ($0.393$)  &($0.887$) & ($0.866$) & ($0.866$)  \\  \cline{3-11}

&& \multirow{2}{*}{$\frac{\pi}{2}$} &$0.269$ &$3.946$ &$0.0235$ & $0.317$ & $0.5$   & \multirow{2}{*}{$1$ ($1$)} & \multirow{2}{*}{$0$ ($0$)} & \multirow{2}{*}{$0$ ($0$)}  \\
&&&($0.272$) &($0.380$) &($0.0241$) & ($0.317$) & ($0.5$) & & & \\ \hline

\multirow{16}{*}{$U^{I, 3}_{\text{PMNS}}$}&\multirow{8}{*}{$\frac{\pi}{18}$} & \multirow{2}{*}{$0$} & $0.0344$ & $27.637$ & \multirow{15}{*}{$0.0201$} & \multirow{5}{*}{$0.308$} & \multirow{7}{*}{$0.601$}   & \multirow{2}{*}{$0$ ($0$)} & \multirow{2}{*}{$0$ ($0$)} & \multirow{2}{*}{$0$ ($0$)}  \\
&&& ($0.0344$) & ($4.238$)  & \multirow{15}{*}{($0.0201$)} & \multirow{5}{*}{($0.308$)} & \multirow{7}{*}{($0.601$)} &&& \\ \cline{3-5} \cline{9-11}

&& \multirow{2}{*}{$\frac{\pi}{6}$} &$0.0399$ & $27.637$  &  &  &   &$0.0431$ & $0.881$ & $0.0279$    \\
&&  & ($0.0399$) & ($4.238$)  &  &  &   & ($0.0431$) & ($0.881$) & ($0.0279$)  \\  \cline{3-5} \cline{9-11}

&& \multirow{2}{*}{$\frac{\pi}{3}$} &$0.0716$ & $27.637$  &  &  &   &$0.134$ & $0.815$ & $0.0868$    \\
&&  & ($0.0716$) & ($4.238$)  &  &  &   &($0.134$) & ($0.815$) & ($0.0868$ )  \\  \cline{3-5} \cline{7-7} \cline{9-11}

&& \multirow{2}{*}{$\frac{\pi}{2}$} & \multirow{2}{*}{$0$($0$)} & $31.219$ &  & $0.340$ &    & \multirow{2}{*}{$0$ ($0$)} & \multirow{2}{*}{$0$ ($0$)} & \multirow{2}{*}{$0$ ($0$)}  \\
&&&& ($7.820$)  &  & ($0.340$) &   &&& \\ \cline{2-5} \cline{7-11}

&\multirow{8}{*}{$\frac{17\pi}{18}$} & \multirow{2}{*}{$0$} & $3.107$ & $5.707$ &  & \multirow{5}{*}{$0.308$} & \multirow{7}{*}{$0.399$}  & \multirow{2}{*}{$0$ ($0$)} & \multirow{2}{*}{$0$ ($0$)} & \multirow{2}{*}{$0$ ($0$)}  \\
&&& ($3.107$) & ($6.374$)  &  & \multirow{5}{*}{($0.308$)} & \multirow{7}{*}{($0.399$)}  &&& \\ \cline{3-5} \cline{9-11}

&& \multirow{2}{*}{$\frac{\pi}{6}$} &$3.102$ & $5.707$  &  &  &   &$0.0431$ & $0.881$ & $0.0279$   \\
&&&  ($3.102$) & ($6.374$)  &  &  &   &($0.0431$) & ($0.881$) & ($0.0279$)   \\  \cline{3-5} \cline{9-11}

&& \multirow{2}{*}{$\frac{\pi}{3}$} &$3.070$ & $5.707$  &  &  &   &$0.134$ & $0.815$ & $0.0868$  \\
&&  & ($3.070$) & ($6.374$)  &  &  &   &($0.134$) & ($0.815$) & ($0.0868$)   \\ \cline{3-5} \cline{7-7} \cline{9-11}

&& \multirow{2}{*}{$\frac{\pi}{2}$} & \multirow{2}{*}{$0$ ($0$)} &$9.289$ &  & $0.340$ &      & \multirow{2}{*}{$0$ ($0$)} & \multirow{2}{*}{$0$ ($0$)} & \multirow{2}{*}{$0$ ($0$)}  \\
&&&& ($9.955$)  & & ($0.340$) &   &&& \\ \hline  \hline
\end{tabular}
\caption{\label{tab:caseI_n12}Results of the $\chi^2$ analysis for $n=1, 2$ in the case I. The $\chi^2$ function has a global minimum $\chi^2_{min}$ at the best fit value $\theta_{bf}$ for $\theta$. We give the values of the mixing angles and CP violation phases for $\theta=\theta_{bf}$. The values given in parentheses denote the results for the IO neutrino mass spectrum. }
\end{table}

Since the Majorana CP violating phases can be predicted in the present framework, we now discuss its phenomenological implications in the neutrinoless double beta ($0\nu\beta\beta$) decay. It is well-known that the $0\nu\beta\beta$ decay process is the most sensitive probe for Majorana neutrinos. Its observation would establish the Majorana nature of neutrinos irrespective of the underlying mass generation mechanism. The $0\nu\beta\beta$ decay rate is proportional to the square of the effective Majorana mass $|m_{ee}|$ which is given by~\cite{Agashe:2014kda}
\begin{equation}
\label{eq:mee}\left|m_{ee}\right|=\left|m_1\cos^2\theta_{12}\cos^2\theta_{13}+m_2\sin^2\theta_{12}\cos^2\theta_{13}e^{i \alpha_{21}}+m_3\sin^2\theta_{13}e^{i\alpha^{\prime}_{31}}\right|\,.
\end{equation}
The values of $\left|m_{ee}\right|$ are dependent on both CP phases $\alpha_{21}$ and $\alpha^{\prime}_{31}\equiv\alpha_{31}-2\delta_{CP}$.
For the mixing pattern $U^{I, 1}_{\text{PMNS}}$, $|m_{ee}|$ is of the form
\begin{eqnarray}
\nonumber&&|m_{ee}|=\frac{1}{3}\Big|2m_1\sin^2\varphi_1+q_1m_2(e^{i \varphi_{2}} \cos\theta-\sqrt{2} \cos\varphi_{1}\sin\theta)^2\\
&&\qquad\qquad+q_2m_3(e^{i \varphi_{2}} \sin\theta+\sqrt{2} \cos\theta \cos\varphi_{1})^2\Big|\,,
\end{eqnarray}
where $q_1, q_2=\pm1$ appears due to the undetermined CP parity of the neutrino states encoded in the matrix $Q_{\nu}$. For another admissible mixing pattern $U^{I, 3}_{\text{PMNS}}$, $|m_{ee}|$ is given by
\begin{eqnarray}
\nonumber&&|m_{ee}|=\frac{1}{3}\Big|2m_3\sin^2\varphi_{1}+q_1m_2( e^{i \varphi_{2}} \cos\theta-\sqrt{2} \cos\varphi_{1}\sin\theta)^2\\
&&\qquad\qquad+q_2m_1(e^{i \varphi_{2}} \sin\theta+\sqrt{2} \cos\theta \cos\varphi_{1})^2\Big|\,.
\end{eqnarray}
The achievable values of the effective mass $|m_{ee}|$ for both $n\rightarrow\infty$ and $n=2$ are plotted in figure~\ref{fig:mee_CaseI}. Here we require the three mixing angle be within their $3\sigma$ allowed values while the neutrino mass-squared splittings are fixed at their best-fit values from Ref.~\cite{Capozzi:2013csa}.
We see that the majority of the experimentally allowed $3\sigma$ region of $|m_{ee}|$ can be reproduced in the limit $n\rightarrow\infty$. In the case of $n=2$, it is remarkable that the effective mass $|m_{ee}|$ obtained from $U^{I, 1}_{\text{PMNS}}$ is found to be around 0.0155eV, 0.0175eV, 0.0279eV, 0.0423eV, or 0.0484eV for IO neutrino mass spectrum. These predictions are beyond the reach of the present $0\nu\beta\beta$ experiments such as GERDA~\cite{Agostini:2013mzu}, EXO-200~\cite{Auger:2012ar, Albert:2014awa} and KamLAND-ZEN~\cite{Gando:2012zm}. However, the proposed facilities nEXO and KamLAND2-Zen~\cite{Piquemal:2013uaa} etc aim to increase the sensitivity to cover the full IO region, such that all of our patterns with this mass spectrum could be tested. For NO the effective mass $|m_{ee}|$ is much smaller than the IO case and it can even vanish for certain values of the lightest neutrino mass because of a cancellation between different terms in Eq.~\eqref{eq:mee}. Obviously exploring the NH region experimentally is beyond the reach of any planned $0\nu\beta\beta$ experiment. Even if the signals of $0\nu\beta\beta$ decays are not observed and the neutrino masses spectrum are measured to be NO by upcoming neutrino oscillation experiments~\cite{An:2015jdp,Kim:2014rfa}, one can still extract useful information on the Majorana phases $\alpha_{21}$ and $\alpha^{\prime}_{31}$ by combining the cosmological data on the absolute neutrino mass scale and the improved measurement of $\theta_{12}$, $\theta_{23}$ and $\delta_{CP}$ from a number of complementary neutrino oscillation experiments.

\begin{figure}[t!]
\begin{center}
\includegraphics[width=0.46\linewidth]{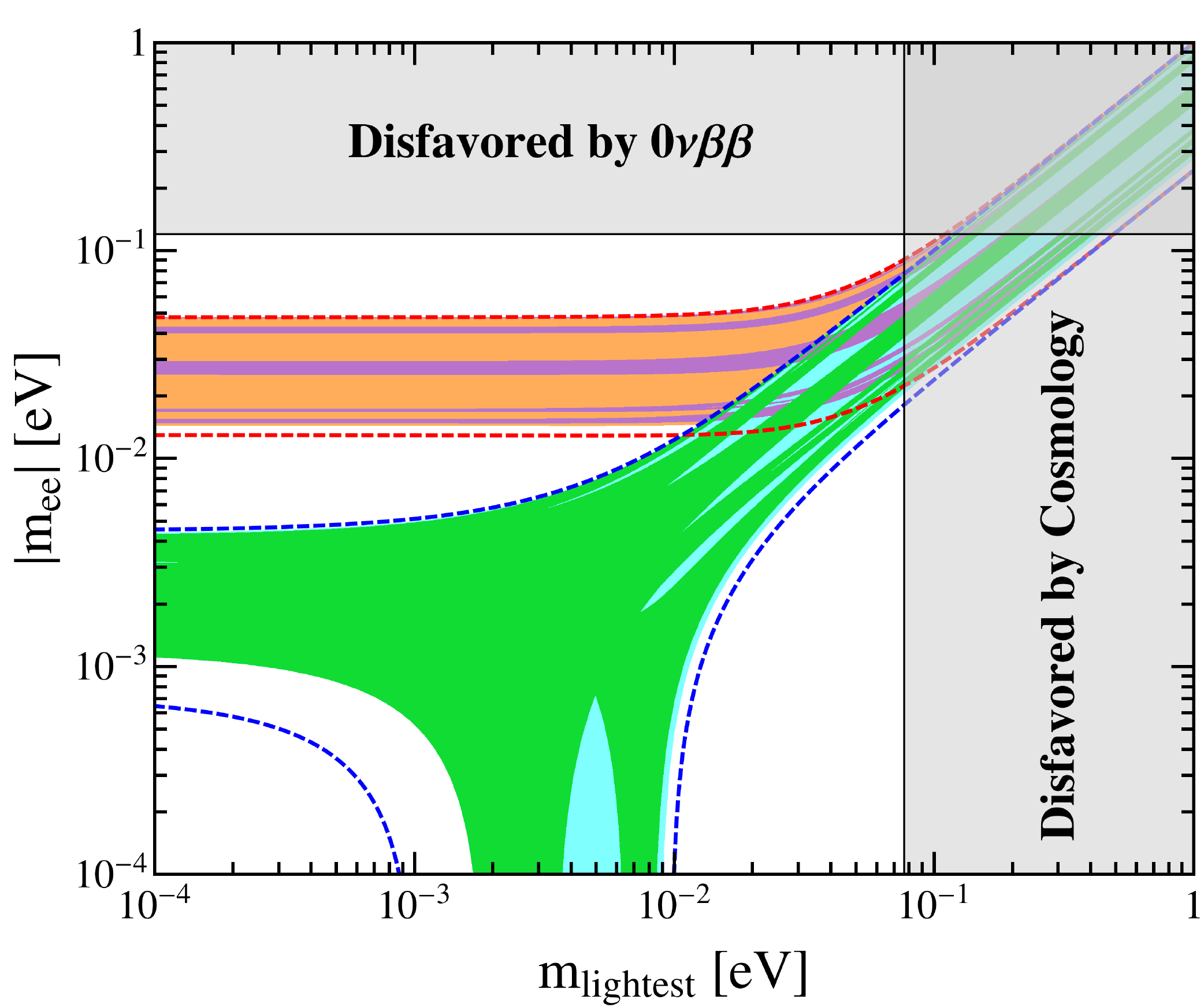}~~~
\includegraphics[width=0.46\linewidth]{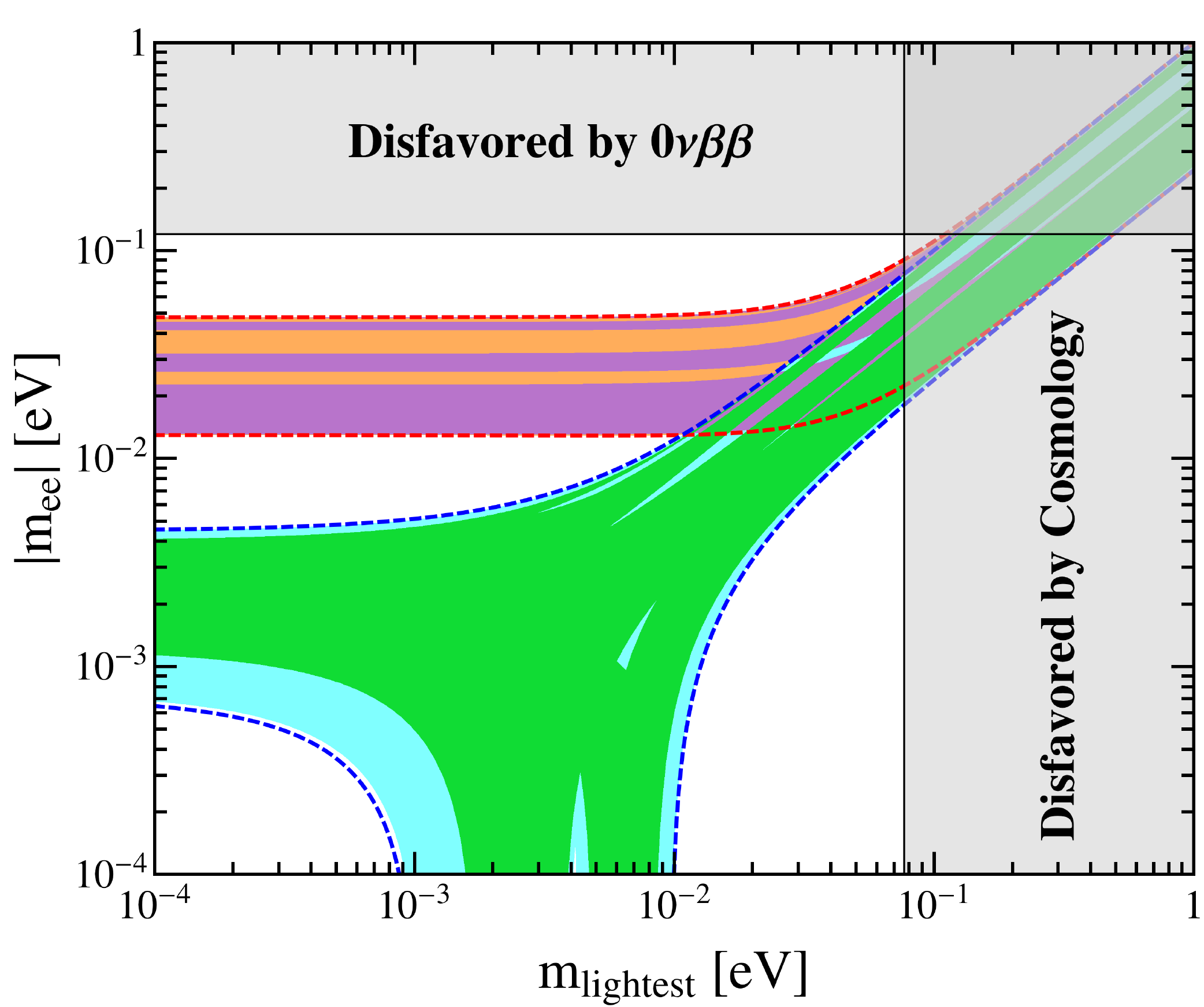}
\caption{\label{fig:mee_CaseI}The possible values of the effective Majorana mass $|m_{ee}|$ as a function of the lightest neutrino mass in the case I. The left and right panels are for the mixing patterns $U^{I, 1}_{\text{PMNS}}$ and $U^{I, 3}_{\text{PMNS}}$ respectively. The red (blue) dashed lines indicate the most general allowed regions for IO (NO) neutrino mass spectrum obtained by varying the mixing parameters over the $3\sigma$ ranges~\cite{Capozzi:2013csa}. The orange (cyan) areas denote the achievable values of $|m_{ee}|$ in the limit of $n\rightarrow\infty$ assuming IO (NO) spectrum. The purple and green regions are the theoretical predictions for the $D^{(1)}_{9n ,3n}$ group with $n=2$. Notice that the purple (green)  region overlaps the orange (cyan) one. The present most stringent upper limits $|m_{ee}|<0.120$ eV from EXO-200~\cite{Auger:2012ar, Albert:2014awa} and KamLAND-ZEN~\cite{Gando:2012zm} is shown by horizontal grey band. The vertical grey exclusion band represents the current bound coming from the cosmological data of $\sum m_i<0.230$ eV at $95\%$ confidence level obtained by the Planck collaboration~\cite{Ade:2013zuv}.
}
\end{center}
\end{figure}

\item[~~(\uppercase\expandafter{\romannumeral2})]
$G_{l}=\left\langle abc^{s}d^{t}\right\rangle$,
$G_{\nu}=Z^{bd^x}_2$,
$X_{\nu\mathbf{r}}=\left\{\rho_{\mathbf{r}}(c^{2\delta+2x+3n\tau}d^{\delta}),
\rho_{\mathbf{r}}(bc^{2\delta+2x+3n\tau}d^{\delta+x})\right\}$  \\

This case differs from the previous one in the residual flavor symmetry $G_{l}$. From table~\ref{tab:cle_diagonal_matrix} and table~\ref{tab:extension_Gch}, we know that the charged lepton diagonalization matrix is exactly $U^{(4)}_{l}$. Since the neutrino mass matrix preserves the same remnant symmetry as case \uppercase\expandafter{\romannumeral1}, the neutrino mass matrix should take the form of Eq.~\eqref{eq:mnu_I}, and it is diagonalized by the unitary transformation $U_{\nu}$ in Eq.~\eqref{eq:unu_bdx}. Using the freedom in exchanging rows and columns, we find the phenomenologically viable lepton mixing matrix is
\begin{equation}
\label{eq:PMNS_case_II_1}
U^{II,1}_{\text{PMNS}}=\frac{1}{2}
\begin{pmatrix}
-\sin\theta-\sqrt{2} e^{i \varphi_{3}} \cos\theta  &~ 1 ~& \cos\theta-\sqrt{2} e^{i \varphi_{3}} \sin\theta \\
 -\sin\theta+\sqrt{2} e^{i\varphi_{3}} \cos\theta &~ 1 ~& \cos\theta+\sqrt{2} e^{i\varphi_{3}} \sin\theta \\
 -\sqrt{2} \sin \theta  &~ -\sqrt{2} ~& \sqrt{2} \cos\theta
\end{pmatrix}Q_{\nu}\,,
\end{equation}
or
\begin{equation}
\label{eq:PMNS_case_II_2}
U^{II,2}_{\text{PMNS}}=\frac{1}{2}
\begin{pmatrix}
-\sin\theta-\sqrt{2} e^{i \varphi_{3}} \cos\theta  &~ 1 ~& \cos\theta-\sqrt{2} e^{i \varphi_{3}} \sin\theta \\
 -\sqrt{2} \sin \theta  &~ -\sqrt{2} ~& \sqrt{2} \cos\theta \\
 -\sin\theta+\sqrt{2} e^{i\varphi_{3}} \cos\theta &~ 1 ~& \cos\theta+\sqrt{2} e^{i\varphi_{3}} \sin\theta
\end{pmatrix}Q_{\nu}\,,
\end{equation}
where
\begin{equation}
\varphi_{3}=\frac{s-t+2x+\delta}{3n}\pi\,,
\end{equation}
and its possible values are
\begin{equation}
\varphi_{3}~(\mathrm{mod}~2\pi)=0, \frac{1}{3n}\pi, \frac{2}{3n}\pi, \ldots, \frac{6n-1}{3n}\pi\,.
\end{equation}
It is easy to check that $U^{II, 1}_{\text{PMNS}}$ as well as $U^{II, 2}_{\text{PMNS}}$ have the symmetry property
\begin{eqnarray}
\nonumber&&U^{II, 1}_{\text{PMNS}}(\theta, \varphi_3+\pi)=U^{II, 1}_{\text{PMNS}}(-\theta, \varphi_3)\text{diag}(-1, 1, 1),\\
\label{eq:UPMNS_sym_caseII}&&U^{II, 2}_{\text{PMNS}}(\theta, \varphi_3+\pi)=U^{II, 2}_{\text{PMNS}}(-\theta, \varphi_3)\text{diag}(-1, 1, 1)\,.
\end{eqnarray}
We see that the second column of the PMNS matrix is $(1, 1, -\sqrt{2})/2$ or $(1, -\sqrt{2}, 1)/2$ in this case. For the mixing pattern $U^{II,1}_{\text{PMNS}}$, the three lepton mixing angles are found to be
\begin{eqnarray}
\nonumber&&\sin^2\theta_{13}=\frac{1}{8}
\left(3-\cos2\theta-2\sqrt{2}\sin2\theta\cos\varphi_3\right),\\
\nonumber&&\sin^2\theta_{12}=\frac{2}{5+\cos2\theta+2\sqrt{2}\sin2\theta\cos\varphi_3},\\
\label{eq:mixing_para_caseII_1st}&&\sin^2\theta_{23}=\frac{3-\cos2\theta+2\sqrt{2}\sin2\theta\cos\varphi_3}{5+\cos2\theta+2\sqrt{2}\sin2\theta\cos\varphi_3}\,,
\end{eqnarray}
which fulfill the following sum rules
\begin{eqnarray}
\nonumber&&\hskip4cm 4\sin^2\theta_{12}\cos^2\theta_{13}=1, \\ &&\hskip-1cm\cos^2\theta_{13}\cos^2\theta_{23}=\frac{\cos2\theta_{13}+2\cos^2\varphi_3\pm2\cos\varphi_3\sqrt{6\sin^2\theta_{13}-8\sin^4\theta_{13}-\sin^2\varphi_3}}{1+8\cos^2\varphi_3}\,.
\end{eqnarray}
Given the $3\sigma$ range of $\theta_{13}$, the solar mixing angle $\theta_{12}$ is determined to lie in the region of $0.254\leq\sin^2\theta_{12}\leq0.258$ which is rather close to its $3\sigma$ lower limit 0.259~\cite{Capozzi:2013csa}. However, this mixing pattern is a good leading order approximation because accordance with the experimental data could be easily achieved in a concrete model after higher order corrections contributions are included. We plot the $1\sigma$, $2\sigma$ and $3\sigma$ contour regions for $\sin^2\theta_{ij}$ with $ij=12, 13, 23$ in the $\varphi_3-\theta$ plane in figure~\ref{fig:caseII_contour_mixing_para}.
Obviously the most stringent constraint comes from the precisely measured reactor angle $\theta_{13}$. Moreover, the three CP rephasing invariants are given by
\begin{eqnarray}
\nonumber && \left|J_{CP}\right|=\frac{1}{8 \sqrt{2}}\left|\sin 2 \theta\sin\varphi_3\right|\,, \\
\nonumber && \left|I_1\right|=\frac{1}{8\sqrt{2}}\left|(\sin2\theta+2\sqrt{2}\cos^2\theta\cos\varphi_3)\sin\varphi_3\right|\,,\\
\label{eq:CP_para_caseII_1st}&&\left|I_2\right|=\frac{1}{8\sqrt{2}}\left|(\sin2\theta-2\sqrt{2}\cos2\theta\cos\varphi_3)\sin\varphi_3\right|\,.
\end{eqnarray}
\begin{figure}[t!]
\begin{center}
\includegraphics[width=0.50\textwidth]{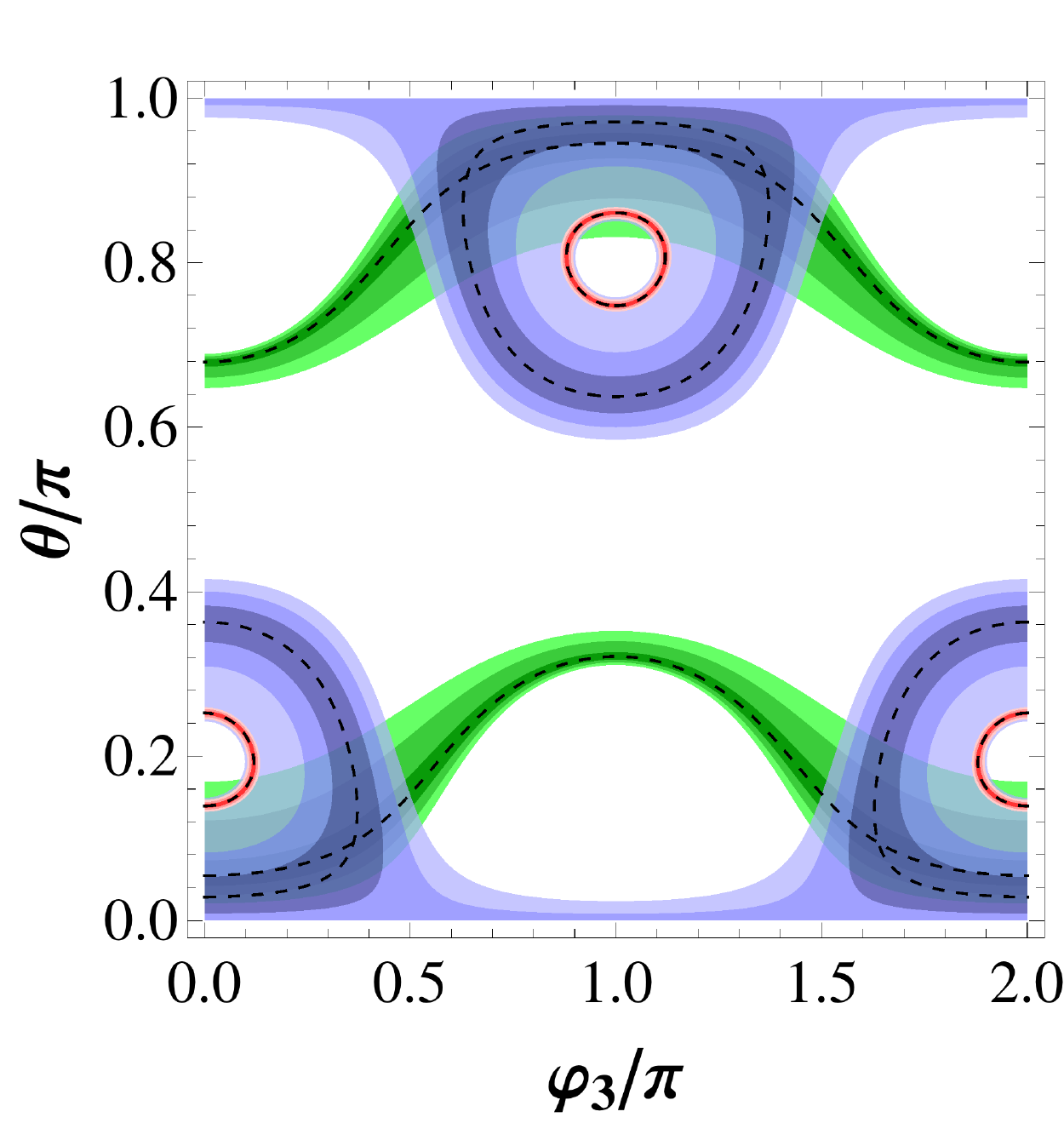}
\caption{\label{fig:caseII_contour_mixing_para}The contour regions of the three mixing angles in the case II. The red, blue and green areas denote the predictions for $\sin^2\theta_{13}$, $\sin^2\theta_{12}$ and $\sin^2\theta_{23}$ respectively.
The allowed $1\sigma$, $2\sigma$ and $3\sigma$ regions of each mixing angle are represented by different shadings. Here we take the $3\sigma$ lower limit of $\sin^2\theta_{12}$ to be 0.254 instead of 0.259 given by Ref.~\cite{Capozzi:2013csa}. The best fit values of the mixing angles are indicated by dotted lines.}
\end{center}
\end{figure}
The three CP violation phases extracted from these invariants depend on $\theta$ and $\varphi_3$. The predictions for $|\sin\delta_{CP}|$, $|\sin\alpha_{21}|$ and $|\sin\alpha^{\prime}_{31}|$ are plotted in figure~\ref{fig:caseII_contour_CP_para}, where the black areas represent the regions in which all three lepton mixing angles are in the experimentally preferred $3\sigma$ ranges. To accommodate the experimental data of mixing angles~\cite{Capozzi:2013csa}, both $\delta_{CP}$ and $\alpha_{21}$ can not be maximal. The values of $|\sin\delta_{CP}|$ and $|\sin\alpha_{21}|$ are bounded from above with $|\sin\delta_{CP}|\leq 0.895$ and $|\sin\alpha_{21}|\leq 0.545$.

\begin{figure}[t!]
\begin{center}
\includegraphics[width=0.495\textwidth]{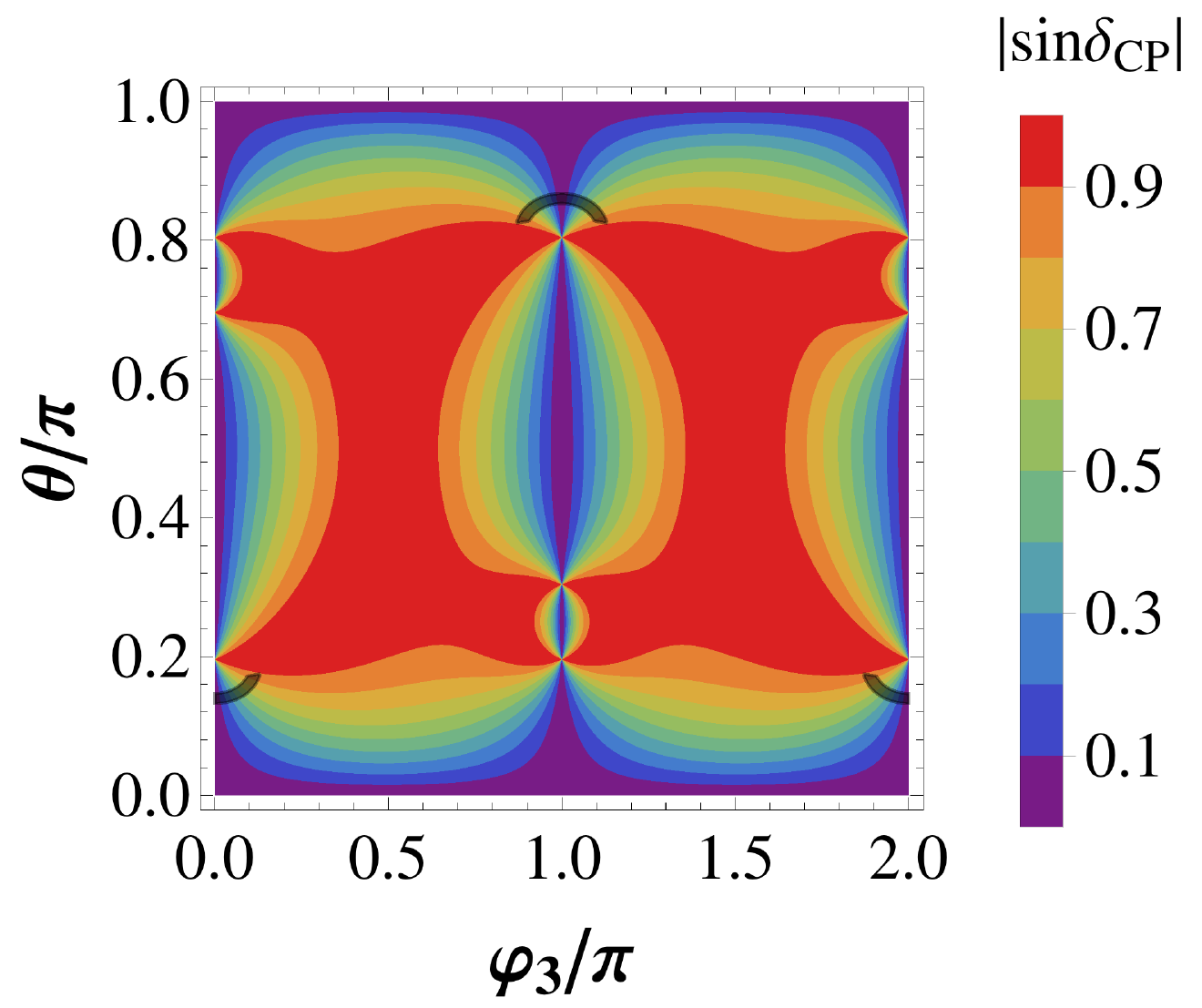}
\includegraphics[width=0.495\textwidth]{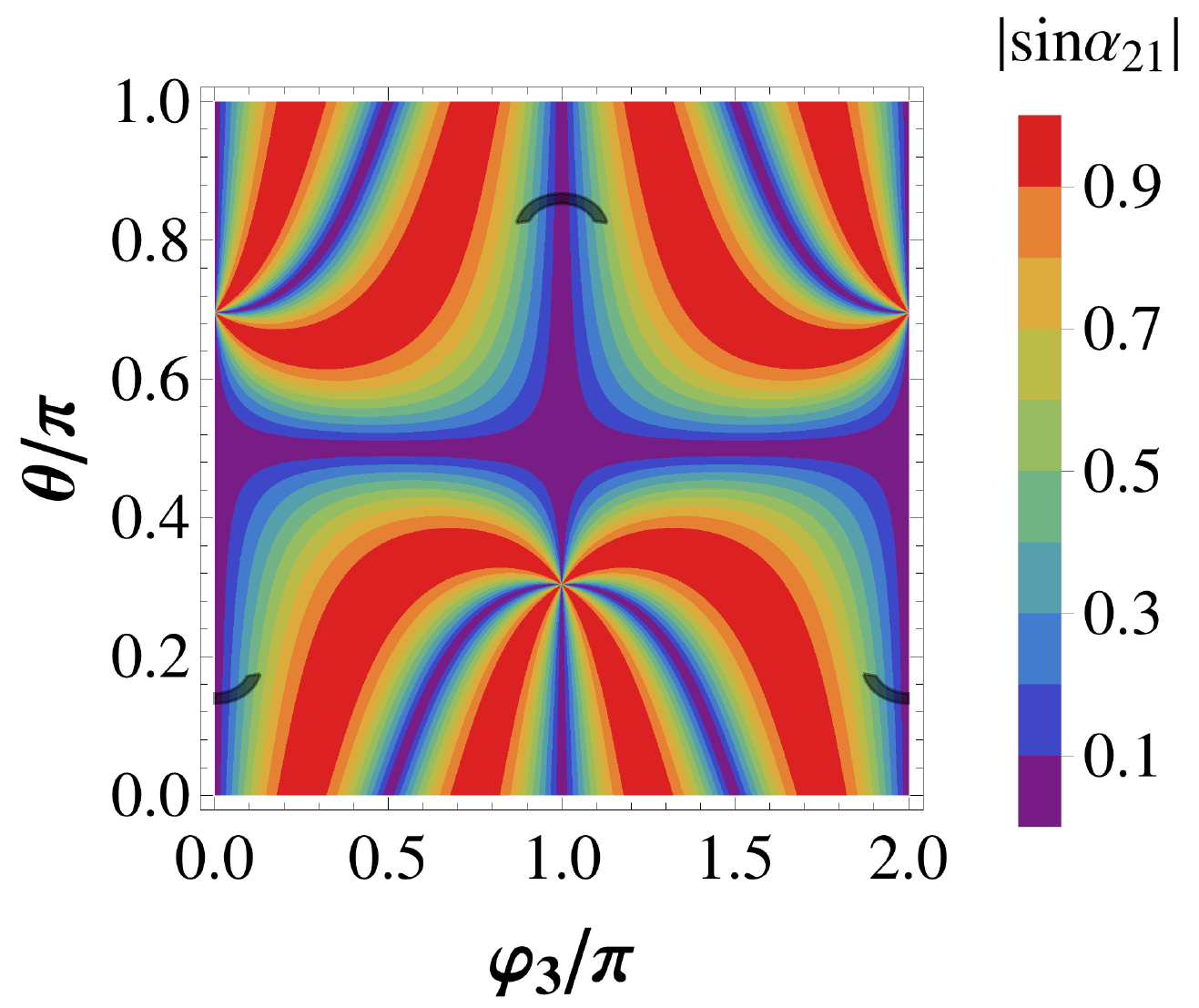}
\includegraphics[width=0.495\textwidth]{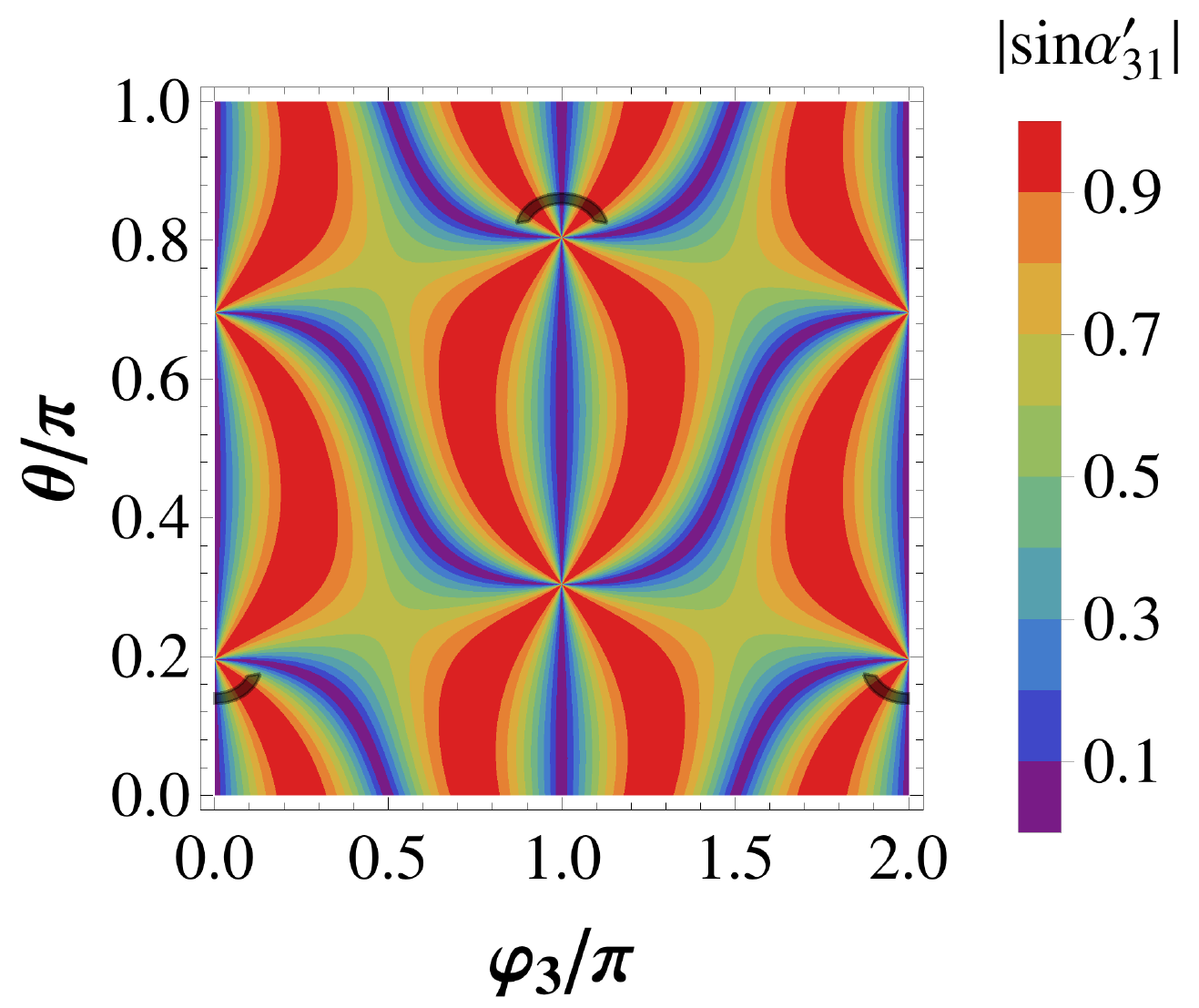}
\caption{\label{fig:caseII_contour_CP_para}The contour plots of $|\sin\delta_{CP}|$, $|\sin\alpha_{21}|$ and $|\sin\alpha^{\prime}_{31}|$ in the $\varphi_3-\theta$ plane in the case II. The black areas represent the regions in which the lepton mixing angles are compatible with experimental data at $3\sigma$ level, and it can be read out from figure~\ref{fig:caseII_contour_mixing_para}.}
\end{center}
\end{figure}

The second PMNS matrix $U^{II, 2}_{\text{PMNS}}$ can be obtained from $U^{II, 1}_{\text{PMNS}}$ by exchanging the second and third rows. Therefore $U^{II, 2}_{\text{PMNS}}$ and $U^{II, 1}_{\text{PMNS}}$ give rise to the same reactor and solar mixing angles and the Majorana phases, while the atmospheric
angle changes from $\theta_{23}$ to $\pi/2-\theta_{23}$ and the Dirac phase changes from $\delta_{CP}$ to $\pi+\delta_{CP}$. The achievable values of the mixing parameters for each $D^{(1)}_{9n, 3n}$ group are displayed in figure~\ref{fig:caseII_mixing_para}.

For the first two smallest $D^{(1)}_{9n, 3n}$ group with $n=1, 2$. The possible values of $\varphi_3$ are $0,\frac{\pi}{3}, \ldots,\frac{5\pi}{3}$ for $n=1$ and  $0,\frac{\pi}{6}, \ldots,\frac{11\pi}{6}$ for $n=2$. We find that agreement with experimental data can be achieved for $\varphi_3=0$ or $\pi$. Due to symmetry relation in Eq.~\eqref{eq:UPMNS_sym_caseII}, $\varphi_3=0$ and $\varphi_3=\pi$ should give rise to the same predictions for the mixing parameters. Therefore it is sufficient to focus on $\varphi_3=0$, and the best fitting results are listed in table~\ref{tab:caseII_n12}. Notice that all the three CP phases are predicted to take CP conserving values $\{\delta_{CP}, \alpha_{21}, \alpha_{31}\}\subseteq\{0, \pi\}$. The same conclusion can be drawn from figure~\ref{fig:caseII_mixing_para}.

\begin{table}[t!]
\centering
\footnotesize
\renewcommand{\tabcolsep}{1.0mm}
\begin{tabular}{|c|c|c|c|c|c|c|c|c|c|}
\hline \hline
\multicolumn{10}{|c|}{Case II}   \\ \hline
\multicolumn{10}{|c|}{$n=1$ and $n=2$}   \\ \hline
 & $\varphi_3$ & $\theta_{bf}$ & $\chi^2_{min}$   & $\sin^2\theta_{13}$ & $\sin^2\theta_{12}$ & $\sin^2\theta_{23}$  & $|\sin\delta_{CP}|$ & $|\sin\alpha_{21}|$ & $|\sin\alpha^{\prime}_{31}|$  \\ \hline

\multirow{2}{*}{$U^{II, 1}_{\text{PMNS}}$} & \multirow{4}{*}{$0$} &
 $0.433$ & $27.807$ & $0.0246$& $0.256$ & $0.578$ &  \multirow{4}{*}{$0$ ($0$)} &  \multirow{4}{*}{$0$ ($0$)} &  \multirow{4}{*}{$0$ ($0$)}  \\

& & $(0.435)$ & $(10.086)$  & $(0.0242)$ & $(0.256)$ & $(0.579)$  &&&   \\  \cline{1-1} \cline{3-7}

\multirow{2}{*}{$U^{II, 2}_{\text{PMNS}}$} & & $0.436$ & $9.865$  & $0.0238$ & $0.256$ &  $0.421$  & &  &   \\
& & $(0.434)$ & $(10.455)$  & $(0.0244)$ & $(0.256)$& $(0.422)$ & & & \\  \hline \hline
\end{tabular}
\caption{\label{tab:caseII_n12}Results of the $\chi^2$ analysis for $n=1, 2$ in the case II. The $\chi^2$ function has a global minimum $\chi^2_{min}$ at the best fit value $\theta_{bf}$ for $\theta$. We give the values of the mixing angles and CP violation phases for $\theta=\theta_{bf}$. The values given in parentheses denote the results for the IO neutrino mass spectrum.}
\end{table}

\begin{figure}[hptb!]
\begin{center}
\includegraphics[width=0.99\textwidth]{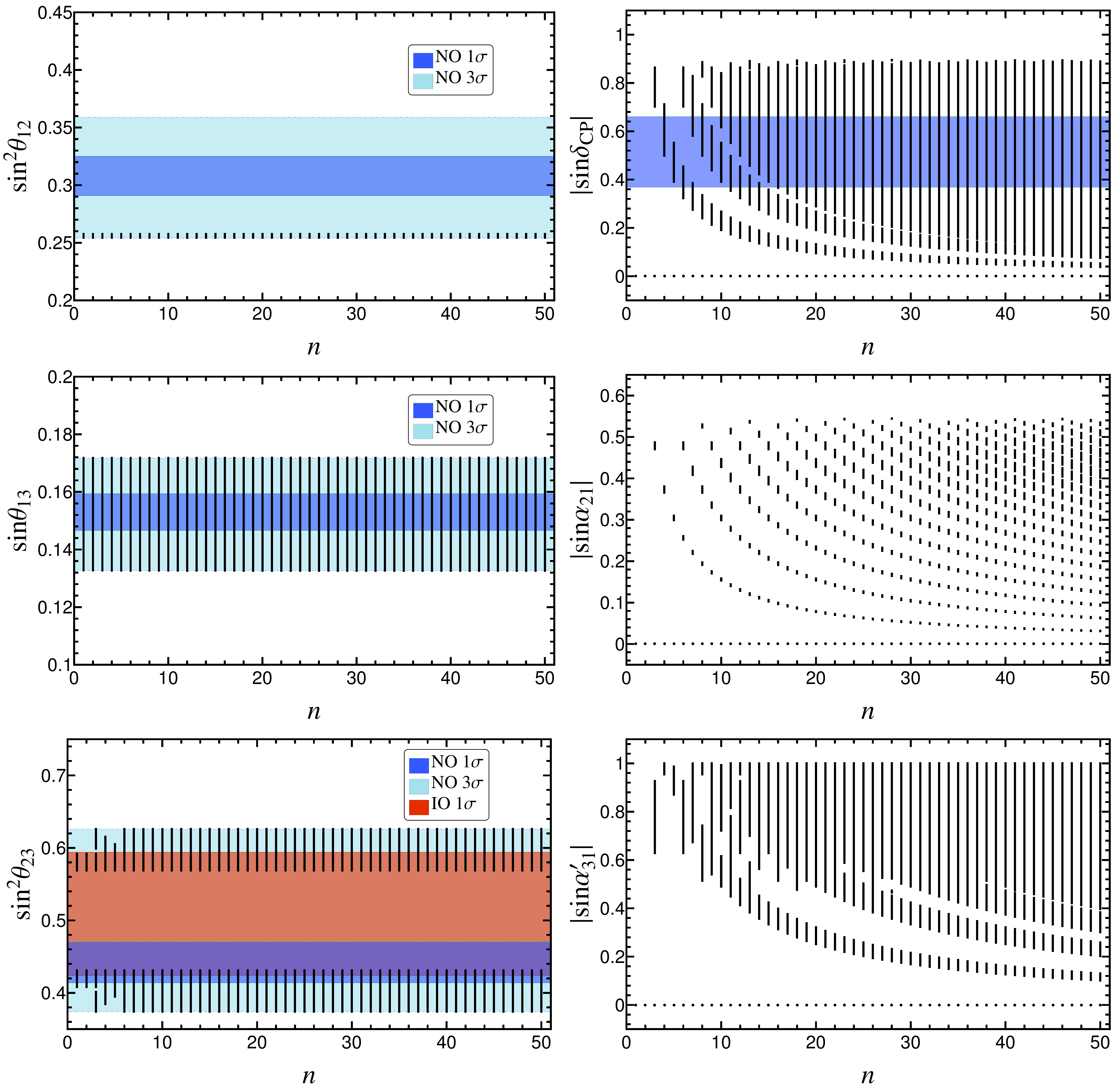}
\caption{\label{fig:caseII_mixing_para}The possible values of $\sin^2\theta_{12}$, $\sin\theta_{13}$, $\sin^2\theta_{23}$, $\left|\sin\delta_{CP}\right|$, $\left|\sin\alpha_{21}\right|$ and $\left|\sin\alpha^{\prime}_{31}\right|$ with respect to $n$ for the mixing pattern $U^{II, 1}_{\text{PMNS}}$ and $U^{II, 2}_{\text{PMNS}}$ in the case II, where the three lepton mixing angles are required to be within the experimentally preferred $3\sigma$ ranges. The $1\sigma$ and $3\sigma$ regions of the three neutrino mixing angles are adapted from global fit~\cite{Capozzi:2013csa}. Here we take the $3\sigma$ lower limit of $\sin^2\theta_{12}$ to be 0.254 instead of 0.259 given by Ref.~\cite{Capozzi:2013csa}.}
\end{center}
\end{figure}

As regards the neutrinoless double beta decay, both $U^{II, 1}_{\text{PMNS}}$ and $U^{II, 2}_{\text{PMNS}}$ yield the same effective Majorana mass:
\begin{equation}
|m_{ee}|=\frac{1}{4}\Big|m_1(\sin\theta+\sqrt{2} e^{i\varphi_{3}} \cos\theta)^2+q_1m_2+q_2m_3(\cos\theta-\sqrt{2} e^{i \varphi_{3}} \sin\theta)^2\Big|
\end{equation}
with $q_1, q_2=\pm1$. We show the predicted values of $|m_{ee}|$ in figure~\ref{fig:mee_CaseII}. Notice that for IO spectrum $|m_{ee}|$ can be either 0.0233eV or 0.0483eV which are accessible to the next generation $0\nu\beta\beta$ experiments. In the case of NO spectrum, $|m_{ee}|$ strongly depends on the lightest neutrino mass and CP parity, and it can be vanishing for certain values of the lightest neutrino mass.

\begin{figure}[t!]
\begin{center}
\includegraphics[width=0.60\linewidth]{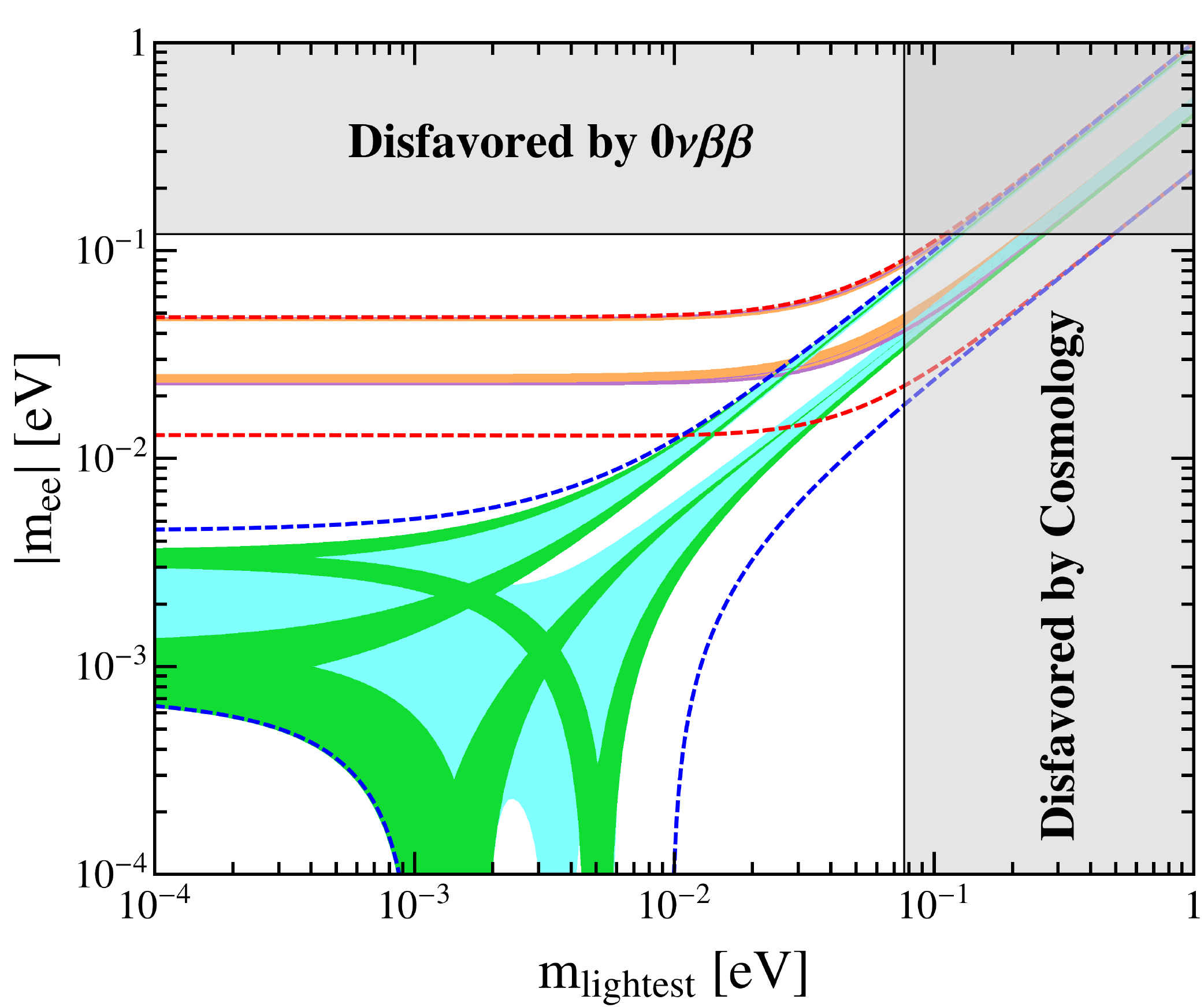}
\caption{\label{fig:mee_CaseII}The possible values of the effective Majorana mass $|m_{ee}|$ as a function of the lightest neutrino mass in the case II. The red (blue) dashed lines indicate the most general allowed regions for IO (NO) neutrino mass spectrum obtained by varying the mixing parameters over the $3\sigma$ ranges~\cite{Capozzi:2013csa}. The orange (cyan) areas denote the achievable values of $|m_{ee}|$ in the limit of $n\rightarrow\infty$ assuming IO (NO) spectrum. The purple and green regions are the theoretical predictions for the $D^{(1)}_{9n ,3n}$ group with $n=2$. Notice that the purple (green)  region overlaps the orange (cyan) one. The present most stringent upper limits $|m_{ee}|<0.120$ eV from EXO-200~\cite{Auger:2012ar, Albert:2014awa} and KamLAND-ZEN~\cite{Gando:2012zm} is shown by horizontal grey band. The vertical grey exclusion band represents the current bound coming from the cosmological data of $\sum m_i<0.230$ eV at $95\%$ confidence level obtained by the Planck collaboration~\cite{Ade:2013zuv}.
}
\end{center}
\end{figure}

\item[~~(\uppercase\expandafter{\romannumeral3})]

$G_{l}=\left\langle ac^{s}d^{t}\right\rangle$, $G_{\nu}=Z^{c^{9n/2}}_2$,
$X_{\nu\mathbf{r}}=\left\{\rho_{\mathbf{r}}(c^{\gamma}d^{\delta})\right\}$\\
In this case, $n$ should be even in order to have a $Z_2$ subgroup generated by $c^{9n/2}$. The neutrino mass matrix invariant under the assumed residual symmetry is found to take the form
\begin{equation}
m_{\nu}=
\begin{pmatrix}
m_{11} e^{-\frac{i2 \pi  \gamma }{9 n}}  & ~m_{12} e^{-\frac{i \pi  (2 \gamma -3 \delta )}{9n}}  ~& 0 \\
m_{12} e^{-\frac{i \pi  (2 \gamma -3 \delta )}{9n}} &~m_{22} e^{-\frac{i 2\pi  (\gamma -3 \delta )}{9 n}}  ~& 0 \\
 0 &~ 0 ~& m_{33} e^{\frac{i2 \pi  (2 \gamma -3 \delta )}{9 n}}
\end{pmatrix}\,,
\end{equation}
where $m_{11}$, $m_{12}$, $m_{13}$ and $m_{22}$ are real. It is diagonalized by the unitary matrix
\begin{equation}
\label{eq:unu_c9n/2}
U_{\nu}=
\begin{pmatrix}
 e^{\frac{i\pi\gamma }{9n}}\cos\theta  &~ e^{\frac{i\pi\gamma }{9n}}\sin\theta ~& 0 \\
 -e^{\frac{i \pi  (\gamma -3 \delta )}{9n}}\sin\theta &~ e^{\frac{i \pi  (\gamma -3 \delta )}{9n}}\cos\theta ~& 0 \\
 0  &~ 0 ~& e^{-\frac{i \pi  (2 \gamma -3 \delta )}{9n}}
\end{pmatrix}Q_{\nu}\,,
\end{equation}
with the rotation angle $\theta$ satisfying
\begin{equation}
\tan2\theta=\frac{2m_{12}}{m_{22}-m_{11}}\,.
\end{equation}
The light neutrino masses are
\begin{eqnarray}
\nonumber&&m_1=\frac{1}{2}\left|m_{11}+m_{22}-\frac{m_{22}-m_{11}}{\cos2\theta}\right|,\\
\nonumber&&m_2=\frac{1}{2}\left|m_{11}+m_{22}+\frac{m_{22}-m_{11}}{\cos2\theta}\right|,\\
\nonumber&&m_3=\left|m_{33}\right|\,.
\end{eqnarray}
As the residual flavor symmetry in the charged lepton sector is $G_{l}=\left\langle ac^{s}d^{t}\right\rangle$, the charged lepton diagonalization matrix is $U^{(3)}_{l}$ shown in table~\ref{tab:cle_diagonal_matrix}. Thus the lepton mixing matrix is determined to be
\begin{equation}
\label{eq:PMNS_case_III}
U^{III}_{\text{PMNS}}=\frac{1}{\sqrt{3}}
\begin{pmatrix}
\cos\theta-e^{i\varphi_4} \sin\theta  &~ 1 ~&\sin\theta+ e^{i\varphi_4} \cos \theta \\
\omega \cos\theta-\omega^2e^{i\varphi_4}\sin\theta  &~ 1 ~& \omega\sin\theta+\omega^2 e^{i\varphi_4}\cos\theta  \\
\omega^2\cos\theta-\omega e^{i\varphi_4} \sin\theta &~ 1 ~&   \omega^2\sin\theta+\omega e^{i\varphi_4} \cos\theta
\end{pmatrix}\text{diag}(e^{i\varphi_5}, 1, e^{i\varphi_5})Q_{\nu}\,,
\end{equation}
where
\begin{equation}
\varphi_4=\frac{2s-6t-3\delta}{9n}\pi,\qquad \varphi_{5}=\frac{2s+3\gamma-3\delta}{9n}\pi\,.
\end{equation}
Notice that $\varphi_4$ and $\varphi_{5}$ are not completely independent, and they can take the following discrete values:
\begin{eqnarray}
\nonumber&&\varphi_4~(\mathrm{mod}~2\pi)=0, \frac{1}{9n}\pi, \frac{2}{9n}\pi, \ldots, \frac{18n-1}{9n}\pi\,,\\
&&\varphi_4-\varphi_{5}~(\mathrm{mod}~2\pi)=0, \frac{1}{3n}\pi, \frac{2}{3n}\pi, \ldots, \frac{6n-1}{3n}\pi\,.
\end{eqnarray}
We easily see that one column of the PMNS matrix is $(1, 1, 1)^{T}/\sqrt{3}$ which can only be the second column vector in order to accommodate the experimental data of lepton mixing angles. The permutations of the PMNS matrix which leave the second column unchanged don't lead to physically different results. From Eq.~\eqref{eq:PMNS_case_III} we can extract the mixing angles
\begin{eqnarray}
\nonumber&&\sin^2\theta_{13}=\frac{1}{3}\left[1+\sin2\theta\cos\varphi_4\right]\,,\\
\nonumber&&\sin^2\theta_{12}=\frac{1}{2-\sin2\theta\cos\varphi_4}\,,\\
\label{eq:mixing_para_caseIII}&&\sin^2\theta_{23}=\frac{1-\sin2\theta\sin\left(\varphi_4+\pi/6\right)}{2-\sin2\theta\cos\varphi_4}\,.
\end{eqnarray}
Then we can derive the following sum rules among the mixing angles
\begin{subequations}
\begin{eqnarray}
\label{eq:corre_caseIII_1}&&\hskip3cm 3\sin^2\theta_{12}\cos^2\theta_{13}=1,\\
\label{eq:corre_caseIII_2}&&
(1-3\sin^2\theta_{13})\tan\varphi_4+\sqrt{3}\cos^2\theta_{13}\cos2\theta_{23}=0\,.
\end{eqnarray}
\end{subequations}
The correlation of Eq.~\eqref{eq:corre_caseIII_1} yields $\sin^2\theta_{12}\simeq0.341$ for the best fit value $\sin^2\theta_{13}=0.0234$~\cite{Capozzi:2013csa}.
Inputting the $3\sigma$ ranges of the atmospheric as well reactor mixing angles in Eq.~\eqref{eq:corre_caseIII_2}, we find the phase difference $\varphi_4$ should vary in the interval
\begin{equation}
\label{eq:varphi4_range}\varphi_4\in\left[0,0.138\pi\right]\cup\left[0.862\pi,1.138\pi\right]\cup\left[1.862\pi,2\pi\right]\,.
\end{equation}
As shown in Eq.~\eqref{eq:mixing_para_caseIII}, all the three mixing angles are expressed in terms of $\varphi_4$ and $\theta$. The contour regions for $\sin^2\theta_{ij}$ in the plane of $\varphi_4$ and $\theta$ are displayed in figure~\ref{fig:caseIII_contour_mixing_para}. One can see that agreement with experimental data can be achieved for appropriate values of $\varphi_4$ and $\theta$.
\begin{figure}[t!]
\begin{center}
\includegraphics[width=0.50\textwidth]{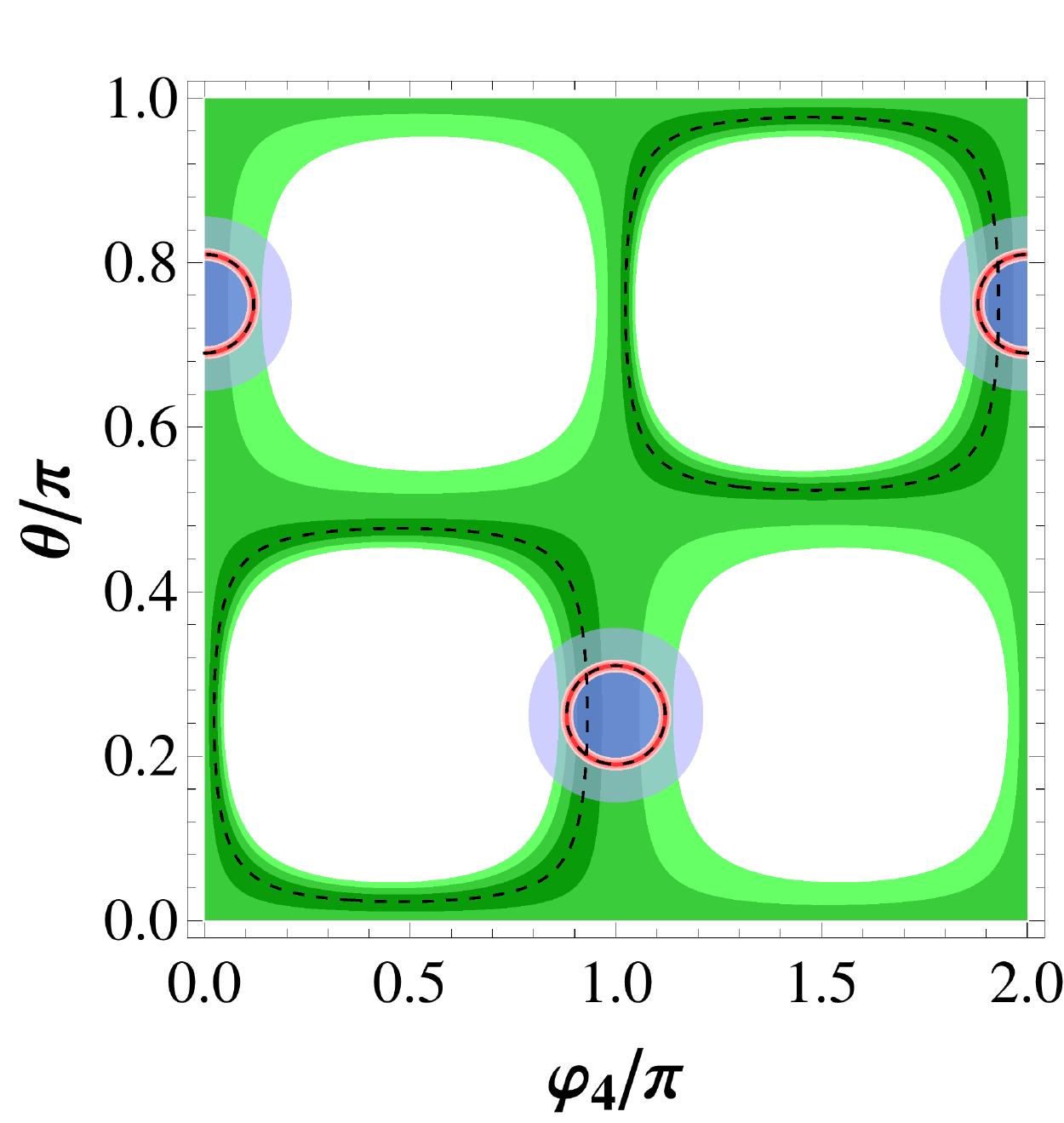}
\caption{\label{fig:caseIII_contour_mixing_para}
The contour regions of the three mixing angles in the case III. The red, blue and green areas denote the predictions for $\sin^2\theta_{13}$, $\sin^2\theta_{12}$ and $\sin^2\theta_{23}$ respectively. The allowed $1\sigma$, $2\sigma$ and $3\sigma$ regions of each mixing angle are represented by different shadings. The best fit values of the mixing angles are indicated by dotted lines. Note that both the $1\sigma$ range and the best fit value of $\theta_{12}$ can not be achieved in this case because of the sum rule in Eq.~\eqref{eq:corre_caseIII_1}.}
\end{center}
\end{figure}
Furthermore, the CP invariants are given by
\begin{eqnarray}
\nonumber && \left|J_{CP}\right|=\frac{1}{6 \sqrt{3}}\left|\cos 2 \theta\right|  \,, \\
\nonumber && \left|I_1\right|=\frac{1}{9}\left|\cos^2\theta \sin2\varphi_5-\sin2\theta \sin(\varphi_4+2\varphi_5)+\sin^2\theta\sin(2\varphi_4+2\varphi_5)\right|  \,, \\
 && \left|I_2\right|= \frac{1}{9}\left|\cos2\theta\sin2\varphi_4\right| \,.
\end{eqnarray}
Thus both $\delta_{CP}$ and $\alpha^{\prime}_{31}$ only depend on $\varphi_4$ and $\theta$ while the Majorana phase $\alpha_{21}$ is dependent on all the three parameters $\varphi_4$, $\varphi_5$ and $\theta$. The predictions for $|\sin\delta_{CP}|$ and $|\sin\alpha^{\prime}_{31}|$ are shown in the plane $\theta$ versus $\varphi_4$ in figure~\ref{fig:caseIII_contour_CP_para}. One can see that all values of the CP phases are possible in the regions where the lepton mixing angles are compatible with the experimental data at $3\sigma$ level. Moreover, the possible values of the mixing angles and CP phases for each $D^{(1)}_{9n, 3n}$ group until $n=50$ are plotted in figure~\ref{fig:caseIII_mixing_para}.
\begin{figure}[hptb]
\begin{center}
\includegraphics[width=0.495\textwidth]{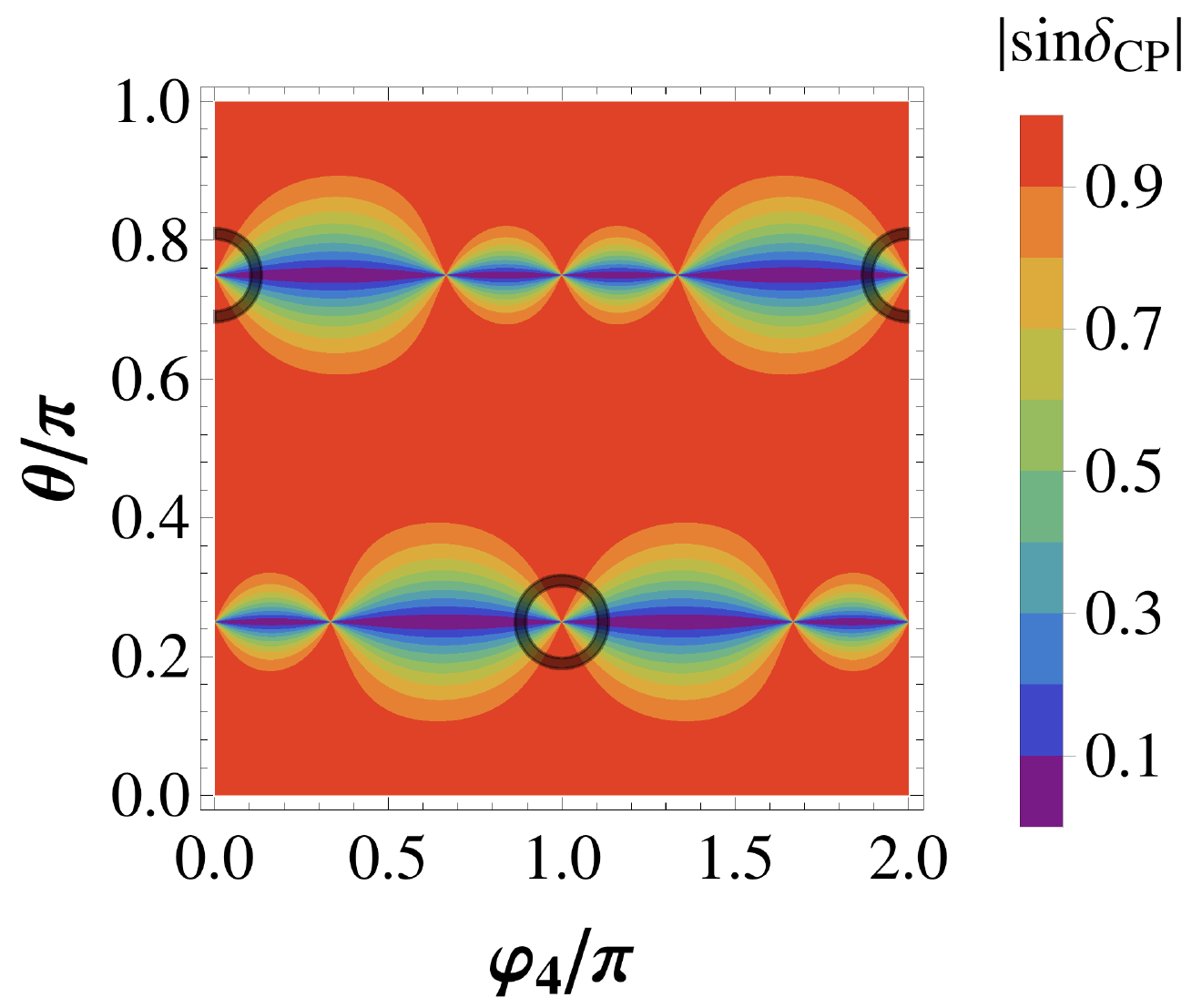}
\includegraphics[width=0.495\textwidth]{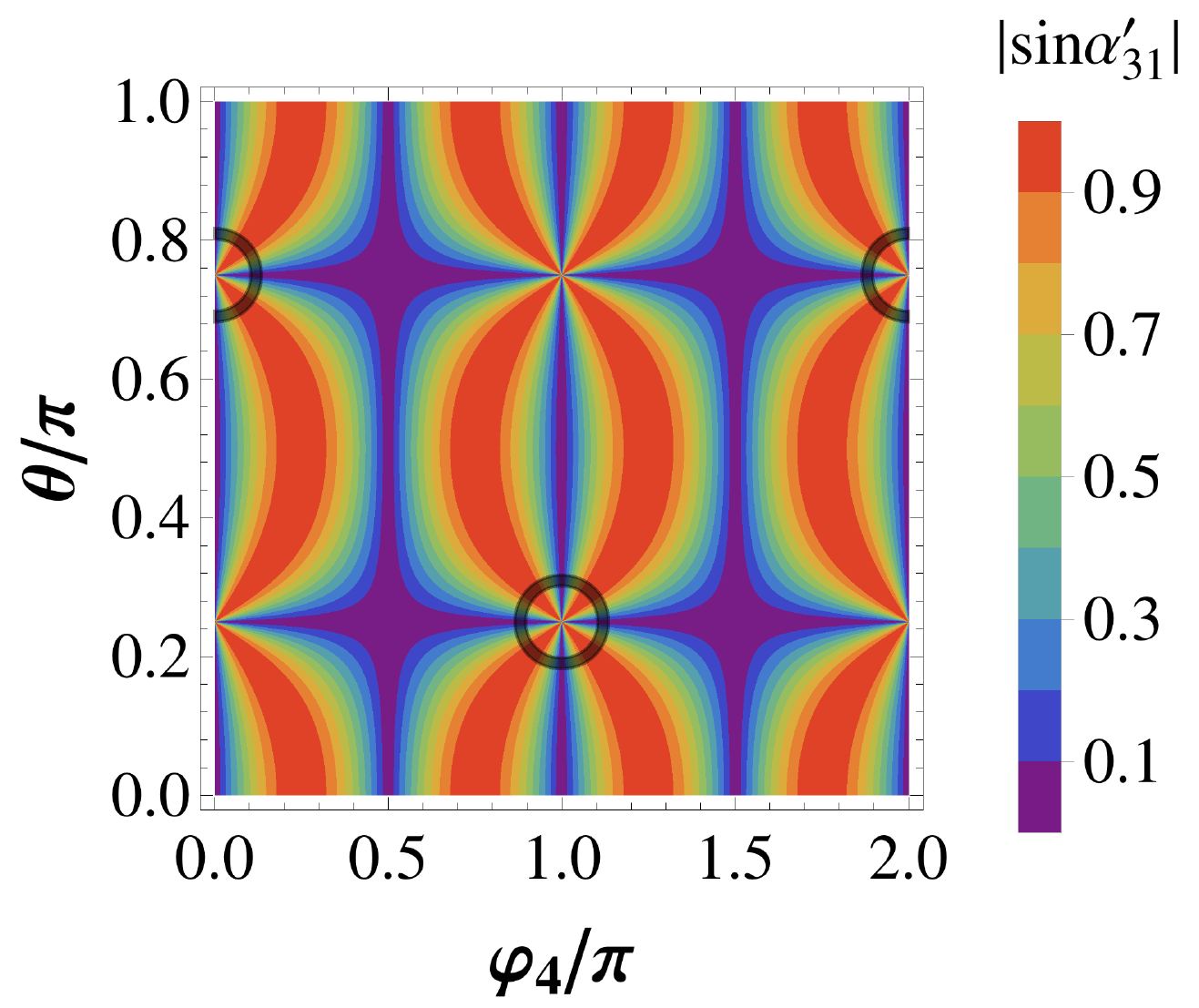}
\caption{\label{fig:caseIII_contour_CP_para}The contour plots of $|\sin\delta_{CP}|$ and $|\sin\alpha^{\prime}_{31}|$ in the case III. The black areas represent the regions in which the lepton mixing angles are compatible with experimental data at $3\sigma$ level, and it can be read out from figure~\ref{fig:caseIII_contour_mixing_para}.}
\end{center}
\end{figure}

\begin{figure}[hptb]
\begin{center}
\includegraphics[width=0.99\textwidth]{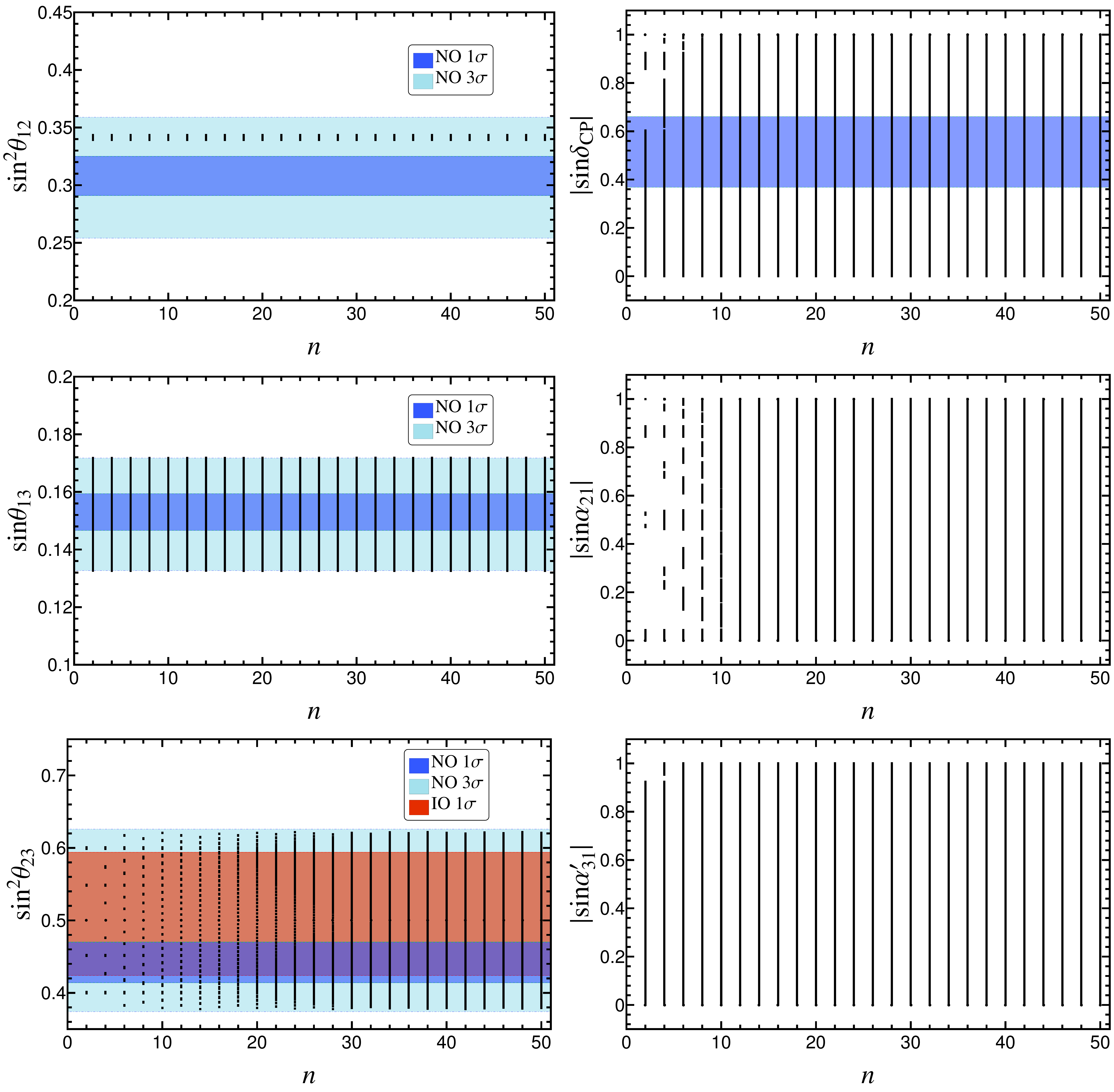}
\caption{\label{fig:caseIII_mixing_para}The possible values of $\sin^2\theta_{12}$, $\sin\theta_{13}$, $\sin^2\theta_{23}$, $\left|\sin\delta_{CP}\right|$, $\left|\sin\alpha_{21}\right|$ and $\left|\sin\alpha^{\prime}_{31}\right|$ with respect to $n$ for the mixing pattern $U^{III}_{\text{PMNS}}$ in the case III, where the three lepton mixing angles are required to be within the experimentally preferred $3\sigma$ ranges. The $1\sigma$ and $3\sigma$ regions of the three neutrino mixing angles are adapted from global fit~\cite{Capozzi:2013csa}.}
\end{center}
\end{figure}

Then we proceed to study the phenomenologically viable mixing patterns which can be derived from the $D^{(1)}_{9n, 3n}$ group with $n=2$. Note that the index $n$ has to be even in this case. We can check that the PMNS matrix given by Eq.~\eqref{eq:PMNS_case_III} has the following symmetry properties
\begin{eqnarray}
\nonumber&&U^{III}_{\text{PMNS}}(\theta, \pi+\varphi_4, \varphi_5)=U^{III}_{\text{PMNS}}(-\theta, \varphi_4, \varphi_5)\text{diag}(1, 1, -1),\\
&&U^{III}_{\text{PMNS}}(\theta, \varphi_4, \pi/2+\varphi_5)=U^{III}_{\text{PMNS}}(\theta, \varphi_4, \varphi_5)\text{diag}(i, 1, i)\,,
\end{eqnarray}
where the diagonal matrix on the right-handed side can be absorbed into $Q_{\nu}$. That is to say, both $U^{III}_{\text{PMNS}}(\theta, \pi+\varphi_4, \varphi_5)$ and $U^{III}_{\text{PMNS}}(\theta, \varphi_4, \pi/2+\varphi_5)$ give rise to the same predictions for the lepton mixing parameters as $U^{III}_{\text{PMNS}}(\theta, \varphi_4, \varphi_5)$ up to redefinition of the free parameter $\theta$. Hence we can take the fundamental intervals of $\varphi_4$ and $\varphi_5$ to be $[0, \pi)$ and $[0, \pi/2)$ respectively. The allowed values of $\varphi_4$ are $0$, $\pi/18$, $\pi/9$, $\ldots$, $17\pi/9$, $35\pi/18$. However, only $\varphi_4~(\mathrm{mod}~\pi)=0$, $\pi/18$, $\pi/9$, $8\pi/9$ and $17\pi/18$ are within the range of Eq.~\eqref{eq:varphi4_range} such that they can give a good fit to the experimental data. The results of the $\chi^2$ analysis are summarized in table~\ref{tab:caseIII_n12}. Notice that the best fitting values of the mixing angles and $|\sin\delta_{CP}|$, $|\sin\alpha^{\prime}_{31}|$ are dependent on $\varphi_4$ while the best fitting value of $|\sin\alpha_{21}|$ depends on $\varphi_4$ as well as $\varphi_5$. The mixing patterns with the same $\varphi_4$ but different $\varphi_5$ are expected to be distinguished by some rare processes which are sensitive to the Majorana phases such as the neutrinoless double decay and the radiative emission of neutrino pair in atoms~\cite{Yoshimura:2006nd}. In this case, the effective Majorana mass $|m_{ee}|$ is predicted to be
\begin{equation}
|m_{ee}|=\frac{1}{3}\Big|m_1(\cos\theta-e^{i\varphi_4} \sin\theta)^2+q_1m_2e^{-2i\varphi_5}+q_2m_3(\sin\theta+ e^{i\varphi_4}\cos \theta)^2\Big|\,,
\end{equation}
where $q_1, q_2=\pm1$. The numerical results are shown in figure~\ref{fig:mee_CaseIII}.

\begin{table}[hptb]
\centering
\footnotesize
\renewcommand{\tabcolsep}{1.5mm}
\begin{tabular}{|c|c|c|c|c|c|c|c|c|c|}
\hline \hline
\multicolumn{10}{|c|}{Case III}   \\ \hline
\multicolumn{10}{|c|}{ $n=2$}   \\ \hline
$\varphi_4$ & $\varphi_5$ & $\theta_{bf}$ & $\chi^2_{min}$   & $\sin^2\theta_{13}$ & $\sin^2\theta_{12}$ & $\sin^2\theta_{23}$  & $|\sin\delta_{CP}|$ & $|\sin\alpha_{21}|$ & $|\sin\alpha^{\prime}_{31}|$  \\ \hline

\multirow{2}{*}{$0$} & \multirow{2}{*}{$0,\frac{\pi}{6},\frac{\pi}{3}$} &$2.168$ &$7.480$ & $0.0233$& $0.341$ & $0.5$ & \multirow{2}{*}{$1$($1$)} &$|\sin2\varphi_5|$  &  \multirow{2}{*}{$0$ ($0$)}  \\
&& ($2.166$) & ($3.987$) & ($0.0238$) & ($0.341$) & ($0.5$) && ($|\sin2\varphi_5|$)&   \\  \hline

\multirow{6}{*}{$\frac{\pi}{18}$} & \multirow{2}{*}{$\frac{\pi}{18}$} &\multirow{5}{*}{$2.189$} &\multirow{5}{*}{$15.247$} & \multirow{5}{*}{$0.0233$} & \multirow{5}{*}{$0.341$} & \multirow{5}{*}{$0.548$} & \multirow{5}{*}{$0.899$} & $0.525$ & \multirow{5}{*}{$0.833$}   \\
&& \multirow{5}{*}{($2.186$)} & \multirow{5}{*}{($4.334$)}  & \multirow{5}{*}{($0.0238$)} & \multirow{5}{*}{($0.341$)} & \multirow{5}{*}{($0.548$)}  &  \multirow{5}{*}{($0.902$)} & ($0.526$) &  \multirow{5}{*}{($0.827$)}  \\ \cline{2-2} \cline{9-9}
&\multirow{2}{*}{$\frac{2\pi}{9}$} &&&&& & & $0.9996$ &  \\
& &&&&& & & ($0.9996$) &  \\  \cline{2-2} \cline{9-9}
&\multirow{2}{*}{$\frac{7\pi}{18}$} &&&&& & & $0.474$ &  \\
& &&&&& & & ($0.474$) &  \\  \hline

\multirow{6}{*}{$\frac{17\pi}{18}$} & \multirow{2}{*}{$\frac{\pi}{9}$} &\multirow{5}{*}{$0.953$} &\multirow{5}{*}{$4.029$} & \multirow{5}{*}{$0.0232$} & \multirow{5}{*}{$0.341$} & \multirow{5}{*}{$0.452$} & \multirow{5}{*}{$0.899$} & $0.474$ & \multirow{5}{*}{$0.834$}   \\
& &\multirow{5}{*}{($0.955$)} & \multirow{5}{*}{($3.894$)}  & \multirow{5}{*}{($0.0238$)} & \multirow{5}{*}{($0.341$)} & \multirow{5}{*}{($0.452$)}  &  \multirow{5}{*}{($0.902$)} & ($0.474$) &  \multirow{5}{*}{($0.827$)}   \\  \cline{2-2} \cline{9-9}
&\multirow{2}{*}{$\frac{5\pi}{18}$} &&&&& & & $0.9996$ &  \\
& &&&&& & & ($0.9996$) &  \\  \cline{2-2} \cline{9-9}
&\multirow{2}{*}{$\frac{4\pi}{9}$} &&&&& & & $0.525$ &  \\
& &&&&& & &  ($0.526$) &  \\  \hline

\multirow{6}{*}{$\frac{\pi}{9}$} & \multirow{2}{*}{$\frac{\pi}{9}$} &\multirow{5}{*}{$2.429$} &\multirow{5}{*}{$28.249$} & \multirow{5}{*}{$0.0234$} & \multirow{5}{*}{$0.341$} & \multirow{5}{*}{$0.600$} & \multirow{5}{*}{$0.400$} & $0.853$ & \multirow{5}{*}{$0.685$}   \\
& &\multirow{5}{*}{($2.433$)} & \multirow{5}{*}{($4.970$)}  & \multirow{5}{*}{($0.0238$)} & \multirow{5}{*}{($0.341$)} & \multirow{5}{*}{($0.600$)}  &  \multirow{5}{*}{($0.422$)} & ($0.852$) &  \multirow{5}{*}{($0.716$)}   \\  \cline{2-2} \cline{9-9}
&\multirow{2}{*}{$\frac{5\pi}{18}$} &&&&& & & $0.879$ &  \\
& &&&&& & & ($0.879$) &  \\  \cline{2-2} \cline{9-9}
&\multirow{2}{*}{$\frac{4\pi}{9}$} &&&&& & & $0.0255$ &  \\
& &&&&& & & ($0.0272$) &  \\  \hline

\multirow{6}{*}{$\frac{8\pi}{9}$} & \multirow{2}{*}{$\frac{\pi}{18}$} &\multirow{5}{*}{$0.714$} &\multirow{5}{*}{$6.432$} & \multirow{5}{*}{$0.0233$} & \multirow{5}{*}{$0.341$} & \multirow{5}{*}{$0.400$} & \multirow{5}{*}{$0.397$} & $0.0253$ & \multirow{5}{*}{$0.680$}   \\
& & \multirow{5}{*}{($0.708$)} & \multirow{5}{*}{($7.023$)}  & \multirow{5}{*}{($0.0239$)} & \multirow{5}{*}{($0.341$)} & \multirow{5}{*}{($0.400$)}  &  \multirow{5}{*}{($0.424$)} & $(0.0274$) &  \multirow{5}{*}{($0.719$)}  \\  \cline{2-2} \cline{9-9}
&\multirow{2}{*}{$\frac{2\pi}{9}$} &&&&& & & $0.878$ &  \\
& &&&&& & & ($0.879$) &  \\  \cline{2-2} \cline{9-9}
&\multirow{2}{*}{$\frac{7\pi}{18}$} &&&&& & & $0.853$ &  \\
& &&&&& & & ($0.852$) &  \\
 \hline \hline
\end{tabular}
\caption{\label{tab:caseIII_n12}Results of the $\chi^2$ analysis for $n=2$ in the case III. The $\chi^2$ function has a global minimum $\chi^2_{min}$ at the best fit value $\theta_{bf}$ for $\theta$. We give the values of the mixing angles and CP violation phases for $\theta=\theta_{bf}$. The values given in parentheses denote the results for the IO neutrino mass spectrum.
Notice that $\theta=\pi/2-\theta_{bf}$ gives rise to the same results for the mixing parameters except $|\sin\alpha_{21}|$, because the PMNS matrix $U^{III}_{\text{PMNS}}$ fulfills $U^{III}_{\text{PMNS}}(\theta, \varphi_4, \varphi_5)=[U^{III}_{\text{PMNS}}(\pi/2-\theta, \varphi_4, -\varphi_4-\varphi_5)]^{\ast}$.}
\end{table}

\begin{figure}[t!]
\begin{center}
\includegraphics[width=0.60\linewidth]{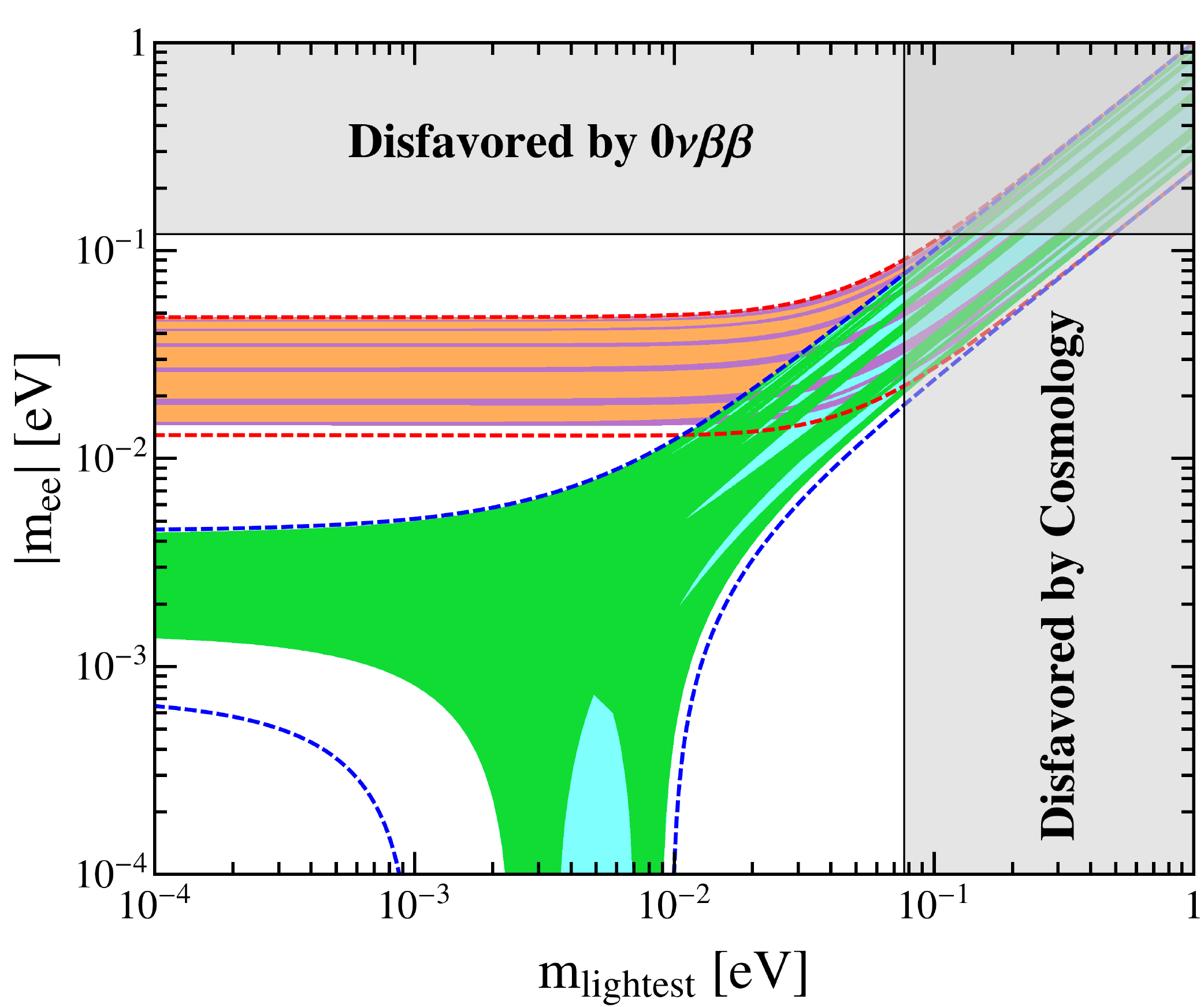}
\caption{\label{fig:mee_CaseIII}The possible values of the effective Majorana mass $|m_{ee}|$ as a function of the lightest neutrino mass in the case III. The red (blue) dashed lines indicate the most general allowed regions for IO (NO) neutrino mass spectrum obtained by varying the mixing parameters over the $3\sigma$ ranges~\cite{Capozzi:2013csa}. The orange (cyan) areas denote the achievable values of $|m_{ee}|$ in the limit of $n\rightarrow\infty$ assuming IO (NO) spectrum. The purple and green regions are the theoretical predictions for the $D^{(1)}_{9n ,3n}$ group with $n=2$. Notice that the purple (green)  region overlaps the orange (cyan) one. The present most stringent upper limits $|m_{ee}|<0.120$ eV from EXO-200~\cite{Auger:2012ar, Albert:2014awa} and KamLAND-ZEN~\cite{Gando:2012zm} is shown by horizontal grey band. The vertical grey exclusion band represents the current bound coming from the cosmological data of $\sum m_i<0.230$ eV at $95\%$ confidence level obtained by the Planck collaboration~\cite{Ade:2013zuv}.
}
\end{center}
\end{figure}

\item[~~(\uppercase\expandafter{\romannumeral4})]

$G_{l}=\left\langle ac^{s}d^{t}\right\rangle$, $G_{\nu}=Z^{c^{9n/2}}_2$,
$X_{\nu\mathbf{r}}=\left\{\rho_{\mathbf{r}}(a^{2}bc^{\gamma})\right\}$  \\
This case differs from the case III in the residual CP transformation of the neutrino sector. The group index $n$ has to be an even integer as well. In the same way, the neutrino mass matrix invariant under the assumed remnant symmetry is determined to be
\begin{equation}
m_{\nu}=\begin{pmatrix}
m_{11} e^{- i \left(\theta+\frac{4 \pi\gamma}{9n}\right)}
&~ m_{12}e^{-\frac{2 i \pi  \gamma }{9 n}}  ~& 0 \\
 m_{12}e^{-\frac{2 i\pi\gamma}{9 n}}  &~ m_{11}e^{i\theta} ~& 0 \\
 0 &~ 0 ~& m_{33}e^{\frac{4i\pi\gamma}{9n}}
\end{pmatrix}\,,
\end{equation}
where $m_{11}$, $m_{12}$, $m_{33}$ and $\theta$ are real parameters. The unitary transformation $U_{\nu}$ which diagonalizes $m_{\nu}$ is of the form
\begin{equation}
U_{\nu}=\frac{1}{\sqrt{2}}
\begin{pmatrix}
 e^{ i \left(\frac{\theta}{2}+\frac{2\pi\gamma}{9n}\right)}  &~ e^{ i \left(\frac{\theta}{2}+\frac{2\pi\gamma }{9n}\right)} ~& 0  \\
 -e^{-i\frac{\theta}{2}} &~ e^{-i\frac{\theta}{2}} ~& 0 \\
 0  & 0 & \sqrt{2}e^{-\frac{2 i \pi  \gamma }{9 n}}
\end{pmatrix}Q_{\nu}\,.
\end{equation}
The light neutrino masses are given by
\begin{equation}
m_1=\left|m_{11}-m_{12}\right|,\quad m_2=\left|m_{11}+m_{12}\right|,\quad m_{3}=\left|m_{33}\right|\,.
\end{equation}
Then we find that the PMNS matrix takes the form
\begin{equation}
\label{eq:PMNS_case_IV}
U^{IV}_{\text{PMNS}}=\frac{1}{\sqrt{3}}\left(
\begin{array}{ccc}
 \sqrt{2}e^{i\varphi_7}\sin\left(\varphi_6+\frac{\theta}{2}\right) ~&~ 1
~&~ \sqrt{2}e^{i\varphi_7}\cos\left(\varphi_6+\frac{\theta}{2}\right) \\
\sqrt{2}e^{i\varphi_7}\cos\left(\varphi_6+\frac{\theta}{2}+\frac{\pi}{6}\right)
~&~1~&~-\sqrt{2}e^{i\varphi_7}\sin\left(\varphi_6+\frac{\theta}{2}+\frac{\pi}{6}\right)\\
-\sqrt{2}e^{i\varphi_7}\cos\left(\varphi_6+\frac{\theta}{2}-\frac{\pi}{6}\right)
~&~1~&~\sqrt{2}e^{i\varphi_7}\sin\left(\varphi_6+\frac{\theta}{2}-\frac{\pi}{6}\right)
\end{array}
\right)Q_{\nu}\,,
\end{equation}
with
\begin{equation}
\varphi_6=\frac{\gamma-s+3t}{9n}\pi ,\qquad
\varphi_7=\frac{\gamma+s-t}{3n}\pi\,.
\end{equation}
Obviously the second column of the PMNS matrix is $\left(1/\sqrt{3}, 1/\sqrt{3}, 1/\sqrt{3}\right)^{T}$ as well. The mixing parameters extracted from Eq.~\eqref{eq:PMNS_case_IV} are:
\begin{eqnarray}
\nonumber&&\sin^2\theta_{13}=\frac{1}{3}\left[1+\cos(\theta+2\varphi_6)\right],\quad
\sin^2\theta_{12}=\frac{1}{2-\cos(\theta+2\varphi_6)}\,,\\
\nonumber&&\sin^2\theta_{23}=\frac{1+\sin\left(\theta+2\varphi_6-\pi/6\right)}{2-\cos(\theta+2\varphi_6)}\,,\\
\label{eq:mixing_parameters_case_XIII}&&\sin\delta_{CP}=\sin\alpha_{31}=0,\qquad
\left|\sin\alpha_{21}\right|=|\sin(2\varphi_7)|\,.
\end{eqnarray}
Notice that the mixing angles depend on the combination $\theta+2\varphi_6$ so that the values of the parameter $\varphi_6$ is irrelevant. Moreover, we can see that the mixing angles fulfill the following sum rules
\begin{equation}
3\cos^2\theta_{13}\sin^2\theta_{12}=1,\qquad
2\sin^2\theta_{23}=1\pm\tan\theta_{13}\sqrt{2-\tan^2\theta_{13}}\,.
\end{equation}
The $3\sigma$ range of the reactor angle $0.0176\leq\sin^2\theta_{13}\leq0.0295$~\cite{Capozzi:2013csa} can be reproduced for
\begin{equation}
\theta+2\varphi_6\in[0.865\pi, 0.896\pi]\cup[1.104\pi, 1.135\pi]\,.
\end{equation}
Thus the solar and atmospheric mixing angles are determined to be within the intervals
\begin{eqnarray}
\nonumber&&\hskip2.3cm 0.339\leq\sin^2\theta_{12}\leq0.343\,,\\
&&0.378\leq\sin^2\theta_{23}\leq0.406,~~\mathrm{or}~~0.594\leq\sin^2\theta_{23}\leq0.622\,,
\end{eqnarray}
which are in accordance with the experimentally measured values. Note that the atmospheric angle $\theta_{23}$ deviates from maximal mixing. These predictions can be test by JUNO~\cite{An:2015jdp} and forthcoming long baseline neutrino oscillation experiments. As regards the CP violating phases, both Dirac phase $\delta_{CP}$ and the Majorana phase $\alpha_{31}$ are conserved while another Majorana phase $\alpha_{21}$ can take the discrete values of $0$, $\frac{2}{3n}\pi$, $\frac{4}{3n}\pi$, $\ldots$, $\frac{6n-2}{3n}\pi$. In this case, the effective Majorana mass $|m_{ee}|$ takes a simple form,
\begin{equation}
|m_{ee}|=\frac{1}{3}\Big|2m_1\sin^2\left(\varphi_6+\theta/2\right)+q_1m_2e^{2i\varphi_7}+2q_2m_3\cos^2\left(\varphi_6+\theta/2\right)\Big|\,.
\end{equation}
The predictions on $|m_{ee}|$ are plotted in figure~\ref{fig:mee_CaseIV}. For the IO spectrum and $n=2$, we find $|m_{ee}|$ can take a few discrete values and these results can be tested in forthcoming $0\nu\beta\beta$ experiments.

\end{description}

\begin{figure}[t!]
\begin{center}
\includegraphics[width=0.60\linewidth]{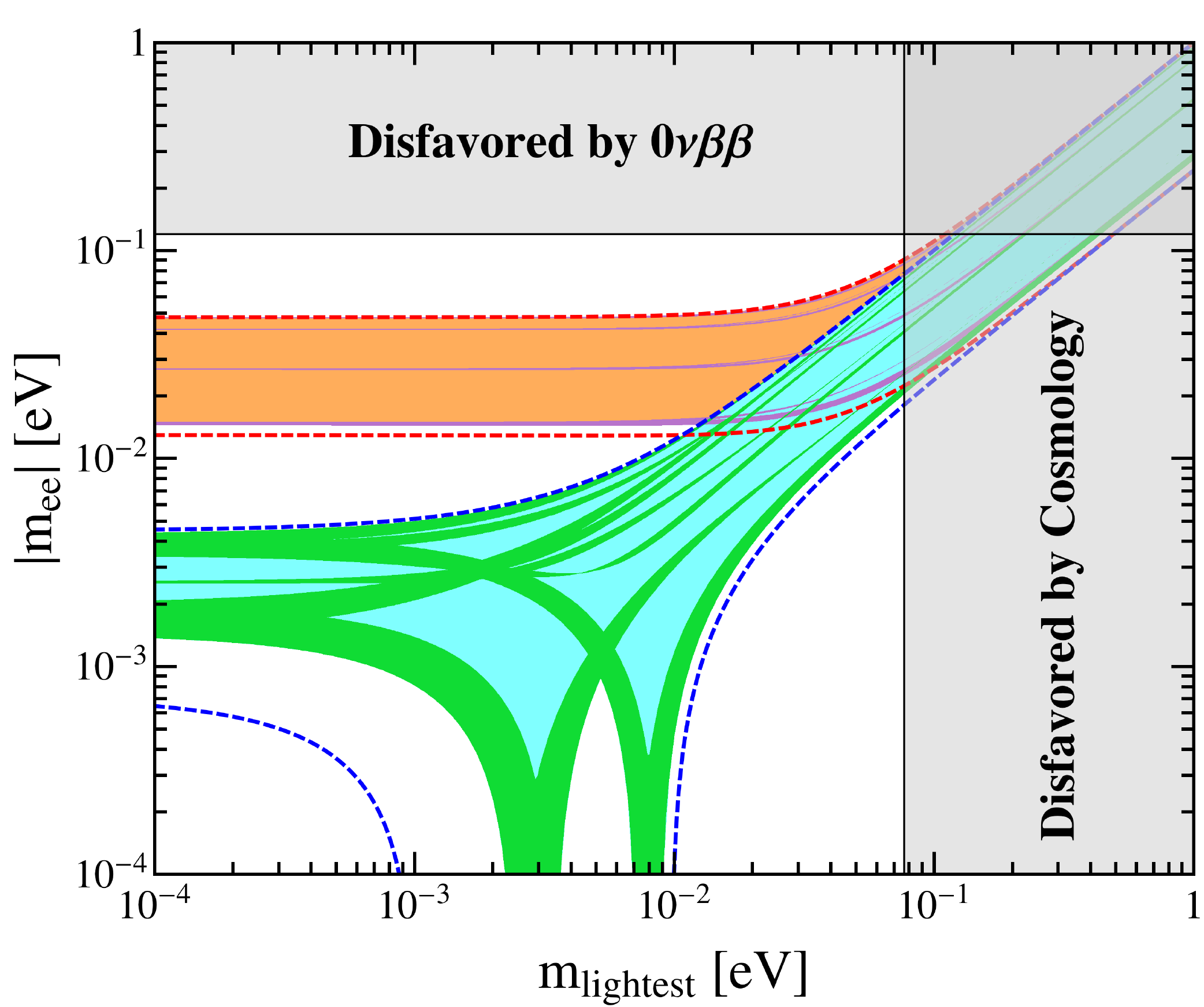}
\caption{\label{fig:mee_CaseIV}The possible values of the effective Majorana mass $|m_{ee}|$ as a function of the lightest neutrino mass in the case IV. The red (blue) dashed lines indicate the most general allowed regions for IO (NO) neutrino mass spectrum obtained by varying the mixing parameters over the $3\sigma$ ranges~\cite{Capozzi:2013csa}. The orange (cyan) areas denote the achievable values of $|m_{ee}|$ in the limit of $n\rightarrow\infty$ assuming IO (NO) spectrum. The purple and green regions are the theoretical predictions for the $D^{(1)}_{9n ,3n}$ group with $n=2$. Notice that the purple (green)  region overlaps the orange (cyan) one. The present most stringent upper limits $|m_{ee}|<0.120$ eV from EXO-200~\cite{Auger:2012ar, Albert:2014awa} and KamLAND-ZEN~\cite{Gando:2012zm} is shown by horizontal grey band. The vertical grey exclusion band represents the current bound coming from the cosmological data of $\sum m_i<0.230$ eV at $95\%$ confidence level obtained by the Planck collaboration~\cite{Ade:2013zuv}.
}
\end{center}
\end{figure}

\section{\label{sec:Z2xCP_charged_lepton}Lepton mixing from a variant of semidirect approach}

\cleqn

In contrast with semidirect approach discussed in section~\ref{sec:Z2xCP_neutrino}, we shall assume that the original symmetry $D^{(1)}_{9n,3n}\rtimes H_{CP}$ is broken down to $Z_2\times CP$ in the charged lepton sector, and the residual symmetry of the neutrino mass matrix is $K_{4}\rtimes H^{\nu}_{CP}$, where $K_4$ is a Klein subgroup of $D^{(1)}_{9n,3n}$. Since each order 2 element of the $D^{(1)}_{9n, 3n}$ group is conjugate to either $bd^{x}$ or $c^{9n/2}$, as shown in Eq.~\eqref{eq:neutrino_conjugate_1} and Eq.~\eqref{eq:neutrino_conjugate_4}, it is sufficient to discuss the representative remnant symmetry $G_{l}=Z^{bd^{x}}_2, Z^{c^{9n/2}}_2$ and $G_{\nu}=K^{(c^{9n/2},d^{3n/2})}_4$, $K^{(d^{3n/2},bd^{x})}_4$, $K^{(c^{9n/2}d^{3n/2},abc^{3y}d^{y})}_4$ and $K^{(c^{9n/2},a^2bc^{3z}d^{2z})}_4$. In this variant of the semidirect approach, the PMNS matrix turns out to depend on only one real continuous parameter besides the discrete parameters specifying the remnant symmetries, and one row of the PMNS matrix would be completely fixed by the assumed remnant symmetries. The fixed row vectors for different representative residual flavor symmetries are listed in table~\ref{tab:PMNS_row}. We find that essentially only one type of residual symmetry with $(G_{\nu}, G_{l})=(K^{(c^{9n/2},a^2bc^{3z}d^{2z})}_4, Z^{bd^{x^{\prime}}}_2)$ is phenomenologically viable in this scenario.

\begin{table}[t!]
\renewcommand{\tabcolsep}{2.0mm}
\centering
\begin{tabular}{|c||c|c|}
\hline \hline
 &  &     \\ [-0.15in]
 &  $G_{l}=Z^{bd^{x^{\prime}}}_2$  &  $G_{l}=Z^{c^{9n/2}}_2$  \\

  &   &      \\ [-0.15in]\hline
 &   &       \\ [-0.15in]

$G_{\nu}=K^{(c^{9n/2},d^{3n/2})}_4$  &  $\frac{1}{\sqrt{2}}\left(\begin{array}{c}
0\\
-1\\
1
\end{array}
\right)^{T}$\quad \xmark  &  $\left(\begin{array}{c}
0\\
0\\
1
\end{array}\right)^{T}$\quad \xmark   \\
 &   &         \\ [-0.15in]\hline
 &   &        \\ [-0.15in]

$G_{\nu}=K^{(d^{3n/2},bd^{x})}_4$  & $\left(\begin{array}{c}
0\\
\cos\left(\frac{x+x'}{3n}\pi\right)\\
-i\sin\left(\frac{x+x'}{3n}\pi\right)
\end{array}
\right)^{T}$\quad \xmark  & $\frac{1}{\sqrt{2}}\left(\begin{array}{c}
0\\
1\\
1
\end{array}
\right)^{T}$\quad \xmark   \\
 &   &         \\ [-0.15in]\hline
 &   &        \\ [-0.15in]

$G_{\nu}=K^{(c^{9n/2}d^{3n/2},abc^{3y}d^{y})}_4$  &  $\frac{1}{2}\left(\begin{array}{c}
-\sqrt{2}\\
1\\
1
\end{array}
\right)^{T}$\quad \cmark  & $\frac{1}{\sqrt{2}}\left(\begin{array}{c}
0\\
1\\
1
\end{array}
\right)^{T}$\quad \xmark    \\
 &   &         \\ [-0.15in]\hline
 &   &        \\ [-0.15in]

$G_{\nu}=K^{(c^{9n/2},a^2bc^{3z}d^{2z})}_4$  & $\frac{1}{2}\left(\begin{array}{c}
-\sqrt{2}\\
1 \\
1
\end{array}\right)^{T}$ \quad \cmark  &  $\left(\begin{array}{c}
1\\
0\\
0
\end{array}
\right)^{T}$ \quad \xmark  \\

&   &      \\ [-0.15in]\hline\hline

\end{tabular}
\caption{\label{tab:PMNS_row}The column vector of the PMNS matrix determined by the residual flavor symmetries $G_{\nu}$ and $G_{l}$, where $x, x^{\prime}, y, z=0, 1, \ldots, 3n-1$. If one (or two) element of the fixed column is vanishing, we would use the notation ``\xmark'' to indicate that it is disfavored by the present experimental data otherwise the notation ``\cmark'' is labelled to indicate that agreement with the experimental data could be achieved. Because two pair of subgroups $(G_{\nu}, G_{l})=(K^{(c^{9n/2}d^{3n/2},abc^{3y}d^{y})}_4,  Z^{bd^{x^{\prime}}}_2)$ and $(K^{(c^{9n/2},a^2bc^{3z}d^{2z})}_4, Z^{bd^{x^{\prime}}}_2)$ with $z=x^{\prime}+y$ are conjugate to each other under the action of the element $bd^{x^{\prime}}$, they lead to the same results for the PMNS matrix after all the admissible residual CP transformations are considered.}
\end{table}

\begin{description}[labelindent=-0.5em, leftmargin=0.1em]

\item[~~(\uppercase\expandafter{\romannumeral5})]

$G_{l}=\left\{1,bd^{x}\right\}$, $X_{l\mathbf{r}}=\left\{\rho_{\mathbf{r}}(c^{2x+2\delta+3n\tau}d^{\delta}), \rho_{\mathbf{r}}(bc^{2x+2\delta+3n\tau}d^{x+\delta})\right\}$,
$G_{\nu}=K^{(c^{9n/2},a^2bc^{3z}d^{2z})}_4$ and
$X_{\nu\mathbf{r}}=\left\{\rho_{\mathbf{r}}(c^{\gamma}d^{-2z}),
\rho_{\mathbf{r}}(a^2bc^{\gamma})\right\}$\\
Here we would like to recall that the residual CP transformations are determined by the restricted consistency conditions in Eqs.~(\ref{eq:res_cons_cond_clep}, \ref{eq:res_cons_cond_nu}). The parameter $n$ should be even in order to have a residual Klein subgroup. The phenomenological constraints of the residual flavor symmetry $G_{\nu}=K^{(c^{9n/2},a^2bc^{3z}d^{2z})}_4$ as well as the residual CP transformation
$X_{\nu\mathbf{r}}=\left\{\rho_{\mathbf{r}}(c^{\gamma}d^{-2z}),
\rho_{\mathbf{r}}(a^2bc^{\gamma})\right\}$ has been studied in section~\ref{sec:direct_approach}. The light neutrino mass matrix $m_{\nu}$ and its diagonalization matrix $U_{\nu}$ are found to be given by Eq.~\eqref{eq:mnu_k4_4} and Eq.~\eqref{eq:unu_k4_4} respectively. Then we proceed to the charged lepton sector. The invariance of the charged lepton mass matrix under the residual symmetry $G_{l}=\left\{1,bd^{x}\right\}$ and $X_{l\mathbf{r}}=\left\{\rho_{\mathbf{r}}(c^{2x+2\delta+3n\tau}d^{\delta}), \rho_{\mathbf{r}}(bc^{2x+2\delta+3n\tau}d^{x+\delta})\right\}$ implies that the hermitian matrix $m^{\dagger}_{l}m_{l}$ has to fulfill the invariant condition of Eq.~\eqref{eq:constr_clep}, i.e.
\begin{eqnarray}
\nonumber&&\hskip1.8cm\rho^{\dagger}_{\mathbf{3}}(bd^{x})m^{\dagger}_{l}m_{l}\rho_{\mathbf{3}}(bd^{x})=m^{\dagger}_{l}m_{l},\\
&&\rho^{\dagger}_{\mathbf{3}}(c^{2x+2\delta+3n\tau}d^{\delta})m^{\dagger}_{l}m_{l}\rho_{\mathbf{3}}(c^{2x+2\delta+3n\tau}d^{\delta})=\big(m^{\dagger}_{l}m_{l}\big)^{*}\,,
\end{eqnarray}
which lead to
\begin{equation}
m^{\dagger}_{l}m_{l}=\left(
\begin{array}{ccc}
 \widetilde{\mathfrak{m}}_{11} &~  \widetilde{\mathfrak{m}}_{12}e^{i\pi\frac{\delta}{3n}} &~ \widetilde{\mathfrak{m}}_{12}e^{i\pi\frac{2x+\delta}{3n}}   \\
 \widetilde{\mathfrak{m}}_{12}e^{-i\pi\frac{\delta}{3n}}  &~  \widetilde{\mathfrak{m}}_{22} &~ \widetilde{\mathfrak{m}}_{23}e^{i\pi\frac{2x}{3n}}\\
 \widetilde{\mathfrak{m}}_{12}e^{-i\pi\frac{2x+\delta}{3n}}  &~ \widetilde{\mathfrak{m}}_{23}e^{-i\pi\frac{2x}{3n}}  &~   \widetilde{\mathfrak{m}}_{22}
\end{array}
\right)\,,
\end{equation}
where $\widetilde{\mathfrak{m}}_{11}$, $\widetilde{\mathfrak{m}}_{12}$, $\widetilde{\mathfrak{m}}_{22}$ and $\widetilde{\mathfrak{m}}_{23}$ are real, and they have dimension of squared mass. It can be diagonalized by the unitary matrix
\begin{equation}
\label{eq:ul_bdx_2}U_{l}=\frac{1}{\sqrt{2}}
\left(
\begin{array}{ccc}
 0 &~ -\sqrt{2} \sin\theta  &~ \sqrt{2} \cos \theta  \\
 -e^{i \pi \frac{2  x}{3 n}} &~ e^{-i \pi  \frac{\delta }{3n}} \cos \theta  &~   e^{-i\pi\frac{\delta}{3n}}\sin\theta  \\
 1 &~ e^{-i \pi  \frac{2 x+\delta }{3 n}} \cos \theta &~   e^{-i \pi \frac{ 2 x+\delta }{3 n}} \sin \theta
\end{array}
\right)\,,
\end{equation}
with the angle $\theta$ satisfying
\begin{equation}
\tan2\theta=\frac{2\sqrt{2}\,\widetilde{\mathfrak{m}}_{12}}{\widetilde{\mathfrak{m}}_{11}-\widetilde{\mathfrak{m}}_{22}-\widetilde{\mathfrak{m}}_{23}}\,.
\end{equation}
The squared charged lepton masses are determined to be of the form
\begin{eqnarray}
\nonumber&&m^2_{l_1}=\widetilde{\mathfrak{m}}_{22}-\widetilde{\mathfrak{m}}_{23}\,,\\
\nonumber&&m^2_{l_2}=\frac{1}{2}\left[\widetilde{\mathfrak{m}}_{11}+\widetilde{\mathfrak{m}}_{22}+\widetilde{\mathfrak{m}}_{23}-\frac{\widetilde{\mathfrak{m}}_{11}-\widetilde{\mathfrak{m}}_{22}-\widetilde{\mathfrak{m}}_{23}}{\cos 2\theta}\right]\,,\\
&&m^2_{l_3}=\frac{1}{2}\left[\widetilde{\mathfrak{m}}_{11}+\widetilde{\mathfrak{m}}_{22}+\widetilde{\mathfrak{m}}_{23}+\frac{\widetilde{\mathfrak{m}}_{11}-\widetilde{\mathfrak{m}}_{22}-\widetilde{\mathfrak{m}}_{23}}{\cos 2\theta}\right]\,.
\end{eqnarray}
Notice that the order of the masses $m^2_{l_1}$, $m^2_{l_2}$ and $m^2_{l_3}$ can not be pinned down by remnant symmetry, therefore the matrix $U_{l}$ in Eq.~\eqref{eq:ul_bdx_2} is determined up to permutations and phases of its column vectors. The lepton flavor mixing originates from the mismatch between the unitary transformations $U_{l}$ in Eq.~\eqref{eq:ul_bdx_2} and $U_{\nu}$ in Eq.~\eqref{eq:unu_k4_4}, and the PMNS matrix can take the form
\begin{equation}
\label{eq:PMNS_case_V_1}U^{V, 1}_{\text{PMNS}}=\frac{1}{2}\left(
\begin{array}{ccc}
 \sin\theta+\sqrt{2}e^{i\varphi_8}\cos\theta ~&~
 \sin\theta-\sqrt{2}e^{i\varphi_8}\cos\theta ~&~
 \sqrt{2}e^{i\varphi_9}\sin\theta   \\
  1 ~&~ 1 ~&~ -\sqrt{2}e^{i\varphi_9} \\
 \cos\theta-\sqrt{2}e^{i\varphi_8}\sin\theta   ~&~
 \cos\theta+\sqrt{2}e^{i\varphi_8}\sin\theta ~&~
 \sqrt{2}e^{i\varphi_9}\cos\theta
\end{array}
\right)\,,
\end{equation}
or
\begin{equation}
\label{eq:PMNS_case_V_2}U^{V, 2}_{\text{PMNS}}=\frac{1}{2}\left(
\begin{array}{ccc}
 \sin\theta+\sqrt{2}e^{i\varphi_8}\cos\theta ~&~
 \sin\theta-\sqrt{2}e^{i\varphi_8}\cos\theta ~&~
 \sqrt{2}e^{i\varphi_9}\sin\theta   \\
 \cos\theta-\sqrt{2}e^{i\varphi_8}\sin\theta   ~&~
 \cos\theta+\sqrt{2}e^{i\varphi_8}\sin\theta ~&~
 \sqrt{2}e^{i\varphi_9}\cos\theta \\
  1 ~&~ 1 ~&~ -\sqrt{2}e^{i\varphi_9}
\end{array}
\right)\,,
\end{equation}
where
\begin{equation}
\varphi_8=-\frac{2z+\delta}{3n}\pi,\qquad
\varphi_9=\frac{2x-4z-\gamma}{3n}\pi\,.
\end{equation}
Obviously the values of both $\varphi_8$ and $\varphi_9$ are integer multiple of $\frac{\pi}{3n}$, i.e.
\begin{equation}
\varphi_8, \varphi_9~(\mathrm{mod}~2\pi)=0, \frac{1}{3n}\pi, \frac{2}{3n}\pi,\ldots, \frac{6n-1}{3n}\pi\,.
\end{equation}
The mixing patterns $U^{V, 1}_{\text{PMNS}}$ and $U^{V, 2}_{\text{PMNS}}$ are related through the exchange of the second and third rows in the PMNS mixing matrix. Other permutations of rows and columns don't lead to new patterns consistent with experimental data. We can extract the following results for the mixing angles,
\begin{eqnarray}
\nonumber&&\sin^2\theta_{13}=\frac{1}{2}\sin^2\theta,\qquad
\sin^2\theta_{12}=\frac{1}{2}-\frac{\sqrt{2}\sin2\theta\cos\varphi_8}{3+\cos2\theta},\\
&&\sin^2\theta_{23}=\frac{2}{3+\cos2\theta}~~\mathrm{for}~~U^{V, 1}_{\text{PMNS}},\qquad \sin^2\theta_{23}=\frac{1+\cos2\theta}{3+\cos2\theta}~~\mathrm{for}~~U^{V, 2}_{\text{PMNS}}\,.
\end{eqnarray}
The $3\sigma$ range of $\sin^2\theta_{13}$ can be reproduced for
\begin{equation}
\theta\in\left[0.060\pi,0.078\pi\right]\cup\left[0.922\pi,0.940\pi\right]\,.
\end{equation}
We can check that the mixing angles fulfill the following sum rules,
\begin{subequations}
\begin{eqnarray}
\label{eq:correlation_case_V_1}&&\qquad\cos2\theta_{12}=\pm2\tan\theta_{13}\sqrt{1-\tan^2\theta_{13}}\cos\varphi_{8}\,,\\
\label{eq:correlation_case_V_2}&&2\cos^2\theta_{13}\sin^2\theta_{23}=1,\quad \text{or}\quad 2\cos^2\theta_{13}\sin^2\theta_{23}=\cos2\theta_{13}\,,
\end{eqnarray}
\end{subequations}
where the ``+'' and ``$-$'' signs are valid for $0<\theta<\pi/2$
and $\pi/2<\theta<\pi$ respectively. In order to accommodate the experimental data on solar and reactor mixing angles, the first sum rule of Eq.~\eqref{eq:correlation_case_V_1} implies that the parameter $\varphi_8$ should vary in the interval
\begin{equation}
\label{eq:varphi8_cons}\varphi_8\in[0, 0.193\pi]\cup[0.807\pi, 1.193\pi]\cup[1.807\pi, 2\pi]\,.
\end{equation}
From the correlation of Eq.~\eqref{eq:correlation_case_V_2}, we can derive
\begin{equation}
\label{eq:theta_23_V}0.509\leq\sin^2\theta_{23}\leq0.515,~~\text{or}~~0.485\leq\sin^2\theta_{23}\leq0.491\,.
\end{equation}
\begin{figure}[t!]
\begin{center}
\includegraphics[width=0.60\textwidth]{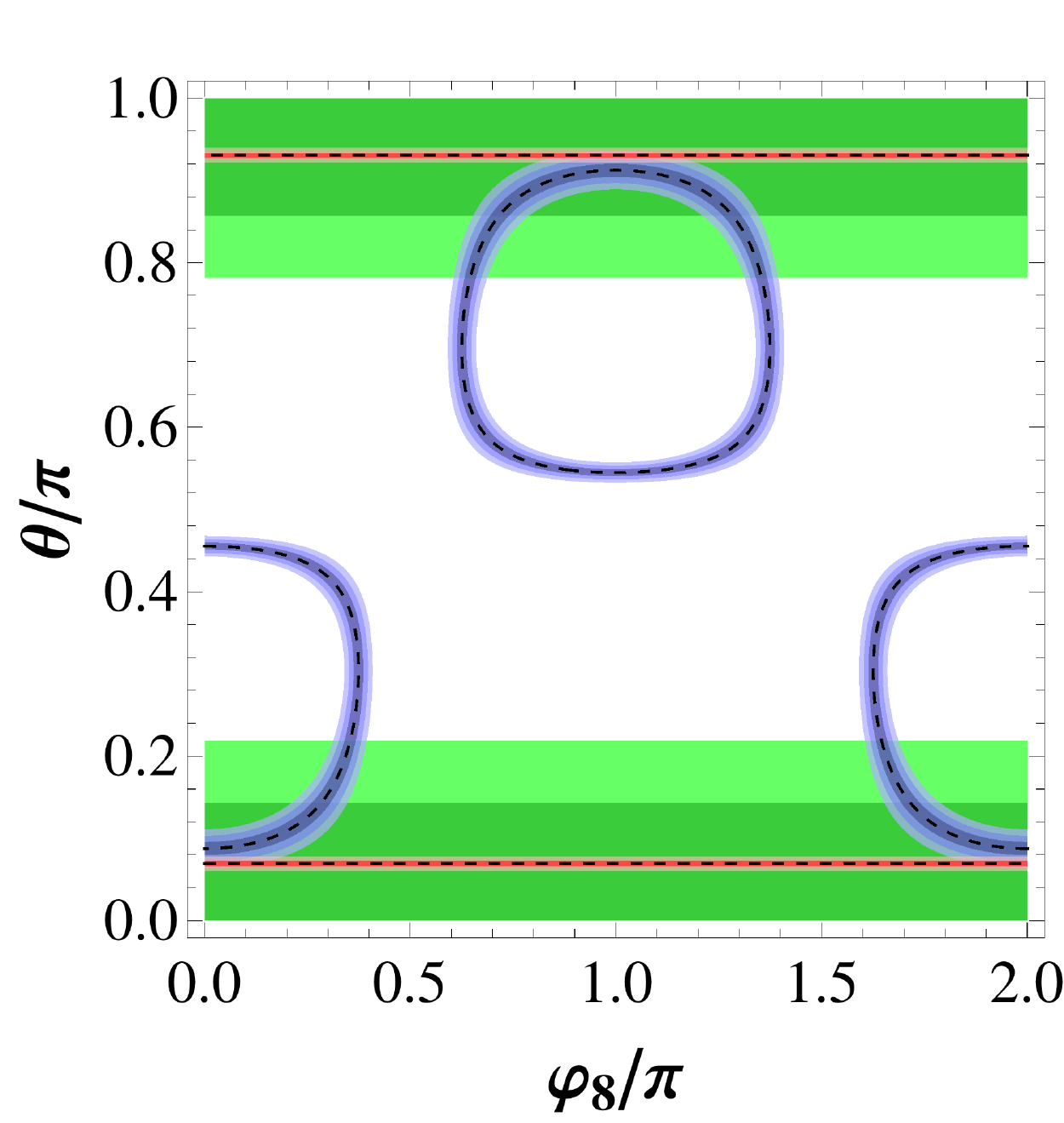}
\caption{\label{fig:caseV_contour_mixing_para}
The contour regions of the three mixing angles in the case V. The red, blue and green areas denote the predictions for $\sin^2\theta_{13}$, $\sin^2\theta_{12}$ and $\sin^2\theta_{23}$ respectively. The allowed $1\sigma$, $2\sigma$ and $3\sigma$ regions of each mixing angle are represented by different shadings. The best fit values of the mixing angles are indicated by dotted lines.
In this case the atmospheric angle $\theta_{23}$ is predicted to be in the interval of Eq.~\eqref{eq:theta_23_V} such that neither the $1\sigma$ range nor its best fit value can be achieved.}
\end{center}
\end{figure}
The contour plots for $\sin^2\theta_{ij}$ is shown in figure~\ref{fig:caseV_contour_mixing_para}. Since both reactor angle $\theta_{13}$ and the atmospheric angle $\theta_{23}$ only depend on the parameter $\theta$, the corresponding contour regions are horizontal bands. There exist three small regions in which all the three mixing angles are within the experimentally preferred
$3\sigma$ ranges. Furthermore, we find the following expressions for the CP invariants,
\begin{eqnarray}
\nonumber && \left|J_{CP}\right|=\frac{1}{8 \sqrt{2}}\left|\sin 2 \theta\sin\varphi_8\right|\,,\qquad \left|I_1\right|=\frac{1}{8 \sqrt{2}}\left|(1+3 \cos2 \theta )\sin2\theta \sin \varphi_8\right|  \,, \\
 && \left|I_2\right|= \frac{\sin^2\theta}{8}\left|\sqrt{2} \sin2\theta\sin (2\varphi_9-\varphi_8)+2\cos ^2\theta \sin2(\varphi_9-\varphi_8)+\sin^2\theta\sin2 \varphi _9\right| \,,
\end{eqnarray}
from which we know that both $\delta_{CP}$ and $\alpha_{21}$ are only dependent on $\theta$ and $\varphi_8$ while the value of $\alpha^{\prime}_{31}$ depends on three parameters $\theta$, $\varphi_8$ and $\varphi_9$. We display the predictions for $|\sin\delta_{CP}|$ and $|\sin\alpha_{21}|$ in the $\varphi_8-\theta$ plane in figure~\ref{fig:caseV_contour_CP_para}. One can see that both $\delta_{CP}$ and $\alpha_{21}$ can not be maximal if the three mixing angles are required to be consistent with the experimental data.
\begin{figure}[t!]
\begin{center}
\includegraphics[width=0.495\textwidth]{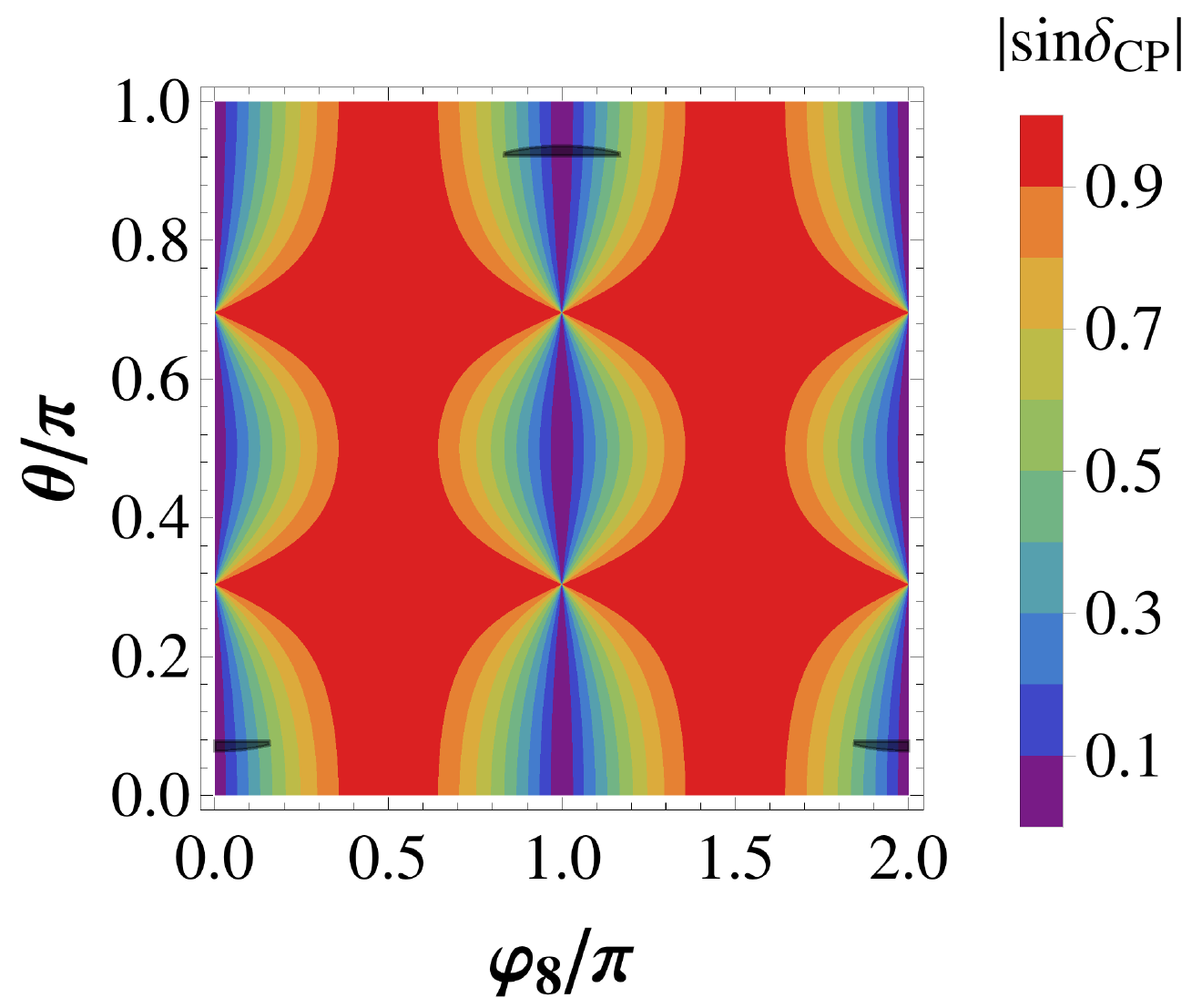}
\includegraphics[width=0.495\textwidth]{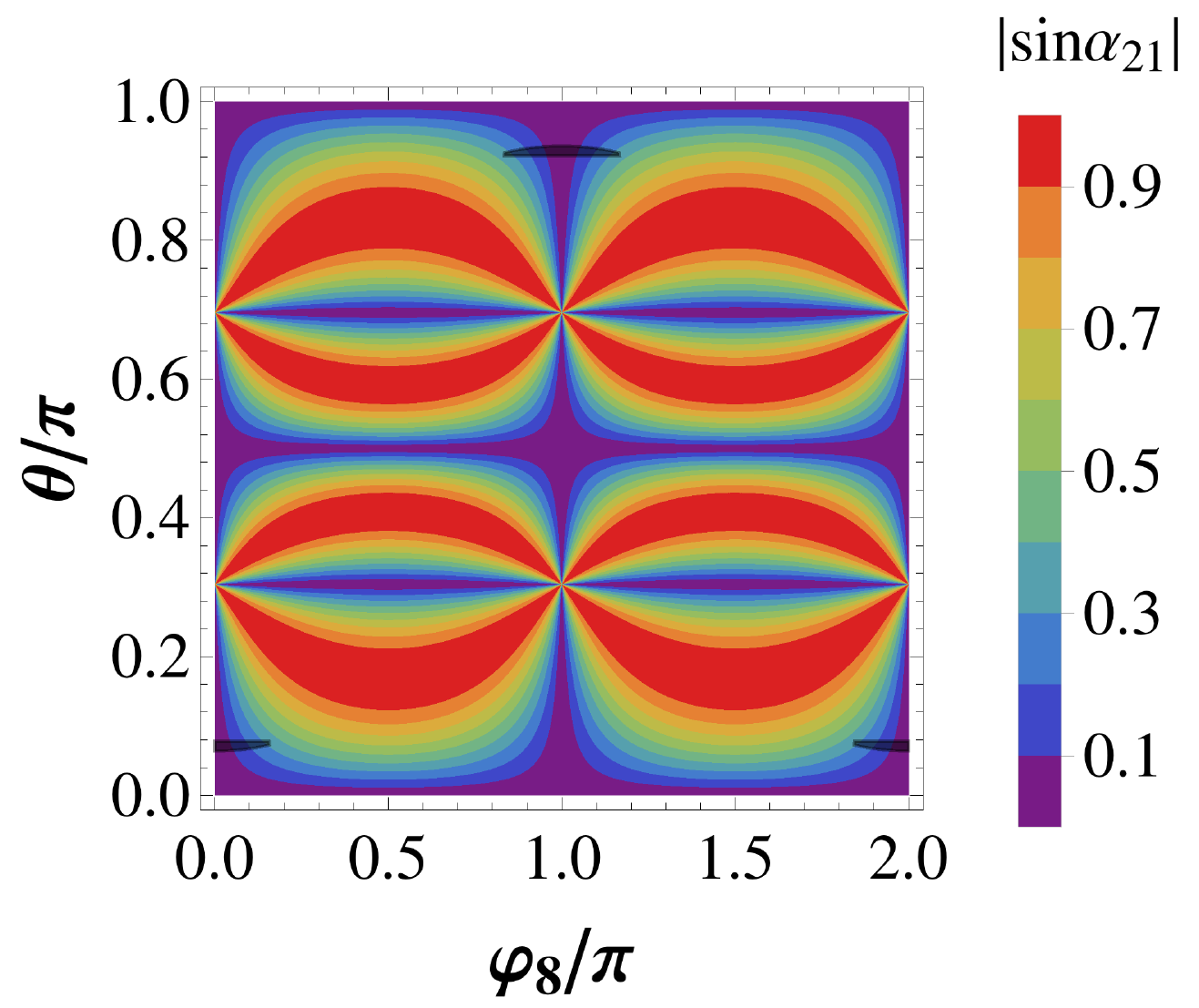}
\caption{\label{fig:caseV_contour_CP_para}The contour plots of $|\sin\delta_{CP}|$ and $|\sin\alpha_{21}|$ in the case V. The black areas represent the regions in which the lepton mixing angles are compatible with experimental data at $3\sigma$ level, and it can be read out from figure~\ref{fig:caseV_contour_mixing_para}.}
\end{center}
\end{figure}
In analogy to previous cases, we numerically study the possible values of the mixing parameters for each $D^{(1)}_{9n, 3n}$ group.
We can read from figure~\ref{fig:caseV_mixing_para} that a bit larger $\theta_{12}$ (still in the $3\sigma$ range) is favored with $0.328\leq\sin^2\theta_{12}\leq0.359$, and the atmospheric angle $\sin^2\theta_{23}$ is predicted to be around 0.487 and 0.513. These results can be testable at forthcoming neutrino oscillation facilities. The same conclusions on CP phases are reached as those from figure~\ref{fig:caseV_contour_CP_para}. We find the upper bounds of $|\sin\delta_{CP}|$ and $|\sin\alpha_{21}|$ are $\left|\sin\delta_{CP}\right|\leq 0.594$ and $\left|\sin\alpha_{21}\right|\leq0.399$ respectively. On the other hand, any value of the Majorana phase $\alpha_{31}$ is possible for large value of $n$.
\begin{figure}[t!]
\begin{center}
\includegraphics[width=0.99\textwidth]{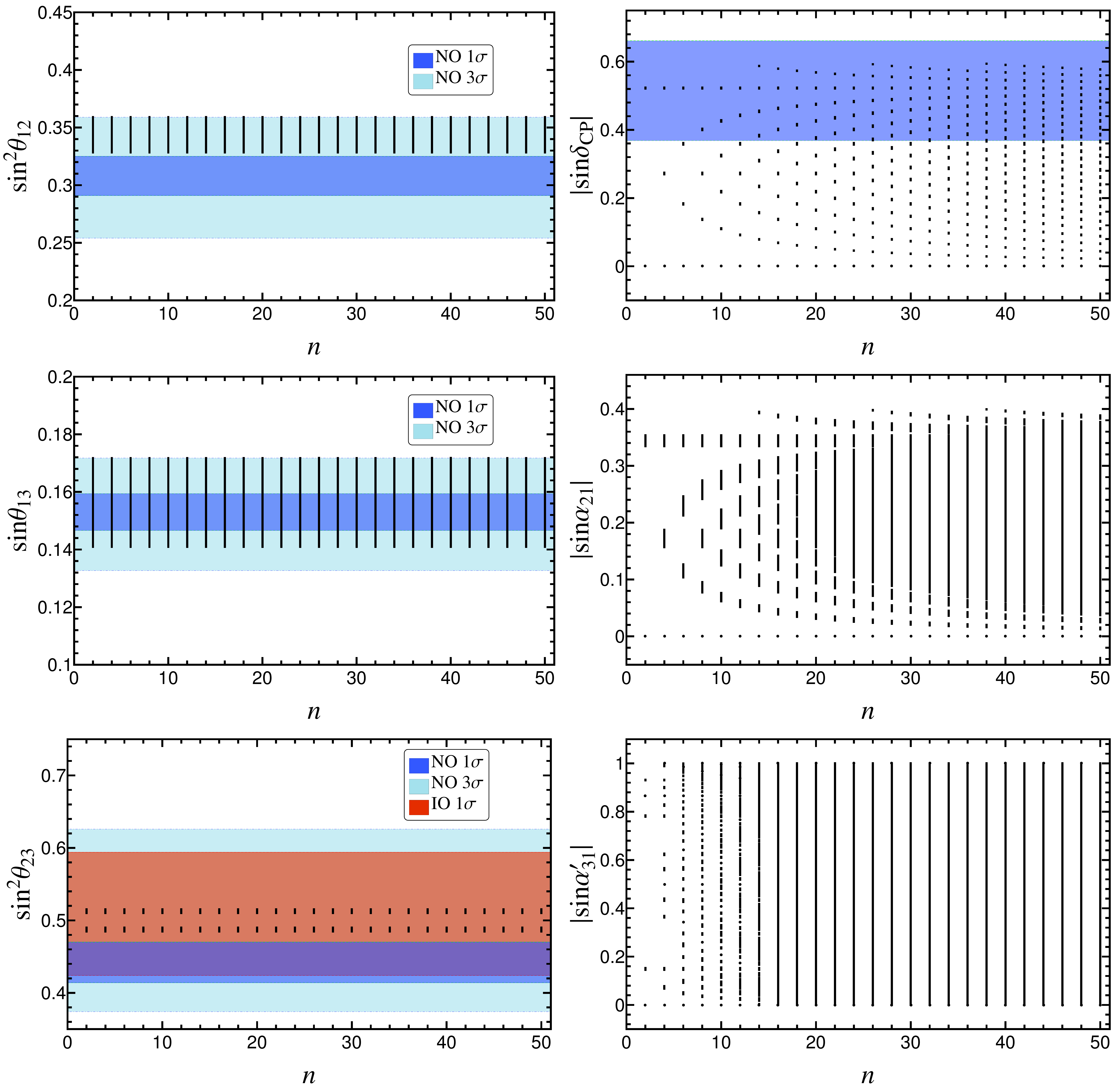}
\caption{\label{fig:caseV_mixing_para}The possible values of $\sin^2\theta_{12}$, $\sin\theta_{13}$, $\sin^2\theta_{23}$, $\left|\sin\delta_{CP}\right|$, $\left|\sin\alpha_{21}\right|$ and $\left|\sin\alpha^{\prime}_{31}\right|$ with respect to $n$ for the mixing pattern $U^{V, 1}_{\text{PMNS}}$ and $U^{V, 2}_{\text{PMNS}}$ in the case V, where the three lepton mixing angles are required to be within the experimentally preferred $3\sigma$ ranges. The $1\sigma$ and $3\sigma$ regions of the three neutrino mixing angles are adapted from global fit~\cite{Capozzi:2013csa}. Note that the group index $n$ should be even in this case.}
\end{center}
\end{figure}

Now we discuss the lepton mixing patterns which can be obtained from the $D^{(1)}_{9n, 3n}$ group with $n=2$.  Note that the smallest $D^{(1)}_{9n, 3n}$ group for $n=1$ doesn't comprise the required Klein subgroup.
The PMNS matrices $U^{V, 1}_{\text{PMNS}}$ and $U^{V, 2}_{\text{PMNS}}$ fulfill the following relations
\begin{subequations}
\label{eq:PMNS_sym_caseV}
\begin{eqnarray}
\label{eq:PMNS_sym_caseV1}&&U^{V}_{\text{PMNS}}(\theta, \varphi_8+\pi, \varphi_9)=\text{diag}(-1, 1, 1)U^{V}_{\text{PMNS}}(-\theta, \varphi_8, \varphi_9),\\
\label{eq:PMNS_sym_caseV2}&&U^{V}_{\text{PMNS}}(\theta, \varphi_8, \varphi_9+\pi/2)=U^{V}_{\text{PMNS}}(\theta, \varphi_8, \varphi_9)\text{diag}(1, 1, i),\\
\label{eq:PMNS_sym_caseV3}&&U^{V}_{\text{PMNS}}(\theta, \pi-\varphi_8, \varphi_9)=\text{diag}(-1, 1, 1)[U^{V}_{\text{PMNS}}(-\theta, \varphi_8, -\varphi_9)]^{\ast}\,,
\end{eqnarray}
\end{subequations}
where $U^{V}_{\text{PMNS}}$ refers to $U^{V, 1}_{\text{PMNS}}$ and $U^{V, 2}_{\text{PMNS}}$. The diagonal matrices on the left-handed and right-handed sides can be absorbed by the charged lepton fields and $Q_{\nu}$ respectively.  Therefore the shifts of $\varphi_8$ into $\varphi_8+\pi$ and $\varphi_9$ into $\varphi_9+\pi/2$ don't lead to physically new results. For $n=2$ the values of $\varphi_8$ and $\varphi_9$ can be $0$, $\pi/6$, $\pi/3$, $\ldots$, $11\pi/6$. Considering the constraint on the parameter $\varphi_8$ given by Eq.~\eqref{eq:varphi8_cons}, we find only $\varphi_8~(\mathrm{mod}~\pi)=0$, $\pi/6$ and $5\pi/6$ can describe the data on lepton mixing. The results of our $\chi^2$ analysis are displayed in table~\ref{tab:caseV_n2}.
Since the mixing angles $\sin^2\theta_{ij}$ and the CP invariants $J_{CP}$ and $I_{1}$ are expressed in terms of $\theta$ and $\varphi_8$, and the parameter $\varphi_9$ only enters into the expression of $I_2$, the relation in Eq.~\eqref{eq:PMNS_sym_caseV3} implies that $\varphi_8$ and $\pi-\varphi_8$ give rise to the same best fitting values of mixing parameters except $|\sin\alpha^{\prime}_{31}|$.
This is exactly the reason why the numerical results for $\varphi_8=\pi/6$ and $\varphi_8=5\pi/6$ are only different in the values of $|\sin\alpha^{\prime}_{31}|$.
Finally we plot the predictions for the effective mass $|m_{ee}|$ with respect to the lightest neutrino mass in figure~\ref{fig:mee_CaseV}.
One sees that the values of $|m_{ee}|$ are rather close to the lower or upper boundary of the $3\sigma$ region for IO.

\begin{table}[t!]
\centering
\footnotesize
\renewcommand{\tabcolsep}{1.8mm}
\begin{tabular}{|c|c|c|c|c|c|c|c|c|c|c|}
\hline \hline
\multicolumn{11}{|c|}{Case V}  \\ \hline
 \multicolumn{11}{|c|}{$n=2$}   \\ \hline
s & $\varphi_8$ & $\varphi_9$  & $\theta_{bf}$ & $\chi^2_{min}$   & $\sin^2\theta_{13}$ & $\sin^2\theta_{12}$ & $\sin^2\theta_{23}$  & $|\sin\delta_{CP}|$ & $|\sin\alpha_{21}|$ & $|\sin\alpha^{\prime}_{31}|$  \\ \hline
\multirow{9}{*}{$U^{V, 1}_{\text{PMNS}}$} &  \multirow{3}{*}{$0$} & $0$ &\multirow{2}{*}{$0.224$} &\multirow{2}{*}{$9.890$} & \multirow{2}{*}{$0.0248$} & \multirow{2}{*}{$0.343$} &\multirow{2}{*}{$0.513$}  & \multirow{3}{*}{$0$ ($0$)} & \multirow{3}{*}{$0$ ($0$)} & $0$ ($0$)  \\  \cline{3-3} \cline{11-11}
 & & $\frac{\pi}{6}$ & \multirow{2}{*}{($0.227$)} & \multirow{2}{*}{($4.409$)}  & \multirow{2}{*}{($0.0253$)} & \multirow{2}{*}{($0.341$)} &\multirow{2}{*}{($0.513$)}  & & & $0.866$ ($0.866$) \\ \cline{3-3} \cline{11-11}
& & $\frac{\pi}{3}$ & & & & & & & & $0.866$ ($0.866$) \\  \cline{2-11}

 & \multirow{3}{*}{$\frac{\pi}{6}$} & $0$ &\multirow{2}{*}{$0.227$} & \multirow{2}{*}{$16.405$}  & \multirow{2}{*}{$0.0253$} & \multirow{2}{*}{$0.362$} & \multirow{2}{*}{$0.513$} & \multirow{2}{*}{$0.520$} & \multirow{2}{*}{$0.326$}  & $0.786$ ($0.785$)  \\  \cline{3-3} \cline{11-11}
 & & $\frac{\pi}{6}$ &\multirow{2}{*}{($0.229$)} & \multirow{2}{*}{($10.772$)}  & \multirow{2}{*}{($0.0258$)} & \multirow{2}{*}{($0.361$)} & \multirow{2}{*}{($0.513$)} & \multirow{2}{*}{($0.521$)} & \multirow{2}{*}{($0.329$)} & $0.142$ ($0.144$) \\ \cline{3-3} \cline{11-11}
& & $\frac{\pi}{3}$ & & & & & & & & $0.928$ ($0.929$) \\  \cline{2-11}

  &  \multirow{3}{*}{$\frac{5\pi}{6}$} & $0$ &\multirow{2}{*}{$2.915$} & \multirow{2}{*}{$16.405$}  & \multirow{2}{*}{$0.0253$} & \multirow{2}{*}{$0.362$} & \multirow{2}{*}{$0.513$} & \multirow{2}{*}{$0.520$} & \multirow{2}{*}{$0.326$} &$0.786$ ($0.785$) \\  \cline{3-3} \cline{11-11}
& & $\frac{\pi}{6}$ & \multirow{2}{*}{($2.913$)} & \multirow{2}{*}{($10.772$)}  & \multirow{2}{*}{($0.0258$)} & \multirow{2}{*}{($0.361$)} & \multirow{2}{*}{($0.513$)} & \multirow{2}{*}{($0.521$)} & \multirow{2}{*}{($0.329$)}  & $0.928$ ($0.929$) \\ \cline{3-3} \cline{11-11}
& & $\frac{\pi}{3}$ & & & & & & & & $0.142$ ($0.144$)\\  \hline

\multirow{9}{*}{$U^{V, 2}_{\text{PMNS}}$} &   \multirow{3}{*}{$0$} & $0$ &\multirow{2}{*}{$0.225$} &\multirow{2}{*}{$6.938$} &\multirow{2}{*}{$0.0250$} & \multirow{2}{*}{$0.342$} &\multirow{2}{*}{$0.487$}  & \multirow{3}{*}{$0$ ($0$)} & \multirow{3}{*}{$0$ ($0$)} & $0$ ($0$)  \\ \cline{3-3} \cline{11-11}
 & &$\frac{\pi}{6}$& \multirow{2}{*}{($0.227$)} & \multirow{2}{*}{($4.288$)} &  \multirow{2}{*}{($0.0253$)} & \multirow{2}{*}{($0.341$)} & \multirow{2}{*}{($0.487$)} & & & $0.866$ ($0.866$) \\ \cline{3-3} \cline{11-11}
& & $\frac{\pi}{3}$ & & & & & & & & $0.866$ ($0.866$) \\ \cline{2-11}

&\multirow{3}{*}{$\frac{\pi}{6}$} & $0$ &\multirow{2}{*}{$0.228$} & \multirow{2}{*}{$13.389$} & \multirow{2}{*}{$0.0255$} & \multirow{2}{*}{$0.362$} & \multirow{2}{*}{$0.487$} & \multirow{2}{*}{$0.520$} & \multirow{2}{*}{$0.328$} & $0.786$($0.785$) \\ \cline{3-3} \cline{11-11}
& &$\frac{\pi}{6}$ & \multirow{2}{*}{($0.229$)} & \multirow{2}{*}{($10.649$)} &  \multirow{2}{*}{($0.0258$)} & \multirow{2}{*}{($0.361$)} & \multirow{2}{*}{($0.487$)} & \multirow{2}{*}{($0.521$)} & \multirow{2}{*}{($0.330$)}   & $0.143$ ($0.144$) \\ \cline{3-3} \cline{11-11}
& & $\frac{\pi}{3}$ & & & & & & & & $0.929$ ($0.929$) \\ \cline{2-11}

& \multirow{3}{*}{$\frac{5\pi}{6}$} & $0$ &\multirow{2}{*}{$2.914$} & \multirow{2}{*}{$13.389$} &  \multirow{2}{*}{$0.0255$} & \multirow{2}{*}{$0.362$} & \multirow{2}{*}{$0.487$} & \multirow{2}{*}{$0.520$} & \multirow{2}{*}{$0.328$}  & $0.786$ ($0.785$)  \\ \cline{3-3} \cline{11-11}
& &$\frac{\pi}{6}$ &\multirow{2}{*}{($2.913$)} & \multirow{2}{*}{($10.649$)} &  \multirow{2}{*}{($0.0258$)} & \multirow{2}{*}{($0.361$)} & \multirow{2}{*}{($0.487$)} & \multirow{2}{*}{($0.521$)} & \multirow{2}{*}{($0.330$)}   & $0.929$ ($0.929$) \\ \cline{3-3} \cline{11-11}
& & $\frac{\pi}{3}$ & & & & & & & & $0.143$ ($0.144$) \\ \cline{3-11}
 \hline \hline
\end{tabular}
\caption{\label{tab:caseV_n2}Results of the $\chi^2$ analysis for $n=2$ in the case V. The $\chi^2$ function has a global minimum $\chi^2_{min}$ at the best fit value $\theta_{bf}$ for $\theta$. We give the values of the mixing angles and CP violation phases for $\theta=\theta_{bf}$. The values given in parentheses denote the results for the IO neutrino mass spectrum. Because of the symmetry relations in Eq.~\eqref{eq:PMNS_sym_caseV}, only the results for $0\leq\varphi_8<\pi$ and $0\leq\varphi_9<\pi/2$ are shown here.}
\end{table}

\begin{figure}[t!]
\begin{center}
\includegraphics[width=0.60\linewidth]{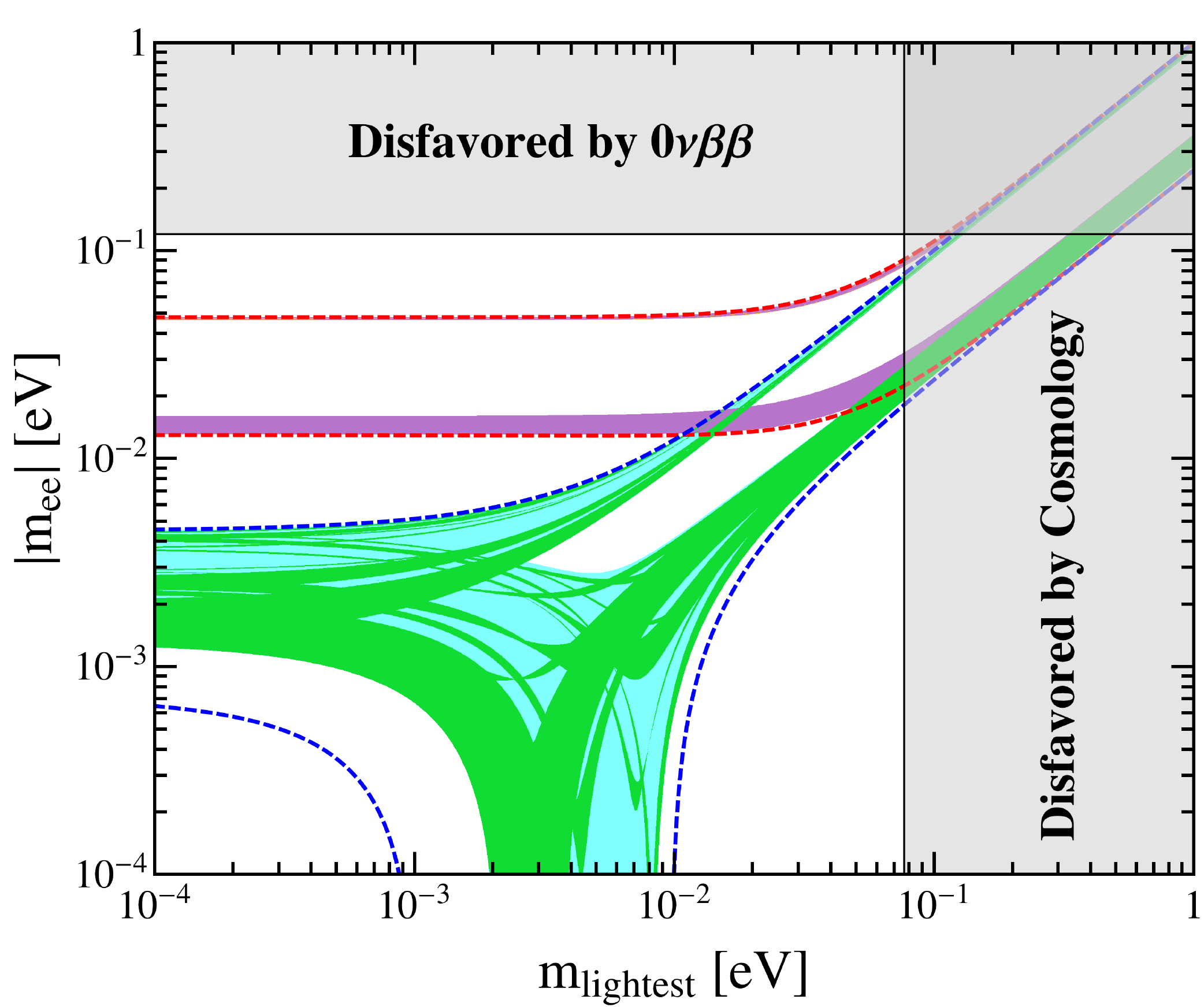}
\caption{\label{fig:mee_CaseV}The possible values of the effective Majorana mass $|m_{ee}|$ as a function of the lightest neutrino mass in the case V. The red (blue) dashed lines indicate the most general allowed regions for IO (NO) neutrino mass spectrum obtained by varying the mixing parameters over the $3\sigma$ ranges~\cite{Capozzi:2013csa}. The orange (cyan) areas denote the achievable values of $|m_{ee}|$ in the limit of $n\rightarrow\infty$ assuming IO (NO) spectrum. The purple and green regions are the theoretical predictions for the $D^{(1)}_{9n ,3n}$ group with $n=2$. Notice that the purple (green)  region overlaps the orange (cyan) one. The present most stringent upper limits $|m_{ee}|<0.120$ eV from EXO-200~\cite{Auger:2012ar, Albert:2014awa} and KamLAND-ZEN~\cite{Gando:2012zm} is shown by horizontal grey band. The vertical grey exclusion band represents the current bound coming from the cosmological data of $\sum m_i<0.230$ eV at $95\%$ confidence level obtained by the Planck collaboration~\cite{Ade:2013zuv}.
}
\end{center}
\end{figure}

\end{description}

\section{\label{sec:Conclusion}Conclusions}

The type D finite subgroup of $SU(3)$ has two independent series: $D^{(0)}_{n, n}\cong\Delta(6n^2)$ and $D^{(1)}_{9n, 3n}\cong(Z_{9n}\times Z_{3n})\rtimes S_3$. The $\Delta(6n^2)$ flavor symmetry with or without CP symmetry and its predictions for the lepton flavor mixing has been discussed in the literature. In the present work, we have performed a comprehensive analysis of the mixing patterns which can be derived from another type D group series $D^{(1)}_{9n ,3n}$ and the generalized CP. The phenomenological consequence of the ``direct'' approach, ``semidirect'' approach and ``variant of semidirect'' approach are studied in a model independent way. The three approaches differ in the residual symmetries preserved by the neutrino and charged lepton sectors.

The mathematical structure of $D^{(1)}_{9n, 3n}$ has been investigated. Using the method of induced representations, we find all the irreducible representations of $D^{(1)}_{9n, 3n}$ group and its character table for arbitrary $n$. We have derived the Kronecker products and constructed the Clebsch-Gordan coefficients. These details would be necessary and particularly useful for
model builders aiming at construction of flavor models based on the group $D^{(1)}_{9n, 3n}$.
Furthermore, we have identified the class-inverting automorphisms of the $D^{(1)}_{9n, 3n}$ group, and show that the corresponding CP transformations are of the same form as the flavor symmetry transformations in our working basis.

In the ``direct'' approach, the original symmetry $D^{(1)}_{9n,3n}\rtimes H_{CP}$ is broken down to  $K_{4}\times H^{\nu}_{CP}$ in the neutrino sector and to $G_{l}\rtimes H^{l}_{CP}$ in the charged lepton sector, where $G_{l}$ is an abelian subgroup which allows to distinguish the three generations of leptons.
In this scenario, all the lepton mixing parameters including the Majorana CP phases are completely fixed by the residual symmetries.
We have considered all the possible residual subgroups $K_4$, $G_{l}$ and the residual CP transformations that can be consistently combined.
We find that the lepton mixing matrices compatible with the data are of the trimaximal form. Both Dirac phase $\delta_{CP}$ and the Majorana phase $\alpha_{31}$ are predicted to be conserved, and the values of the Majorana phase $\alpha_{21}$ are $0$, $\frac{2}{3n}\pi$, $\frac{4}{3n}\pi$, $\ldots$, $\frac{6n-2}{3n}\pi$.

In contrast with the ``direct'' approach, the residual symmetry preserved by the neutrino mass matrix is $Z_2\times H^{\nu}_{CP}$ in the ``semidirect'' approach. Since the remnant flavor symmetry of the neutrino sector is $Z_2$ instead of $K_4$, it would fix only one column of the PMNS matrix.
Taking into account the remnant CP transformations further, all the lepton mixing angles as well as the CP violating phases would be predicted in terms of a continuous free parameter $\theta$ besides the parameters characterizing the residual symmetries. We find that only four types of mixing patterns named as cases I, II, III and IV can accommodate the experimental data on lepton mixing angles for certain values of the continuous parameter $\theta$ and the discrete parameter $\varphi_i$ determined by the postulated residual symmetries. For cases III and IV, the residual $Z_2$ subgroup is chosen to be generated by the element $c^{9n/2}$ such that the group index $n$ has to be even. We have performed a detailed analytical and numerical analysis. It is remarkable that either the solar mixing angle $\theta_{12}$ or the atmospheric mixing angle $\theta_{23}$ is bounded within certain
intervals for arbitrary $n$. As a consequence, these predictions can be testable by the next generation of reactor neutrino experiments and long baseline experiments. The admissible values of the mixing angles and CP phases for each $D^{(1)}_{9n,3n}$ group until $n=50$ have been studied. Interestingly enough, the first two smallest $D^{(1)}_{9n,3n}$ groups with $n=1, 2$ already allow a good fit to the data on lepton mixing angles, and the CP violating phases can be conserved, maximal or some other irregular values.
Moreover, the phenomenological predictions for the neutrinoless double beta decay are exploited.

In the so-called ``variant of semidirect'' approach, the remnant symmetries of the neutrino and the charged lepton mass matrices are assumed to be $K_{4}\times H^{\nu}_{CP}$ and $Z_{2}\times H^{l}_{CP}$ respectively.
We find only one type of mixing pattern named as case V is phenomenologically viable in this scenario. One row of the PMNS matrix is determined to be $(1/2, 1/2, -e^{i\varphi_9}/\sqrt{2})$. The solar mixing angle is predicted to lie in the interval $0.328\leq\sin^2\theta_{12}\leq0.359$, and the atmospheric angle is in the range of $0.510\leq\sin^2\theta_{23}\leq0.515$ or $0.485\leq\sin^2\theta_{23}\leq0.490$. Moreover, both Dirac phase and the Majorana phase $\alpha_{21}$ are bounded from above  $\left|\sin\delta_{CP}\right|\leq0.594$ and $\left|\sin\alpha_{21}\right|\leq0.399$ respectively.

In our framework, the obtained results for lepton flavor mixing only depend on the structure of flavor symmetry group and the postulated residual symmetries, and they are independent of the breaking mechanism that how the required vacuum alignment needed to achieve the remnant symmetries is dynamically realized. It would be interesting to construct concrete models in which the breaking of the symmetry group to the residual symmetries are spontaneous due to the non-vanishing vacuum expectation values of some
flavon fields.

\section*{Acknowledgements}

This work is supported by the National Natural Science Foundation of China under Grant Nos. 11275188, 11179007 and 11522546.

\section*{Appendix}
\begin{appendix}

\section{\label{app:group_theory}Group theory of $D^{(1)}_{9n,3n}\cong(Z_{9n}\times Z_{3n})\rtimes S_{3}$}\label{sclass}
\cleqn
The group $D^{(1)}_{9n,3n}$ for a generic integer $n$ is a non-Abelian finite subgroup of $SU(3)$ of type D~\cite{Grimus:2013apa}. Its order is $162n^2$. It is isomorphic to the semidirect product of the $S_3$, the smallest non-Abelian finite group, with
$(Z_{9n} \times Z_{3n})$, i.e. $D^{(1)}_{9n,3n}\cong(Z_{9n}\times Z_{3n})\rtimes S_{3}$. The $D^{(1)}_{9n,3n}$ group can be defined in terms of four generators $a$, $b$, $c$ and $d$ fulfilling the following relations~\cite{Grimus:2013apa,Yao:2015dwa}:
\begin{eqnarray}
\nonumber&&\qquad\quad a^3=b^2=(ab)^2=c^{9n}=d^{3n}=1,\quad cd=dc,\\
\label{eq:multiplication_rules}&&aca^{-1}=cd,\quad ada^{-1}=c^{-3}d^{-2},\quad bcb^{-1}=cd,\quad bdb^{-1}=d^{-1}\,.
\end{eqnarray}
One can see that $a$ and $b$ generate $S_3$, and $c$ and $d$ generate the $Z_{9n}$ and $Z_{3n}$ subgroups respectively. Any group element $g\in D^{(1)}_{9n, 3n}$ can be written as a product of powers of the generators $a$, $b$, $c$ and $d$,
\begin{equation}
\label{eq:genform1}
g=a^{\alpha}b^{\beta}c^{\gamma}d^{\delta}\,,
\end{equation}
with $\alpha=0, 1, 2$, $\beta=0, 1$, $\gamma=0, 1, \ldots, 9n-1$ and $\delta=0, 1,\ldots, 3n-1$. From the multiplication rules in Eq.~\eqref{eq:multiplication_rules}, the following useful relations can be obtained,
\begin{eqnarray}
\nonumber&&ca=ac^{-2}d^{-1},\qquad da=ac^3d,\qquad ba=a^2b,\quad ca^2=a^2cd,\\
\label{eq:useful_relations}&& da^2=a^2c^{-3}d^{-2},\qquad ba^2=ab,\qquad cb=bcd,\qquad db=bd^{-1}\,.
\end{eqnarray}
Utilizing Eqs.~(\ref{eq:genform1}, \ref{eq:useful_relations}), we find that the elements of $D^{(1)}_{9n, 3n}$ group belong to the following conjugacy classes:
\begin{eqnarray}\label{eq:classes}
\nonumber 1 &:&1\mathcal{C}_{1}=\left\{1\right\},\\
\nonumber 2 &:&1\mathcal{C}_{1}^{(\nu)}=\left\{c^{\nu}\right\},\quad \nu=\mbox{$3n,6n$}, \\
\nonumber 9n-3 &:& 3\mathcal{C}_{1}^{(\rho)}=\left\{c^{\rho}, c^{\rho}d^{\rho}, c^{-2\rho}d^{-\rho}\right \},\quad\rho\neq \mbox{$0,3n,6n$},  \\
\nonumber\frac{27n(n-1)+6}{6}&:&6\mathcal{C}_{1}^{(\rho,\sigma)}=\left\{c^{\rho}d^{\sigma}, c^{\rho-3\sigma}d^{\rho-2\sigma}, c^{3\sigma-2\rho}d^{\sigma-\rho}, c^{\rho}d^{\rho-\sigma}, c^{3\sigma-2\rho}d^{2\sigma-\rho}, c^{\rho-3\sigma}d^{-\sigma} \right \},\\
\nonumber 3 &:& 18n^{2}\mathcal{C}_{2}^{(\tau)}=\{ac^{\tau+3x}d^{y}, a^2c^{\tau+3x}d^{\tau+3x-y}|\;x,y=0,1,\ldots,3n-1\},~\tau=0,1,2 \\
\nonumber 9n &:& 9n\mathcal{C}_{3}^{(\rho)}=\left\{bc^{\rho}d^{x}, a^2bc^{\rho-3x}d^{\rho-2x}, abc^{\rho-3x}d^{-x}| x=0,1,\ldots,3n-1 \right\}, \rho=0,1,\ldots,9n-1\,,
\end{eqnarray}
where the quantity on the left of the colon denotes the number of classes and the quantity on the right of the colon refers to the number of elements contained in the classes. The parameters $\rho$ and $\sigma$ in the conjugacy class $6\mathcal{C}_{1}^{(\rho,\sigma)}$ can take the values
$\rho=0,1,\ldots,9n-1$, $\sigma=0,1,\ldots, 3n-1$, and the following possibilities are excluded,
\begin{equation}
\rho-2\sigma=0~\text{mod}(3n),\qquad  \rho-\sigma=0~\text{mod}(3n),\qquad \sigma=0\,.
\end{equation}
As a result, the $D^{(1)}_{9n, 3n}$ group totally has $1+2+(9n-3)+\frac{27n(n-1)+6}{6}+3+9n=(3n+1)(3n+8)/2$ different conjugacy classes. Furthermore, we can check that the center of the $D^{(1)}_{9n,3n}$ group is $Z(D^{(1)}_{9n,3n})=\left\{1, c^{3n}, c^{6n}\right\}$.

\subsection{\label{sec:representation}Irreducible representations of $D^{(1)}_{9n, 3n}$ group}

Now we proceed to construct all the irreducible representations of
the $D^{(1)}_{9n,3n}$ group. Firstly we concentrate on the one-dimensional representations in which all generators are represented by pure numbers and they are commutable with each other. From Eq.~\eqref{eq:multiplication_rules}, we see that
\begin{equation}
a=d=1,\quad b^2=1,\quad c^3=1\,.
\end{equation}
Hence $D^{(1)}_{9n,3n}$ group has six singlet representations given by
\begin{equation}
\label{eq:one_rep}
\begin{array}{lll}
\mathbf{1_{0,0}} &:& a=b=c=d=1  \,,\\
\mathbf{1_{0,1}} &:& a=b=d=1,\quad c=\omega\,,\\
\mathbf{1_{0,2}} &:& a=b=d=1,\quad c=\omega^{2}  \,,\\
\mathbf{1_{1,0}} &:& a=c=d=1, \quad b=-1  \,,  \\
\mathbf{1_{1,1}} &:& a=d=1, \quad b=-1, \quad c=\omega  \,, \\
\mathbf{1_{1,2}} &:& a=d=1, \quad b=-1, \quad c=\omega^{2}  \,,
\end{array}
\end{equation}
for any integer $n$, where $\omega\equiv e^{2\pi i/3}$. These one-dimensional representations differ in the values of the generators $b$ and $c$, and they can be neatly written as
\begin{equation}
\mathbf{1}_{i,j}~:~ a=d=1,\quad b=(-1)^i,\quad c=\omega^j,~~ \text{with}~~i=0,1,~j=0,1,2\,.
\end{equation}
As far as we know, the representations of the $D^{(1)}_{9n,3n}$ group has not been worked out in the literature. It is a nontrivial task. In the following, we shall use the method of induced representations to build the remaining irreducible representations. The induced representation can be commonly found in the literature. In the following, we first briefly review the basic idea of the induced representation. Let $\mathcal{G}$ be a finite group and $\mathcal{H}$ any subgroup of $\mathcal{G}$ with index $n$.
The index of $\mathcal{H}$ in $\mathcal{G}$ is the number of cosets of $\mathcal{H}$ in $\mathcal{G}$, i.e. $n=|\mathcal{G}|/|\mathcal{H}|$ where $|\mathcal{G}|$ and $|\mathcal{H}|$ denote the order of $\mathcal{G}$ and $\mathcal{H}$ respectively. We denote $x_1$, $x_{2}$, $\ldots$, $x_n$ as a full set of representatives in $\mathcal{G}$ of the cosets in $\mathcal{G}/\mathcal{H}$, i.e.
\begin{equation}
\mathcal{G}/\mathcal{H}=x_{1}(x_{1}\equiv 1)\mathcal{H}~\oplus~x_2\mathcal{H}~\oplus~\cdots~\oplus ~x_{n}\mathcal{H}\,.
\end{equation}
Furthermore, let $\varrho$ be a $d$-dimensional irreducible representation of $\mathcal{H}$ with $\varrho: H\rightarrow GL(V)$, where $V$ is the representation space of dimension $d$ and $GL(V)$ is the group of non-singular linear maps on $V$.
Supposing $\left\{e_{1},\ldots, e_{d}\right\}$ is a basis of the vector space $V$, the action of any element $h\in\mathcal{H}$ on the basis vector $e_{i}$ is
\begin{equation}
h:~e_{i}~\mapsto~ \varrho(h)_{ji}e_{j}\,.
\end{equation}
The induced representation can be thought of as acting on the following space:
\begin{equation}
W=\bigoplus^{n}_{i=1}x_{i}V\,,
\end{equation}
where each $x_{i}V$ is an isomorphic copy of the vector space $V$. The basis vector of the space $W$ can be taken to be
\begin{equation}
x_{k}e_{i}\equiv e_{k,i}\,,\quad\text{with}\quad k=1, 2,\ldots,n,~~i=1, 2,\ldots,d\,.
\end{equation}
According to the definition of coset, any $g\in\mathcal{G}$ will then send each $x_{k}$ to a unique $x_{m}h$ with $h\in\mathcal{H}$ such that $gx_{k}=x_{m}h$ where $k, m=1, 2,\ldots,n$.
In the induced representation, an element $g\in\mathcal{G}$ acts on the vector space $W$ as follows
\begin{equation}
\label{eq:def_induced_rep}g:~e_{k, i}~\mapsto~ge_{k, i}=gx_{k}e_{i}=x_{m}he_{i}
=\varrho(h)_{ji}x_me_{j}
=\varrho(h)_{ji}e_{m,j}\,.
\end{equation}
Thus we see that $\mathcal{G}$ acts linearly on $W$, and its action is thus represented by a $(dn\times dn)$ matrix. Notice that the induced representation is not necessarily irreducible.

We now apply this method to the group $D^{(1)}_{9n,3n}=\mathcal{G}$, and take the subgroup to be $\mathcal{H}=Z_{9n}\times Z_{3n}$. The index of $\mathcal{H}$ in $\mathcal{G}$ is $n=6$. Since $\mathcal{H}$ is an abelian
subgroup, its irreducible representations $\varrho$ can only be one-dimensional.
$e_1$ is the basis for the representation space of $\mathcal{H}$, the generators $c$ and $d$ act on $e_1$ as follows
\begin{equation}
\label{eq:eigenvalue_cd}
c e_1=\eta^le_{1},\qquad d e_{1}=\eta^{-3k}e_{1}\,,
\end{equation}
where $\eta=e^{\frac{2\pi i}{9n}}$, the values of the parameters $l$ and $k$ are $l=0, 1, \dots, 9n-1$ and $k=0, 1, \dots, 3n-1$. The six representative elements of the coset $\mathcal{G}/\mathcal{H}$ can be chosen to be
\begin{equation}
x_1=1,\quad x_2=a^2,\quad x_3=a,\quad x_4=b,\quad x_5=ab,\quad x_6=a^2b\,.
\end{equation}
As a consequence, we can obtain the basis of the vector space $W$ on which the induced representation is defined,
\begin{eqnarray}
\nonumber&& e_{1}\equiv x_1e_1=e_1,~\quad e_{2}\equiv x_2e_1=a^{2}e_1,
\quad e_{3}\equiv x_3e_1=ae_1,\\
\label{eq:six_basis}&&e_{4}\equiv x_4e_1=be_1, \quad e_{5}\equiv x_5e_1=abe_1,\quad e_{6}\equiv x_6e_1=a^{2}be_{1}\,.
\end{eqnarray}
According to Eq.~\eqref{eq:def_induced_rep}, the actions of the generators $a$, $b$, $c$ and $d$ on the above six basis vectors can be straightforwardly derived by utilizing the useful identities in Eq.~\eqref{eq:useful_relations} :
\begin{equation}
\label{eq:linear1}
\begin{array}{lll}
ae_{1}=e_{3}, ~~&~~  ae_{2}=e_{1}, ~~&~~   ae_{3}=e_{2},  \\
ae_{4}=e_{5}, ~~&~~  ae_{5}=e_{6}, ~~&~~  ae_{6}=e_{4}, \\
be_{1}=e_{4}, ~~&~~  be_{2}=e_{5}, ~~&~~  be_{3}=e_{6},  \\
be_{4}=e_{1}, ~~&~~ be_{5}=e_{2}, ~~&~~  be_{6}=e_{3}, \\
ce_{1}=\eta^{l}e_{1}, ~~&~~  ce_{2}=\eta^{l-3k}e_{2},  ~~&~~ ce_{3}=\eta^{-2l+3k}e_{3},  \\
ce_{4}=\eta^{l-3k}e_{4},  ~~&~~  ce_{5}=\eta^{-2l+3k}e_{5},  ~~&~~ ce_{6}=\eta^{l}e_{6}, \\
de_{1}=\eta^{-3k}e_{1},  ~~&~~  de_{2}=\eta^{-3l+6k}e_{2},  ~~&~~ de_{3}=\eta^{3l-3k}e_{3},  \\
de_{4}=\eta^{3k}e_{4},  ~~&~~ de_{5}=\eta^{3l-6k}e_{5},  ~~&~~ de_{6}=\eta^{-3l+3k}e_{6}\,.
\end{array}
\end{equation}
Then we can read out the representation matrices as follows
\begin{equation}
\label{eq:6drep_induced}
\mathbf{6}_{(l, k)}:\quad a=\left(\begin{array}{cc} a_{1}& 0 \\  0 & a_{2} \end{array}\right),\qquad
b=\left(\begin{array}{cc} 0    & \mathbb{1}_{3} \\  \mathbb{1}_{3}  & 0 \end{array}\right),\qquad
c=\left(\begin{array}{cc} c_{1} & 0 \\ 0 & c_{2}
\end{array}\right),\qquad
d=\left(\begin{array}{cc} d_{1} & 0 \\ 0 & d_{2}
\end{array}\right)\,,
\end{equation}
where $\mathbb{1}_{3}$ refers to a $3\times3$ unit matrix, and the different submatrices are given by
\begin{eqnarray}
\nonumber&&a_{1}=\left(\begin{array}{ccc}
0 & 1 & 0 \\
0 & 0 & 1 \\
1 & 0 & 0
\end{array}\right),\qquad\quad
a_{2}=\left(\begin{array}{ccc}
0 & 0 & 1 \\
1 & 0 & 0 \\
0 & 1 & 0
\end{array}\right)\,, \\
\nonumber&&\hskip-0.4in
c_{1}=\left(\begin{array}{ccc}
\eta^{l}& 0 & 0 \\
0 & \eta^{l-3k} & 0 \\
0 & 0 & \eta^{-2l+3k}
\end{array}\right),\qquad\quad
c_{2}=\left(\begin{array}{ccc}
\eta^{l-3k} & 0 & 0   \\
0 & \eta^{-2l+3k} & 0 \\
0 & 0 & \eta^{l}
\end{array}\right)\,, \\
\label{eq:6drep_subMatrix}&&\hskip-0.4in d_{1}=\left(\begin{array}{ccc}
\eta^{-3k}& 0 & 0 \\
0 & \eta^{-3l+6k} & 0 \\
0 & 0 & \eta^{3l-3k}
\end{array}\right),\qquad
d_{2}=d^{-1}_{1}=\left(\begin{array}{ccc}
\eta^{3k}& 0 & 0 \\
0 & \eta^{3l-6k} & 0 \\
0 & 0 & \eta^{-3l+3k}
\end{array}\right)\,.
\end{eqnarray}
The above different representations labelled by $(l, k)$ may be equivalent. If we perform the similarity transformations generated by
\begin{equation}\label{eq:simtra1}
S=\left(\begin{array}{cc}
a_{1} & 0 \\
0 & a_{1} \\
\end{array}\right), \qquad
T=\left(\begin{array}{cc}
0 & t \\
t & 0 \\
\end{array}\right),\qquad S^{3}=T^{2}=(ST)^{2}=1\,,
\end{equation}
where
\begin{equation}
t=\left(\begin{array}{ccc}
0 & 0 & 1\\
0 & 1 & 0\\
1 & 0 & 0\\
\end{array}\right)\,,
\end{equation}
the representations matrices for $a$ and $b$ are kept intact while the diagonal elements of both $c$ and $d$ are interchanged. As a result, the same representation is labeled in six different ways
\begin{equation}
\label{eq:6pairs}
\left(\begin{array}{c} l    \\ k
\end{array}\right),~
\left(\begin{array}{c} l-3k \\ l-2k
\end{array}\right),~
\left(\begin{array}{c}-2l+3k    \\ -l+k
\end{array}\right),~
\left(\begin{array}{c} l   \\ l-k
\end{array}\right),~
\left(\begin{array}{c} -2l+3k  \\ -l+2k
\end{array}\right),~
\left(\begin{array}{c} l-3k   \\ -k
\end{array}\right)\,.
\end{equation}
The six pairs above can be compactly written into the form
\begin{equation}
M^{p}_{s}\left(\begin{array}{c} l \\ k
\end{array}\right)\,,
\end{equation}
where
\begin{equation}
\label{eq:defineM}
M^{p}_{s}=\left(\begin{array}{cc}
1 ~&~ -3 \\
1 ~&~ -2
\end{array}\right)^{p}
\left(\begin{array}{cc}
1 ~&~ 0 \\
1 ~&~ -1
\end{array}\right)^{s},\quad\mathrm{with}\quad p=0, 1, 2, ~s=0, 1\,.
\end{equation}
Now we proceed to study whether the six-dimensional representations constructed by the induced representation method are irreducible or not by the famous \textbf{Mackey theorem} in math~\cite{Mackey:1951,Simon:1996,Serre:1977}. If any one of them is reducible, we further decompose it into the direct sum of the irreducible representations of $D^{(1)}_{9n, 3n}$ group.

\textbf{Theorem (Mackey's Irreducibility Criterion):} Let $\mathcal{H}\subset\mathcal{G}$ and $\varrho$ be a representation of $\mathcal{H}$. For $s\in\mathcal{G}$, we define
\begin{equation}
\mathcal{H}_{s}\equiv\mathcal{H}\cap s\mathcal{H}s^{-1}, \qquad \varrho^{s}(h)\equiv\varrho(s^{-1}hs)
\end{equation}
for $h\in \mathcal{H}_{s}$ such that $\varrho^{s}$ is a representation of $\mathcal{H}_{s}$. Then the induced representation $\text{Ind}^{\mathcal{G}}_{\mathcal{H}}(\varrho)$ is irreducible if and only if
\begin{description}
\item[(1)] {$\varrho$ is irreducible.}
\item[(2)]{For all $s\in \mathcal{G}\setminus\mathcal{H}$, $\varrho^{s}$ and $\text{Res}^{\mathcal{H}_{s}}_{\mathcal{H}}(\varrho)$ are disjoint.}
\end{description}
Here $\text{Ind}^{\mathcal{G}}_{\mathcal{H}}(\varrho)$ denotes a representation of $G$ and it is induced from a representation $\varrho$ on a subgroup $\mathcal{H}$. $\text{Res}^{\mathcal{H}_{s}}_{\mathcal{H}}(\varrho)$ is the restriction of the representation $\varrho$ on $\mathcal{H}$ to $\mathcal{H}_{s}$. The notation $\mathcal{G}\setminus\mathcal{H}$ denotes the group elements in $\mathcal{G}$ but not in $\mathcal{H}$. Two representations $\varrho$ and $\varrho^{\prime}$ of a group are said to be disjoint if and only if they contain no equivalent subrepresentations, equivalently if and only if their characters are orthogonal.
From this theorem, it is easy to further obtain a useful corollary.\\
\textbf{Corollary:} Suppose $\mathcal{H}$ is a normal subgroup of $\mathcal{G}$, then we have $\mathcal{H}_{s}=\mathcal{H}$ and $\text{Res}^{\mathcal{H}_{s}}_{\mathcal{H}}(\varrho)=\varrho$. In order that $\text{Ind}^{\mathcal{G}}_{\mathcal{H}}(\varrho)$ be irreducible, it is necessary and sufficient that $\varrho$ is irreducible and not isomorphic to any of its conjugate $\rho^{s}$ for $s\notin H$.\\

This implies that the representation $\text{Ind}^{\mathcal{G}}_{\mathcal{H}}(\varrho)$ would be reducible if there is a $s\in\mathcal{G}\setminus\mathcal{H}$ leading to $\varrho^s(\mathcal{H})\cong \varrho(\mathcal{H})$ for normal subgroup $\mathcal{H}$. The corollary can be exploited to determine whether the six-dimensional representations $\mathbf{6}_{(l,k)}$ of the $D^{(1)}_{9n,3n}$ group in Eq.~\eqref{eq:6drep_induced} are reducible or not. The subgroup $\mathcal{H}=Z_{9n}\times Z_{3n}$ is a normal subgroup of $D^{(1)}_{9n,3n}$, and it is abelian such that its irreducible representation $\varrho$ is one-dimensional and specified by Eq.~\eqref{eq:eigenvalue_cd}. From the above corollary of the Mackey theorem, we know that the six-dimensional representation $\mathbf{6}_{(l,k)}$ is reducible if and only if $\varrho^{s}(\mathcal{H})$ and $\varrho(\mathcal{H})$ are equivalent representations for an element $s\in D^{(1)}_{9n,3n} \setminus \mathcal{H}$. In order to obtain the conditions in which the six-dimensional representations $\mathbf{6}_{(l,k)}$ is reducible, we only need to consider the value of $s$ is $b$, $ab$, $a^2b$, $a$ and $a^2$ respectively. The results are collected in table~\ref{tab:re_six_irrep}.

\begin{table}[t!]
\centering
\begin{tabular}{|c|c|c|} \hline \hline

$s$  & $\varrho^s(\mathcal{H})\cong \varrho(\mathcal{H})$ &  \texttt{Reducible conditions}   \\ \hline

$b$  & $\varrho(b^{-1}cb)=\varrho(c)$, $\varrho(b^{-1}db)=\varrho(d)$ &  $3k=0$~$(\text{mod}~9n)$  \\ \hline

$ab$  & $\varrho((ab)^{-1}c(ab))=\varrho(c)$, $\varrho((ab)^{-1}d(ab))=\varrho(d)$ &  $3l-3k=0$ $\text{mod}(9n)$  \\ \hline

$a^2b$  & $\varrho\left((a^2b)^{-1}c(a^2b)\right)=\varrho(c)$, $\varrho((a^2b)^{-1}d(a^2b))=\varrho(d)$ &  $3l-6k=0$ $\text{mod}(9n)$  \\ \hline

$a$($a^2$) & $\varrho(a^{-1}ca)=\varrho(c)$, $\varrho(a^{-1}da)=\varrho(d)$ &  $3k=0$, $3l=0$ $\text{mod}(9n)$ \\ \hline \hline
\end{tabular}
\caption{\label{tab:re_six_irrep} The reducible conditions for the six-dimensional representations $\mathbf{6}_{(l, k)}$. The one-dimensional representation $\varrho$ of $\mathcal{H}$ is given by Eq.~\eqref{eq:eigenvalue_cd}, i.e. $\varrho(c)=\eta^{l}$ and $\varrho(d)=\eta^{-3k}$. The values of parameters $l$ and $k$ are $l=0, 1, \cdots, 9n-1$ and $k=0, 1, \cdots, 3n-1$.}
\end{table}

\begin{itemize}

\item {\textbf{Six-dimensional representations}}

Six-dimensional representations of the $D^{(1)}_{9n, 3n}$ group have been constructed by the method of induced representation, as shown in Eq.~\eqref{eq:6drep_induced} and Eq.~\eqref{eq:6drep_subMatrix}. From table~\ref{tab:re_six_irrep}, we find that the induced representation $\mathbf{6}_{(l, k)}$ in Eq.~\eqref{eq:6drep_induced} would be reducible when any of the following conditions are met
\begin{equation}
\begin{cases}
9n ~~:~~k=0,~~l=0, 1, \ldots, 9n-1,   \\
9n-3 ~~:~~3l-3k=0~\text{mod}(9n),~~ k\neq 0\\
9n-3 ~~:~~3l-6k=0~\text{mod}(9n),~~ k\neq 0\\
\end{cases}\,,
\end{equation}
where the quantity on the left of the colon is the number of $(l, k)$ values of the properties on the right of the colon. Excluding these values for $l$ and $k$, there should be $9n\times3n-9n-(9n-3)-(9n-3)=27n(n-1)+6$  different pairs of $(l, k)$. Furthermore, taking into account the over counting issue shown in Eq.~\eqref{eq:6pairs}, we essentially have $\frac{27n(n-1)+6}{6}$ six-dimensional irreducible representations, and the representation matrices of the generators are given in Eq.~\eqref{eq:6drep_induced}.

\item {\textbf{Three-dimensional representations}}

Once the conditions $3k=0$, $3l-3k=0~\text{mod}(9n)$ or $3l-6k=0~\text{mod}(9n)$ are fulfilled, the six-dimensional induced representation $6_{(l, k)}$ could be decomposed into the direct sum of three and two-dimensional representations. Firstly we concentrate on the case of $3k=0$ or equivalently $k=0$ . From Eq.~\eqref{eq:linear1} we see that the eigenvalues of $c$ on the three pairs of basis vectors $(e_1, e_4)$, $(e_2, e_6)$ and $(e_3, e_5)$ are $\eta^{l}$, $\eta^l$ and $\eta^{-2l}$ respectively, and the eigenvalues of $d$ on the three pairs vectors $(e_1, e_4)$, $(e_2, e_6)$ and $(e_3, e_5)$ are $1$, $\eta^{-3l}$ and $\eta^{3l}$ respectively. Hence we recombine the six vectors $e_1, \ldots, e_6$ into
\begin{equation}
e^{\prime}_1=x_1e_1+y_1e_4, \quad e^{\prime}_2=x_2e_2+y_2e_6, \quad e^{\prime}_3=x_3e_3+y_3e_5\,.
\end{equation}
In the case of $3k=0$ and $l\neq0,3n,6n$, the three vectors $e^{\prime}_1$, $e^{\prime}_2$ and $e^{\prime}_3$ can be distinguished from each other by the actions of $c$ and $d$, and they must be closed under the action of $a$ and $b$. Considering the effect of $a$, we find
\begin{equation}
ae^{\prime}_1=e^{\prime}_3, \quad ae^{\prime}_2=e^{\prime}_1, \quad ae^{\prime}_3=e^{\prime}_2\,,
\end{equation}
which yields
\begin{equation}
x_1=x_2=x_3\equiv x, \qquad y_1=y_2=y_3\equiv y\,.
\end{equation}
Furthermore, closeness under the action of $b$ implies
\begin{equation}
be^{\prime}_1=\pm e^{\prime}_1, \quad be^{\prime}_2=\pm e^{\prime}_3, \quad be^{\prime}_3=\pm e^{\prime}_2\,.
\end{equation}
In the case of $be^{\prime}_1=e^{\prime}_1$, i.e. $x=y$, then the normalized basis vectors of a three-dimensional subspace can be chosen to be
\begin{equation}\label{eq:three_basis1}
e^{\prime}_{1}=\frac{1}{\sqrt{2}}(e_{1}+e_{4}),\quad
e^{\prime}_{2}=\frac{1}{\sqrt{2}}(e_{2}+e_{6})=a^{2}e^{\prime}_{1},
\quad  e^{\prime}_{3}=\frac{1}{\sqrt{2}}(e_{3}+e_{5})=ae^{\prime}_{1}\,.
\end{equation}
In the case of $be^{\prime}_1=-e^{\prime}_1$, i.e. $x=-y$, we define the three normalized orthogonal vectors as
\begin{equation}\label{eq:three_basis2}
e^{\prime}_{4}=\frac{1}{\sqrt{2}}(e_{1}-e_{4}),  \quad
e^{\prime}_{5}=\frac{1}{\sqrt{2}}(e_{2}-e_{6})=a^{2}e^{\prime}_{4}, \quad  e^{\prime}_{6}=\frac{1}{\sqrt{2}}(e_{3}-e_{5})=ae^{\prime}_{4}\,.
\end{equation}
It is easy to check that $e^{\prime}_{4}$, $e^{\prime}_{5}$ and $e^{\prime}_{6}$ span another three-dimensional subspace. The basis transformation matrix from $e_{i}$ to $e^{\prime}_{i}$ is denoted by $\Omega$, i.e.
\begin{equation}
e^{\prime}_{i}=\sum^{6}_{j=1}{e_{j}\Omega_{ji}}\,,
\end{equation}
where the similarity transformation $\Omega$ reads
\begin{equation}
\Omega=\frac{1}{\sqrt{2}}\left(\begin{array}{cc}
\mathbb{1}_{3} & \mathbb{1}_{3}  \\
\varpi  & -\varpi
\end{array}\right), \qquad
\varpi=
\left(\begin{array}{ccc}
1&0&0\\
0&0&1\\
0&1&0
\end{array}\right),
\end{equation}
In the new basis, the representation matrices of the generators are of the following form
\begin{equation}
\begin{array}{cc}
a^{\prime}\equiv\Omega^{-1}a\Omega=\left(\begin{array}{cc}
a_{\mathbf{3}_{l, 0}} & 0 \\ 0 & a_{\mathbf{3}_{l, 1}}
\end{array}\right), ~~&~~
b^{\prime}\equiv\Omega^{-1}b\Omega=\left(\begin{array}{cc}
b_{\mathbf{3}_{l, 0}} & 0 \\ 0 & b_{\mathbf{3}_{l, 1}}
\end{array}\right), \\[-7pt]
                    \\[4pt]
c^{\prime}\equiv\Omega^{-1}c\Omega=\left(\begin{array}{cc}
c_{\mathbf{3}_{l, 0}} & 0 \\ 0 & c_{\mathbf{3}_{l, 1}}
\end{array}\right), ~~&~~
d^{\prime}\equiv\Omega^{-1}d\Omega=\left(\begin{array}{cc}
d_{\mathbf{3}_{l, 0}} & 0 \\ 0 & d_{\mathbf{3}_{l, 1}}
\end{array}\right)\,,
\end{array}
\end{equation}
where
\begin{eqnarray}
\nonumber&& a_{\mathbf{3}_{l, 0}}=\left(\begin{array}{ccc}
0 &1 &0 \\
0&0&1 \\
1&0&0\end{array}\right),~
b_{\mathbf{3}_{l, 0}}=\left(\begin{array}{ccc}
1 &0 &0 \\
0&0&1 \\
0&1&0\end{array}\right),~
c_{\mathbf{3}_{l, 0}}=\left(\begin{array}{ccc}
\eta^{l}&0 &0 \\
0&\eta^{l}&0 \\
0&0&\eta^{-2l}
\end{array}\right),~
d_{\mathbf{3}_{l, 0}}=\left(\begin{array}{ccc}
1 &0 &0 \\
0&\eta^{-3l}&0 \\
0&0&\eta^{3l}
\end{array}\right),\\
\label{eq:3dimrep1}&&a_{\mathbf{3}_{l, 1}}~=~a_{\mathbf{3}_{l, 0}},\qquad
b_{\mathbf{3}_{l, 1}}~=~-b_{\mathbf{3}_{l, 0}},\qquad
c_{\mathbf{3}_{l, 1}}~=~c_{\mathbf{3}_{l, 0}},\qquad
d_{\mathbf{3}_{l, 1}}~=~d_{\mathbf{3}_{l, 0}}\,.
\end{eqnarray}
This means that the six-dimensional representation $\mathbf{6}_{l,0}$ breaks up into two three-dimensional irreducible representation $\mathbf{3}_{l, 0}$ and $\mathbf{3}_{l, 1}$ which differ in the overall sign of $b$. Notice that the values of $l=0$, $3n$, $6n$ should be excluded, since both triplet representations $\mathbf{3}_{l, 0}$ and $\mathbf{3}_{l, 1}$ then could be decomposed into one-dimensional and two-dimensional representation.

Next we proceed to consider the case of $3l-3k=0~\text{mod}(9n)$ and $l\neq0, 3n, 6n$. We can construct the eigenstates of the generators $c$ and $d$ as follows
\begin{equation}
e^{\prime}_1=x_1e_1+y_1e_5, \quad e^{\prime}_2=x_2e_2+y_2e_4, \quad e^{\prime}_3=x_3e_3+y_3e_6\,.
\end{equation}
Note that $e^{\prime}_1$,  $e^{\prime}_2$ and $e^{\prime}_3$ are mapped into each other under the action of $a$ and $b$, Taking into account the normalization condition further, we have
$x_1=x_2=x_3=y_1=y_2=y_3=1/\sqrt{2}$ such that
\begin{equation}\label{eq:three_basis3}
e^{\prime}_{1}=\frac{1}{\sqrt{2}}(e_{1}+e_{5}), \quad
e^{\prime}_{2}=\frac{1}{\sqrt{2}}(e_{2}+e_{4})=a^{2}e^{\prime}_{1},
\quad  e^{\prime}_{3}=\frac{1}{\sqrt{2}}(e_{3}+e_{6})=ae^{\prime}_{1}\,,
\end{equation}
or $x_1=x_2=x_3=-y_1=-y_2=y_3=-1/\sqrt{2}$ which leads to
\begin{equation}\label{eq:three_basis4}
e^{\prime}_{4} =\frac{1}{\sqrt{2}}(e_{1}-e_{5}),  \quad
e^{\prime}_{5}=\frac{1}{\sqrt{2}}(e_{2}-e_{4})=a^{2}e^{\prime}_{4},
\qquad  e^{\prime}_{6}=\frac{1}{\sqrt{2}}(e_{3}-e_{6})=ae^{\prime}_{4}\,.
\end{equation}
It is straightforward to check the following equations are fulfilled
\begin{eqnarray}
\nonumber&&\hskip-0.25in be^{\prime}_1=e^{\prime}_2,\quad be^{\prime}_2=e^{\prime}_1, \quad be^{\prime}_3=e^{\prime}_3,\quad ce^{\prime}_1=\eta^{l}e^{\prime}_1,\quad ce^{\prime}_2=\eta^{-2l}e^{\prime}_2,\quad ce^{\prime}_3
=\eta^{l}e^{\prime}_3,\\
\nonumber&&\hskip-0.25in de^{\prime}_1=\eta^{-3l}e^{\prime}_1,\quad de^{\prime}_2=\eta^{3l}e^{\prime}_2,\quad de^{\prime}_3=e^{\prime}_3,\quad be^{\prime}_4=-e^{\prime}_5,\quad be^{\prime}_5=-e^{\prime}_4, \quad be^{\prime}_6=-e^{\prime}_6,\\
&&\hskip-0.25in ce^{\prime}_4=\eta^{l}e^{\prime}_4,~~ ce^{\prime}_5=\eta^{-2l}e^{\prime}_5,~~ ce^{\prime}_6
=\eta^{l}e^{\prime}_6,~~ de^{\prime}_4=\eta^{-3l}e^{\prime}_4,~~ de^{\prime}_5=\eta^{3l}e^{\prime}_5,~~ de^{\prime}_6=e^{\prime}_6\,.
\end{eqnarray}
As a result, the induced representation $\mathbf{6}_{l,\,l\, \text{mod}(3n)}$ for $l\neq0$, $3n$, $6n$ can be split into two three-dimensional representations. The unitary transformation from the $e_{i}$ basis to the $e^{\prime}_{i}$ basis is
\begin{equation}
e^{\prime}_{i}=\sum^{6}_{j=1}\Omega_{ji}e_{j}\,,
\end{equation}
with
\begin{equation}
\Omega=\frac{1}{\sqrt{2}}\left(\begin{array}{cc}
\mathbb{1}_{3} & \mathbb{1}_{3}  \\
\varpi & -\varpi
\end{array}\right), \qquad
\varpi=
\begin{pmatrix}
0&1&0\\
1&0&0\\
0&0&1
\end{pmatrix}\,.
\end{equation}
Performing the similarity transformation $\Omega$, the representation matrices in the new basis are given by
\begin{equation}
\begin{array}{cc}
a^{\prime}=\Omega^{-1}a\Omega=\left(\begin{array}{cc}
a_{\mathbf{3}_{l, 2}} & 0 \\ 0 & a_{\mathbf{3}_{l, 3}}
\end{array}\right), ~~&~~
b^{\prime}=\Omega^{-1}b\Omega=\left(\begin{array}{cc}
b_{\mathbf{3}_{l, 2}} & 0 \\ 0 & b_{\mathbf{3}_{l, 3}}
\end{array}\right), \\[-7pt]
                    \\[4pt]
c^{\prime}=\Omega^{-1}c\Omega=\left(\begin{array}{cc}
c_{\mathbf{3}_{l, 2}} & 0 \\ 0 & c_{\mathbf{3}_{l, 3}}
\end{array}\right), ~~&~~
d^{\prime}=\Omega^{-1}d\Omega=\left(\begin{array}{cc}
d_{\mathbf{3}_{l, 2}} & 0 \\ 0 & d_{\mathbf{3}_{l, 3}}
\end{array}\right)\,,
\end{array}
\end{equation}
where
\begin{eqnarray}
\nonumber&&
a_{\mathbf{3}_{l, 2}}=\left(\begin{array}{ccc}
0 &1 &0 \\
0&0&1 \\
1&0&0\end{array}\right),~
b_{\mathbf{3}_{l, 2}}=\left(\begin{array}{ccc}
0 &1 &0 \\
1&0&0 \\
0&0&1\end{array}\right),~
c_{\mathbf{3}_{l, 2}}=\left(\begin{array}{ccc}
\eta^{l}&0 &0 \\
0&\eta^{-2l}&0 \\
0&0&\eta^{l}\end{array}\right),~
d_{\mathbf{3}_{l, 2}}=\left(\begin{array}{ccc}
\eta^{-3l} &0 &0 \\
0&\eta^{3l}&0 \\
0&0&1\end{array}\right), \\
\label{eq:3dimrep2}&& a_{\mathbf{3}_{l, 3}}~=~a_{\mathbf{3}_{l, 2}},\qquad
b_{\mathbf{3}_{l, 3}}~=~-b_{\mathbf{3}_{l, 2}},\qquad
c_{\mathbf{3}_{l, 3}}~=~c_{\mathbf{3}_{l, 2}},\qquad
d_{\mathbf{3}_{l, 3}}~=~d_{\mathbf{3}_{l, 2}}\,.
\end{eqnarray}
Therefore $\mathbf{6}_{l,\,l\, \text{mod}(3n)}$ for $l\neq0$, $3n$, $6n$ is the direct sum of three-dimensional irreducible representations $\mathbf{3}_{l, 3}$ and $\mathbf{3}_{l, 4}$.

Finally we consider the case of $3l-6k=0~\text{mod}(9n)$ with $k\neq0$ and $l\neq0, 3n, 6n$.
In the same fashion as previous cases, we first recombine the basis vectors into
\begin{equation}
e^{\prime}_1=x_1e_1+y_1e_6, \quad e^{\prime}_2=x_2e_2+y_2e_5, \quad e^{\prime}_3=x_3e_3+y_3e_4\,,
\end{equation}
which are eigenstates of both $c$ and $d$ and fulfill
\begin{equation}
ce^{\prime}_1=\eta^{-2l^{\prime}}e^{\prime}_1,~ ce^{\prime}_2=\eta^{l^{\prime}}e^{\prime}_2,~ ce^{\prime}_3=\eta^{l^{\prime}}e^{\prime}_3,~ de^{\prime}_1=\eta^{3l^{\prime}}e^{\prime}_1,~
de^{\prime}_2=e^{\prime}_2,~
de^{\prime}_3=\eta^{-3l^{\prime}}e^{\prime}_3\,,
\end{equation}
for any values of $x_i$ and $y_i$ ($i=1, 2, 3$), where $l^{\prime}=l-3k=3n-k, 6n-k, 9n-k$ with $k=0, 1, \ldots, 3n-1$ such that the value of $l^{\prime}$ can be $0, 1, \ldots, 9n-1$.
Taking into account the action of the remaining two generators $a$ and $b$, we find two three-dimensional subspaces would be generated. The basis of the first subspace can be chosen to be
\begin{equation}
\label{eq:three_basis3}
e^{\prime}_{1}=\frac{1}{\sqrt{2}}(e_{1}+e_{6}), \quad
e^{\prime}_{2}=\frac{1}{\sqrt{2}}(e_{2}+e_{5})=a^{2}e^{\prime}_{1},
\quad  e^{\prime}_{3}=\frac{1}{\sqrt{2}}(e_{3}+e_{4})=ae^{\prime}_{1}\,,
\end{equation}
The basis vectors of the second three-dimensional subspace are
\begin{equation}
\label{eq:three_basis4}
e^{\prime}_{4}=\frac{1}{\sqrt{2}}(e_{1}-e_{6}), \quad
e^{\prime}_{5}=\frac{1}{\sqrt{2}}(e_{2}-e_{5})=a^{2}e^{\prime}_{4},
\quad  e^{\prime}_{6}=\frac{1}{\sqrt{2}}(e_{3}-e_{4})=ae^{\prime}_{4}\,,
\end{equation}
We can read out the unitary basis transformation
\begin{equation}
\Omega=\frac{1}{\sqrt{2}}\left(\begin{array}{cc}
\mathbb{1}_{3} & \mathbb{1}_{3} \\
\varpi &  -\varpi
\end{array}\right), \qquad
\varpi=
\left(\begin{array}{ccc}
0 & 0 & 1\\
0 & 1 & 0\\
1 & 0 & 0
\end{array}\right)\,.
\end{equation}
The representation matrices for the generators $a$, $b$, $c$ and $d$ transform as
\begin{equation}
\begin{array}{cc}
a^{\prime}=\Omega^{-1}a\Omega=\left(\begin{array}{cc}
a_{\mathbf{3}_{l^{\prime}, 4}} & 0 \\
0 & a_{\mathbf{3}_{l^{\prime}, 5}}
\end{array}\right), ~~&~~
b^{\prime}=\Omega^{-1}b\Omega=\left(\begin{array}{cc}
b_{\mathbf{3}_{l^{\prime}, 4}} & 0 \\ 0 & b_{\mathbf{3}_{l^{\prime}, 5}}
\end{array}\right), \\[-7pt] \\[4pt]
c^{\prime}=\Omega^{-1}c\Omega=\left(\begin{array}{cc}
c_{\mathbf{3}_{l^{\prime}, 4}} & 0 \\ 0 & c_{\mathbf{3}_{l^{\prime}, 5}}
\end{array}\right), ~~&~~
d^{\prime}=\Omega^{-1}d\Omega=\left(\begin{array}{cc}
d_{\mathbf{3}_{l^{\prime}, 4}} & 0 \\ 0 & d_{\mathbf{3}_{l^{\prime}, 5}}
\end{array}\right)\,,
\end{array}
\end{equation}
where
\begin{eqnarray}
\nonumber&& a_{\mathbf{3}_{l^{\prime}, 4}}=\left(\begin{array}{ccc}
0  & 1  & 0 \\
0  & 0  & 1 \\
1  & 0  & 0 \end{array}\right),~
b_{\mathbf{3}_{l^{\prime}, 4}}=\left(\begin{array}{ccc}
0  & 0  &  1 \\
0  & 1  &  0 \\
1  & 0  &  0\end{array}\right),~
c_{\mathbf{3}_{l^{\prime}, 4}}=\left(\begin{array}{ccc} \eta^{-2l^{\prime}} & 0 & 0 \\
0  &  \eta^{l^{\prime}} & 0 \\
0  &  0  &  \eta^{l^{\prime}}\end{array}\right),~
d_{\mathbf{3}_{l^{\prime}, 4}}=\left(\begin{array}{ccc}
\eta^{3l^{\prime}}   &  0  &  0 \\
0   &   1  &  0 \\
0   &   0  &\eta^{-3l^{\prime}}
\end{array}\right), \\
\label{eq:3dimrep3}&&
a_{\mathbf{3}_{l^{\prime}, 5}}~=~a_{\mathbf{3}_{l^{\prime}, 4}},\qquad
b_{\mathbf{3}_{l^{\prime}, 5}}~=~-b_{\mathbf{3}_{l^{\prime}, 4}},\qquad
c_{\mathbf{3}_{l^{\prime}, 5}}~=~c_{\mathbf{3}_{l^{\prime}, 4}},\qquad
d_{\mathbf{3}_{l^{\prime}, 5}}~=~d_{\mathbf{3}_{l^{\prime}, 4}}\,.
\end{eqnarray}
Note that both triplet representations $\mathbf{3}_{l^{\prime}, 4}$ and $\mathbf{3}_{l^{\prime}, 5}$ would be reducible for $l^{\prime}=0, 3n, 6n$.

So far we have obtained six three-dimensional irreducible representations $\mathbf{3}_{l, 0}$, $\mathbf{3}_{l, 1}$, $\mathbf{3}_{l, 2}$, $\mathbf{3}_{l, 3}$, $\mathbf{3}_{l, 4}$, $\mathbf{3}_{l, 5}$. However, only two of them are inequivalent because they are related with each other by similarity transformations as follows :
\begin{equation}
\begin{array}{llll}
a_{\mathbf{3}_{l, 2}}=U^{\dagger} a_{\mathbf{3}_{l, 0}}U, ~&~ b_{\mathbf{3}_{l, 2}}=U^{\dagger} b_{\mathbf{3}_{l, 0}}U, ~&~
c_{\mathbf{3}_{l, 2}}=U^{\dagger} c_{\mathbf{3}_{l, 0}}U, ~&~ d_{\mathbf{3}_{l, 2}}=U^{\dagger} d_{\mathbf{3}_{l, 0}}U, \\
a_{\mathbf{3}_{l, 3}}=U^{\dagger} a_{\mathbf{3}_{l, 1}}U, ~&~ b_{\mathbf{3}_{l, 3}}=U^{\dagger} b_{\mathbf{3}_{l, 1}}U, ~&~
c_{\mathbf{3}_{l, 3}}=U^{\dagger} c_{\mathbf{3}_{l, 1}}U, ~&~ d_{\mathbf{3}_{l, 3}}=U^{\dagger} d_{\mathbf{3}_{l, 1}}U, \\
a_{\mathbf{3}_{l, 4}}=U a_{\mathbf{3}_{l, 0}}U^{\dagger}, ~&~ b_{\mathbf{3}_{l, 4}}=U b_{\mathbf{3}_{l, 0}}U^{\dagger}, ~&~
c_{\mathbf{3}_{l, 4}}=U c_{\mathbf{3}_{l, 0}}U^{\dagger}, ~&~ d_{\mathbf{3}_{l, 4}}=U d_{\mathbf{3}_{l, 0}}U^{\dagger}, \\
a_{\mathbf{3}_{l, 5}}=U a_{\mathbf{3}_{l, 1}}U^{\dagger}, ~&~ b_{\mathbf{3}_{l, 5}}=U b_{\mathbf{3}_{l, 1}}U^{\dagger}, ~&~
c_{\mathbf{3}_{l, 5}}=U c_{\mathbf{3}_{l, 1}}U^{\dagger}, ~&~
d_{\mathbf{3}_{l, 5}}=U d_{\mathbf{3}_{l, 1}}U^{\dagger}\,,
\end{array}
\end{equation}
where the unitary transformation $U$ is
\begin{equation}
U=\left(\begin{array}{ccc}
0   &   0   &  1 \\
1   &   0   &  0 \\
0   &   1   &  0
\end{array}\right)\,.
\end{equation}
Hence we conclude that the $D^{(1)}_{9n, 3n}$ group totally has $2(9n-3)$ inequivalent three-dimensional irreducible representations which can be chosen to be $\mathbf{3}_{l, 0}$ and $\mathbf{3}_{l, 1}$ with $l\neq0, 3n, 6n$.

\item {\textbf{Two-dimensional representations}}

In the following, we shall show that both triplet representations $\mathbf{3}_{l, 0}$ and $\mathbf{3}_{l, 1}$ for $l=0, 3n, 6n$ would be reduced into the direct sum of one-dimensional representation and two-dimensional representation. Firstly we concentrate on $\mathbf{3}_{l, 0}$. In this case, the three basis vectors $e^{\prime}_1$, $e^{\prime}_2$ and $e^{\prime}_3$ in Eq.~\eqref{eq:three_basis1} can not be distinguished by the actions of $c$ and $d$, the eigenstates of the generator $a$ are
\begin{equation}
e^{\prime\prime}_1=\frac{1}{\sqrt{3}}(e^{\prime}_1+e^{\prime}_2+e^{\prime}_3), \quad
e^{\prime\prime}_2=\frac{1}{\sqrt{3}}(e^{\prime}_1+\omega e^{\prime}_2+\omega^2e^{\prime}_3), \quad
e^{\prime\prime}_3=\frac{1}{\sqrt{3}}(e^{\prime}_1+\omega^2e^{\prime}_2+\omega e^{\prime}_3)\,,
\end{equation}
with $ae^{\prime\prime}_1=e^{\prime\prime}_1$, $ae^{\prime\prime}_2=\omega e^{\prime\prime}_2$ and $ae^{\prime\prime}_3=\omega^2e^{\prime\prime}_3$. Under the action of the generator $b$, $e^{\prime\prime}_1$ is mapped into itself and $e^{\prime\prime}_2$ and $e^{\prime\prime}_3$ are interchanged.
Therefore the representation space of $\mathbf{3}_{l, 0}$ is split into one-dimensional subspace proportional to $e^{\prime\prime}_1$ and two-dimensional invariant subspaces spanned by $e^{\prime\prime}_2$ and $e^{\prime\prime}_3$. However, the representation matrix for $b$ is off-diagonal in the two-dimensional representation. In the present work, we would like to work in a basis where the representation matrix of $b$ is diagonal in the doublet representation such that all the relevant clebsch-gordan coefficients are real, as shown in Appendix~\ref{sec:App_CGC}. Consequently we choose the basis vectors as follows
\begin{eqnarray}
\nonumber&&e^{\prime\prime\prime}_1=e^{\prime\prime}_1=\frac{1}{\sqrt{3}}(e^{\prime}_1+e^{\prime}_2+e^{\prime}_3),\\
\nonumber&&e^{\prime\prime\prime}_2=\frac{1}{\sqrt{2}}(e^{\prime\prime}_2+e^{\prime\prime}_3)=\frac{1}{\sqrt{6}}(2e^\prime_1-e^\prime_2-e^\prime_3),\\
&&e^{\prime\prime\prime}_3=\frac{i}{\sqrt{2}}(e^{\prime\prime}_2-e^{\prime\prime}_3)=\frac{-1}{\sqrt{2}}(e^\prime_2-e^\prime_3)\,.
\end{eqnarray}
Then we can read out the unitary basis transformation matrix as
\begin{equation}
\mathcal{S}=\frac{1}{\sqrt{6}}\left(\begin{array}{ccc}
 \sqrt{2} &~ 2 ~& 0 \\
 \sqrt{2} &~ -1 ~& -\sqrt{3} \\
 \sqrt{2} &~ -1 ~& \sqrt{3} \\
\end{array}\right)\,,
\end{equation}
with $e^{\prime\prime\prime}_{i}=\sum^{3}_{j=1}e^{\prime}_{j}\mathcal{S}_{ji}$. In this set of new basis, the representation matrices for the generators $a$, $b$, $c$ and $d$ are
\begin{equation}
\begin{array}{ll}
a^{\prime\prime}=\mathcal{S}^{-1}a_{\mathbf{3}_{l, 0}}\mathcal{S}=\left(\begin{array}{cc}
1 & 0 \\   0 &  a_{\mathbf{2}_{0}}
\end{array}\right), ~~&~~
b^{\prime\prime}=\mathcal{S}^{-1}b_{\mathbf{3}_{l, 0}}\mathcal{S}=\left(\begin{array}{cc}
1 & 0 \\  0 &  b_{\mathbf{2}_{0}}
\end{array}\right), \\[-7pt] \\[4pt]
c^{\prime\prime}=\mathcal{S}^{-1}c_{\mathbf{3}_{l, 0}}\mathcal{S}=\eta^{l}\mathbb{1}_{3}, ~~&~~
d^{\prime\prime}=\mathcal{S}^{-1}d_{\mathbf{3}_{l, 0}}\mathcal{S}=\mathbb{1}_{3},
\end{array}
\end{equation}
with
\begin{equation}\label{eq:a2_b2}
a_{\mathbf{2}_{0}}=\frac{1}{2}\left(\begin{array}{cc}
-1 &~  -\sqrt{3} \\ \sqrt{3} &~-1
\end{array}\right), \quad
b_{\mathbf{2}_{0}}=\left(\begin{array}{cc}
1 &~  0 \\ 0 &~-1
\end{array}\right)\,,
\end{equation}
Note that $\eta^{l}=1, \omega,\omega^{2}$ for $l=0, 3n, 6n$ respectively.

Now we turn to another set of reducible triplet representations $\mathbf{3}_{l, 1}$ with $l=0, 3n, 6n$. In the same way as previous case, the new basis vectors are taken to be
\begin{equation}
e^{\prime\prime}_4=\frac{1}{\sqrt{3}}(e^\prime_4+e^\prime_5+e^\prime_6)\,,  \quad
e^{\prime\prime}_5=\frac{i}{\sqrt{2}}(e^\prime_5-e^\prime_6)\,,  \quad
e^{\prime\prime}_6=\frac{i}{\sqrt{6}}(2e^\prime_4-e^\prime_5-e^\prime_6)\,,
\end{equation}
where $e^{\prime}_4$, $e^{\prime}_5$ and $e^{\prime}_6$ are specified by Eq.~\eqref{eq:three_basis2}. The unitary transformation for this basis change is
\begin{equation}
\mathcal{S}=\frac{1}{\sqrt{6}}
\left(\begin{array}{ccc}
 \sqrt{2} &~ 0 ~& 2i \\
 \sqrt{2} &~ i\sqrt{3} ~& -i \\
 \sqrt{2} &~ -i\sqrt{3} ~& -i \\
\end{array}\right)\,.
\end{equation}
The corresponding representation matrices are given by
\begin{equation}
\begin{array}{ll}
a^{\prime\prime}=\mathcal{S}^{-1}a_{\mathbf{3}_{l, 1}}\mathcal{S}=\begin{pmatrix}
1 & 0 \\   0 &  a_{\mathbf{2}_{0}}
\end{pmatrix}, ~~&~~
b^{\prime\prime}=\mathcal{S}^{-1}b_{\mathbf{3}_{l,1}}\mathcal{S}=\begin{pmatrix}
-1 & 0 \\  0 &  b_{\mathbf{2}_{0}}
\end{pmatrix}, \\[-7pt]       \\[4pt]
c^{\prime\prime}=\mathcal{S}^{-1}c_{\mathbf{3}_{l, 1}}\mathcal{S}=\eta^{l}\mathbb{1}_{3}, ~~&~~
d^{\prime\prime}=\mathcal{S}^{-1}d_{\mathbf{3}_{l, 1}}\mathcal{S}=\mathbb{1}_{3}\,,
\end{array}
\end{equation}
where $a_{\mathbf{2}_{0}}$ and $b_{\mathbf{2}_{0}}$ are shown in Eq.~\eqref{eq:a2_b2}.
Hence by performing similarity transformation on the reducible triplet
representations $\mathbf{3}_{l, 0}$ and $\mathbf{3}_{l, 1}$ for $l=0, 3n, 6n$, we can obtain three inequivalent two-dimensional irreducible representations and six one-dimensional representations given in Eq.~\eqref{eq:one_rep}. The three two-dimensional representations differ in the representation matrix of $c$:
\begin{eqnarray}
\nonumber&& a_{\mathbf{2}_{0}}=\frac{1}{2}\left(\begin{array}{cc}
-1 &~  -\sqrt{3} \\ \sqrt{3} &~-1
\end{array}\right),
\quad
b_{\mathbf{2}_{0}}=\left(\begin{array}{cc}
1 &~  0 \\ 0 &~-1
\end{array}\right),
\quad
c_{\mathbf{2}_{0}}=d_{\mathbf{2}_{0}}=\mathbb{1}_{2}, \\
\nonumber&& a_{\mathbf{2}_{1}}= a_{\mathbf{2}_{0}},
\quad
b_{\mathbf{2}_{1}}= b_{\mathbf{2}_{0}},
\quad
c_{\mathbf{2}_{1}}=\omega\mathbb{1}_{2},\quad d_{\mathbf{2}_{1}}=\mathbb{1}_{2},\\
\label{eq:2_dimension}&& a_{\mathbf{2}_{2}}= a_{\mathbf{2}_{0}},
\quad
b_{\mathbf{2}_{2}}= b_{\mathbf{2}_{0}},
\quad
c_{\mathbf{2}_{2}}=\omega^{2}\mathbb{1}_{2},\quad d_{\mathbf{2}_{2}}=\mathbb{1}_{2}\,,
\end{eqnarray}
which can also be sententiously written as
\begin{equation}
a_{\mathbf{2}_{i}}=\frac{1}{2}\begin{pmatrix}
-1 &~  -\sqrt{3} \\
\sqrt{3} &~-1
\end{pmatrix},
~
b_{\mathbf{2}_{i}}= \begin{pmatrix}
1 &~  0 \\
0 &~-1
\end{pmatrix},
~
c_{\mathbf{2}_{i}}=\omega^i\mathbb{1}_{2},~ d_{\mathbf{2}_{i}}=\mathbb{1}_{2}, ~i=0, 1, 2\,.
\end{equation}
There are no more irreducible representations as we see that the number of irreducible representations is already equal to the number
of conjugacy classes:
\begin{equation}
\frac{27n(n-1)+6}{6}+2(9n-3)+3+6=1+2+(9n-3)+\frac{27n(n-1)+6}{6}+3+9n\,.
\end{equation}
Furthermore, we find that the sum of the squares of the dimensions of the irreducible representations is really equal to the order of the group, i.e.
\begin{equation}
\frac{27n(n-1)+6}{6}\times6^2+2(9n-3)\times3^2+3\times2^2+6\times1^2=162n^2\,.
\end{equation}
We can derive the $D^{(1)}_{9n,3n}$ character table by taking
traces over the relevant representation matrices. The results are displayed in table~\ref{tab:characterAa}, where $\tilde{M}^{p}_{s}$ refers to
\begin{equation}
\label{eq:Mtilde_nota}\tilde{M}^{p}_{s}\equiv\mbox{$
\left(\begin{array}{cc}
1 ~&~ 1 \\
-3 ~&~ -2
\end{array}\right)
^{p}
\left(\begin{array}{cc}
1 ~&~ 0 \\
-3 ~&~ -1
\end{array}\right)
^{s}$},\quad \mathrm{with} ~~ p=0,1,2, ~~s=0,1,
\end{equation}

With the character table, it is easy to calculate the Kronecker products of two irreducible representations of the $D^{(1)}_{9n, 3n}$ group as follows:
\begin{eqnarray}
\nonumber&&\mathbf{1}_{i,j}\otimes\mathbf{1}_{p,q}=\mathbf{1}_{f, g},\quad \mathbf{1}_{i,j}\otimes\mathbf{2}_{q}=\mathbf{2}_{g},\quad \mathbf{1}_{i,j}\otimes\mathbf{3}_{l, p}=\mathbf{3}_{l+3nj, f},\quad \mathbf{1}_{i,j}\otimes\mathbf{6}_{(l, k)}=\mathbf{6}_{(l+3nj, k)},\\
\nonumber&&\mathbf{2}_{j}\otimes\mathbf{2}_{q}=\mathbf{1}_{0,g}\oplus\mathbf{1}_{1,g}\oplus\mathbf{2}_{g},\quad
\mathbf{2}_{j}\otimes\mathbf{3}_{l, i}=\mathbf{3}_{l+3nj, 0}\oplus\mathbf{3}_{l+3nj, 1},\\
\nonumber&&\mathbf{2}_{j}\otimes\mathbf{6}_{(l, k)}~=\mathbf{6}_{(l+3nj, k)}\oplus\mathbf{6}_{(l+3nj,k)},\qquad \mathbf{3}_{l, i}\otimes\mathbf{3}_{ l^{\prime}, p}~=\mathbf{3}_{l+l^{\prime}, f}\oplus\mathbf{6}_{(l+l^{\prime}, l^{\prime})}
\end{eqnarray}

\end{itemize}

\begin{table}[t]
\begin{center}
\begin{tabular}{|c|c|c|c|c|c|c|}
\hline
\hline
  & $1\mathcal{C}_1$ &
$1\mathcal{C}_1^{(\nu)}$&$3\mathcal{C}_1^{(\rho)}$ & $6\mathcal{C}_1^{(\rho,\sigma)}$ &
$18n^2\mathcal{C}_2^{(\tau)}$ & $\!\!9n\mathcal{C}_3^{(\rho)}\!\!$\\
\hline
 $\mathbf{1_{0,0}}$  & $1$ & $1$ & $1$ & $1$ & $1$ & $1$   \\
 $\mathbf{1_{0,1}}$ & $1$ & $1$ & $\omega^{\rho}$ & $\omega^{\rho}$ & $\omega^{\tau}$ & $\omega^{\rho}$  \\
 $\mathbf{1_{0,2}}$  & $1$ & $1$ & $\omega^{2\rho}$ & $\omega^{2\rho}$ & $\omega^{2\tau}$ & $\omega^{2\rho}$   \\
 $\mathbf{1_{1,0}}$ & $1$ & $1$ & $1$ & $1$ & $1$ & $-1$  \\
 $\mathbf{1_{1,1}}$ & $1$ & $1$ & $\omega^{\rho}$ & $\omega^{\rho}$ & $\omega^{\tau}$ & $-\omega^{\rho}$  \\
 $\mathbf{1_{1,2}}$  & $1$ & $1$ & $\omega^{2\rho}$ & $\omega^{2\rho}$ & $\omega^{2\tau}$ & $-\omega^{2\rho}$ \\
 $\mathbf{2_0}$ & $2$ & $2$ & $2$ & $2$ &  $-1$ & $0$  \\
 $\mathbf{2_1}$ & $2$ & $2$ & $2\omega^{\rho}$ &
$2\omega^{\rho}$ & $-\omega^{\tau}$ & $0$  \\
 $\mathbf{2_2}$ & $2$ & $2$ & $2\omega^{2\rho}$ &
$2\omega^{2\rho}$ & $-\omega^{2\tau}$ & $0$  \\
 $\mathbf{3}_{l, 0}$ & $3$ & $ 3\eta^{l\nu}$
& $\sum_{p}\eta^{(\rho,0)\tilde{M}^{p}_{0}
\mbox{\tiny $\begin{pmatrix} l \\0 \end{pmatrix}$}} $
&  $\sum_{p}\eta^{(\rho,\sigma)\tilde{M}^{p}_{0}
\mbox{\tiny $\begin{pmatrix} l \\0 \end{pmatrix}$}}$
& $0$ & $\!\!\eta^{l\rho}\!\!$   \\
 $\mathbf{3}_{l, 1}$ & $3$ &  $3\eta^{l\nu}$
&  $\sum_{p}\eta^{(\rho,0)\tilde{M}^{p}_{0}
\mbox{\tiny $\begin{pmatrix} l \\0 \end{pmatrix}$}}
$ & $\sum_{p}\eta^{(\rho,\sigma)\tilde{M}^{p}_{0}
\mbox{\tiny $\begin{pmatrix} l \\0 \end{pmatrix}$}}$
& $0$ & $\!\!-\eta^{l\rho}\!\!$   \\
 $\mathbf{6}_{(l, k)}$  & $6$ &
$6\eta^{l\nu}$ &$\sum_{p,s}\eta^{(\rho,0)\tilde{M}^{p}_{s}
\mbox {\tiny $\begin{pmatrix} l\\-3k \end{pmatrix}$}} $ &
$\sum_{p,s}\eta^{(\rho,\sigma)\tilde{M}^{p}_{s}
\mbox{ \tiny $\begin{pmatrix} l\\-3k \end{pmatrix}$}}$ &   0  & $0$ \\ \hline \hline
\end{tabular}
\caption{\label{tab:characterAa}The character table of the $D^{(1)}_{9n,3n}$ group. The different conjugacy classes are presented in Eq.~\eqref{eq:classes}. The notation $\tilde{M}^{p}_{s}$ is explained in Eq.~\eqref{eq:Mtilde_nota}.
}
\end{center}
\end{table}

\subsection{\label{sec:App_CGC}Clebsch-Gordan Coefficients of the $D^{(1)}_{9n,3n}$ group}

In the following, we shall decompose the product of two irreducible representations
into a sum of irreducible representations of $D^{(1)}_{9n,3n}$. Under the action of the generators $a$, $b$ and $c$, different $D^{(1)}_{9n,3n}$ vector multiplets transform as follows:
\begin{eqnarray}
\nonumber\mathbf{6}_{(l, k)} & :&~~
\left(\begin{array}{c} \alpha_1 \\ \alpha_2 \\ \alpha_3 \\ \alpha_4 \\ \alpha_5 \\ \alpha_6
\end{array}\right)\stackrel{a}{\mapsto}
\left(\begin{array}{c} \alpha_2 \\ \alpha_3 \\ \alpha_1 \\ \alpha_6 \\ \alpha_4 \\ \alpha_5
\end{array}\right),~~~\stackrel{b}{\mapsto}
\left(\begin{array}{c} \alpha_4 \\ \alpha_5 \\ \alpha_6 \\ \alpha_1 \\ \alpha_2 \\ \alpha_3
\end{array}\right),~~~\stackrel{c}{\mapsto}
\left(\begin{array}{c} \eta^{l}\alpha_1 \\ \eta^{l-3k}\alpha_2 \\ \eta^{3k-2l}\alpha_3   \\ \eta^{l-3k}\alpha_4 \\ \eta^{3k-2l}\alpha_5 \\ \eta^{l}\alpha_6
\end{array}\right),  \\
\nonumber\mathbf{3}_{l, 0} &:&~~
\left(\begin{array}{c}\alpha_1 \\ \alpha_2 \\ \alpha_3
\end{array}\right)\stackrel{a}{\mapsto}
\left(\begin{array}{c} \alpha_2 \\ \alpha_3 \\ \alpha_1
\end{array}\right),~~~~\stackrel{b}{\mapsto}
\left(\begin{array}{c} \alpha_1 \\ \alpha_3 \\ \alpha_2
\end{array}\right),~~~\stackrel{c}{\mapsto}
\left(\begin{array}{c} \eta^{l}\alpha_1 \\ \eta^{l}\alpha_2 \\ \eta^{-2l}\alpha_3
\end{array}\right), \\
\nonumber\mathbf{3}_{l, 1} &:&~~
\left(\begin{array}{c} \alpha_1 \\ \alpha_2 \\ \alpha_3
\end{array}\right)\stackrel{a}{\mapsto}
\left(\begin{array}{c} \alpha_2 \\ \alpha_3 \\ \alpha_1
\end{array}\right),~~~~\stackrel{b}{\mapsto}
\left(\begin{array}{c} -\alpha_1 \\ -\alpha_3 \\ -\alpha_2
\end{array}\right),~~\stackrel{c}{\mapsto}
\left(\begin{array}{c} \eta^{l}\alpha_1 \\ \eta^{l}\alpha_2 \\ \eta^{-2l} \alpha_3  \end{array}\right),  \\
\nonumber\mathbf{2_0} &:&~~
\left(\begin{array}{c} \alpha_1 \\ \alpha_2
\end{array}\right)\stackrel{a}{\mapsto}
\frac{1}{2}\left(\begin{array}{c}
-\alpha_1-\sqrt{3}\alpha_{2} \\ \sqrt{3}\alpha_{1}-\alpha_2
\end{array}\right),~~\stackrel{b}{\mapsto}
\left(\begin{array}{c} \alpha_1 \\ -\alpha_2
\end{array}\right),~~\stackrel{c}{\mapsto}
\left(\begin{array}{c} \alpha_1 \\ \alpha_2
\end{array}\right),\\
\nonumber\mathbf{2_1} &:&~~
\left(\begin{array}{c} \alpha_1 \\ \alpha_2
\end{array}\right)\stackrel{a}{\mapsto}
\frac{1}{2}\left(\begin{array}{c}
-\alpha_1-\sqrt{3}\alpha_{2} \\ \sqrt{3}\alpha_{1}-\alpha_2
\end{array}\right),~~\stackrel{b}{\mapsto}
\left(\begin{array}{c} \alpha_1 \\ -\alpha_2
\end{array}\right),~~\stackrel{c}{\mapsto}
\omega\left(\begin{array}{c} \alpha_1 \\ \alpha_2
\end{array}\right), \\
\label{eq:repsa}\mathbf{2_2} &:&~~
\left(\begin{array}{c} \alpha_1 \\ \alpha_2
\end{array}\right)\stackrel{a}{\mapsto}
\frac{1}{2}\left(\begin{array}{c}
-\alpha_1-\sqrt{3}\alpha_{2} \\ \sqrt{3}\alpha_{1}-\alpha_2
\end{array}\right),~~\stackrel{b}{\mapsto}
\left(\begin{array}{c} \alpha_1 \\ -\alpha_2
\end{array}\right),~~\stackrel{c}{\mapsto}
\omega^{2}\left(\begin{array}{c} \alpha_1 \\ \alpha_2
\end{array}\right)\,,
\end{eqnarray}
where the action of the generator $d$ is not considered because it can be expressed in terms of $a$, $b$ and $c$, as shown in Eq.~\eqref{eq:multiplication_rules}. Starting from these set of transformations rules, we can build a set of terms which define a space of an irreducible representation. Henceforth all Clebsch-Gordan (CG) coefficients would be reported in the form of $\alpha\otimes\beta$. We shall use $\alpha_{i}$ to denote the elements of first representation and $\beta_{i}$ stands for the elements of the second representation of the tensor product. Moreover, we shall denote $f\equiv i+p~(\text{mod}~2)$ and $g\equiv j+q~(\text{mod}~3)$ for simplicity of notation.

\begin{itemize}
\item[$\bullet$]{$\mathbf{1}_{i,j}\otimes\mathbf{1}_{p,q}~=\mathbf{1}_{f, g}$}\\
\begin{equation}
\mathbf{1}_{f, g}\sim\alpha\beta\,.
\end{equation}

\item[$\bullet$]{$\mathbf{1}_{i,j}\otimes\mathbf{2}_{q}~=\mathbf{2}_{g}$}\\
\begin{equation}
i=0~:~\mathbf{2}_{g}~\sim\left(\begin{array}{c}
\alpha\beta_1 \\
\alpha\beta_2
\end{array}\right), ~\qquad
i=1~:~ \mathbf{2}_{g}~\sim\left(\begin{array}{c}
 \alpha\beta_2 \\
 -\alpha\beta_1
\end{array}\right)\,.
\end{equation}

\item[$\bullet$]{$\mathbf{1}_{i,j}\otimes\mathbf{3}_{l, p}~=\mathbf{3}_{l+3nj, f}$}\\
\begin{equation}
\mathbf{3}_{l+3nj, f}~\sim\left(\begin{array}{c}
\alpha\beta_1 \\
\alpha\beta_2 \\
\alpha\beta_3
\end{array}\right)\,.
\end{equation}

\item[$\bullet$]{$\mathbf{1}_{i,j}\otimes\mathbf{6}_{(l, k)}~=\mathbf{6}_{(l+3nj, k)}$} \\
\begin{equation}
\mathbf{6}_{(l+3nj, k)}~\sim\left(\begin{array}{c}
\alpha\beta_1 \\
\alpha\beta_2 \\
\alpha\beta_3 \\
(-1)^{i}\alpha\beta_4 \\
(-1)^{i}\alpha\beta_5 \\
(-1)^{i}\alpha\beta_6 \\
\end{array}\right)\,.
\end{equation}

\item[$\bullet$]{$\mathbf{2}_{j}\otimes\mathbf{2}_{q}~=\mathbf{1}_{0,g}\oplus\mathbf{1}_{1,g}\oplus\mathbf{2}_{g}$}\\
\begin{equation}
\mathbf{1}_{0,g}\sim\alpha_1\beta_1+\alpha_2\beta_2,\qquad
\mathbf{1}_{1,g}\sim\alpha_1\beta_2-\alpha_2\beta_1,\qquad
\mathbf{2}_{g}\sim\left(\begin{array}{c}
\alpha_1\beta_1-\alpha_2\beta_2 \\
-\alpha_1\beta_2-\alpha_2\beta_1\\
\end{array}\right)\,.
\end{equation}

\item[$\bullet$]{$\mathbf{2}_{j}\otimes\mathbf{3}_{l, i}~=\mathbf{3}_{l+3nj, i}\oplus\mathbf{3}_{l+3nj, m}$, where $m=i+1$~ $(\text{mod}~2)$} \\
\begin{equation}
 \mathbf{3}_{l+3nj, i}~\sim\left(\begin{array}{c}
2\alpha_{1}\beta_{1}\\
 -(\alpha_{1}+\sqrt{3}\alpha_{2})\beta_{2} \\
(-\alpha_{1}+\sqrt{3}\alpha_{2})\beta_{3} \\
\end{array}\right),  \qquad
\mathbf{3}_{l+3nj, m}~\sim\left(\begin{array}{c}
2\alpha_{2}\beta_{1}\\
(\sqrt{3}\alpha_{1}-\alpha_{2})\beta_{2} \\
-(\sqrt{3}\alpha_{1}+\alpha_{2})\beta_{3} \\
\end{array}\right)\,.
\end{equation}

\item[$\bullet$]{$\mathbf{2}_{j}\otimes\mathbf{6}_{(l, k)}~=\mathbf{6}_{(l+3nj, k)}\oplus\mathbf{6}_{(l+3nj,k)}$}\\
\begin{equation}
\mathbf{6}_{(l+3nj, k)}~\sim
\begin{pmatrix}
2\alpha_{1}\beta_{1}\\
-(\alpha_{1}+\sqrt{3}\alpha_{2})\beta_{2} \\
(-\alpha_{1}+\sqrt{3}\alpha_{2})\beta_{3} \\
2\alpha_{1}\beta_{4}\\
(-\alpha_1+\sqrt{3}\alpha_2)\beta_{5} \\
-(\alpha_1+\sqrt{3}\alpha_2)\beta_{6} \\
\end{pmatrix}, \qquad
\mathbf{6}_{(l+3nj, k)}~\sim
\begin{pmatrix}
2\alpha_{2}\beta_{1}\\
(\sqrt{3}\alpha_{1}-\alpha_{2})\beta_{2} \\
-(\sqrt{3}\alpha_{1}+\alpha_{2})\beta_{3} \\
-2\alpha_{2}\beta_{4}\\
(\sqrt{3}\alpha_{1}+\alpha_{2})\beta_{5} \\
-(\sqrt{3}\alpha_{1}-\alpha_{2})\beta_{6} \\
\end{pmatrix}\,.
\end{equation}

\item[$\bullet$]{$\mathbf{3}_{l, i}\otimes\mathbf{3}_{ l^{\prime}, p}~=\mathbf{3}_{l+l^{\prime}, f}\oplus\mathbf{6}_{(l+l^{\prime}, l^{\prime})}$}\\
\begin{equation}
\begin{array}{ll}
\mathbf{3}_{l+l^{\prime}, f}~\sim\left(\begin{array}{c}
\alpha_{1}\beta_{1}\\
\alpha_{2}\beta_{2} \\
\alpha_{3}\beta_{3} \\
\end{array}\right),
~~&~~
\mathbf{6}_{(l+l^{\prime}, l^{\prime})}~\sim\left(\begin{array}{c}
\alpha_{1}\beta_{2}\\
\alpha_{2}\beta_{3} \\
\alpha_{3}\beta_{1} \\
(-1)^{i-p}\alpha_{1}\beta_{3}\\
(-1)^{i-p}\alpha_{3}\beta_{2} \\
(-1)^{i-p}\alpha_{2}\beta_{1} \\
\end{array}\right)\,.
\end{array}
\end{equation}

\item[$\bullet$]{$\mathbf{3}_{l, i}\otimes\mathbf{6}_{(l^{\prime}, k^{\prime})}~
=\mathbf{6}_{(l+l^{\prime}, k^{\prime})}\oplus
\mathbf{6}_{(l+l^{\prime}-3k^{\prime}, l^{\prime}-2k^{\prime})}
\oplus\mathbf{6}_{(l-2l^{\prime}+3k^{\prime}, k^{\prime}-l^{\prime})}$}\\
\begin{equation*}
\begin{array}{lll}
\mathbf{6}_{(l+l^{\prime}, k^{\prime})}\sim\begin{pmatrix}
\alpha_{1}\beta_{1}\\
\alpha_{2}\beta_{2} \\
\alpha_{3}\beta_{3} \\
(-1)^{i}\alpha_{1}\beta_{4}\\
(-1)^{i}\alpha_{3}\beta_{5} \\
(-1)^{i}\alpha_{2}\beta_{6}
\end{pmatrix},
&~ \mathbf{6}_{(l+l^{\prime}-3k^{\prime}, l^{\prime}-2k^{\prime})}\sim\begin{pmatrix}
\alpha_{1}\beta_{2}\\
\alpha_{2}\beta_{3} \\
\alpha_{3}\beta_{1} \\
(-1)^{i}\alpha_{1}\beta_{5}\\
(-1)^{i}\alpha_{3}\beta_{6} \\
(-1)^{i}\alpha_{2}\beta_{4}
\end{pmatrix},
~&\mathbf{6}_{(l-2l^{\prime}+3k^{\prime}, k^{\prime}-l^{\prime})}
\sim\begin{pmatrix}
\alpha_{1}\beta_{3}\\
\alpha_{2}\beta_{1} \\
\alpha_{3}\beta_{2} \\
(-1)^{i}\alpha_{1}\beta_{6}\\
(-1)^{i}\alpha_{3}\beta_{4} \\
(-1)^{i}\alpha_{2}\beta_{5}
\end{pmatrix}\,.
\end{array}
\end{equation*}

\item[$\bullet$]{$\mathbf{6}_{(l, k)}\otimes\mathbf{6}_{(l^{\prime}, k^{\prime})}~
=\mathbf{6}_{(l+l^{\prime}, k+k^{\prime})}\oplus
\mathbf{6}_{(l+l^{\prime}-3k^{\prime}, k-2k^{\prime}+l^{\prime})}
\oplus\mathbf{6}_{(l-2l^{\prime}+3k^{\prime}, k+k^{\prime}-l^{\prime})}$}\\
\hskip2cm$\oplus\mathbf{6}_{(l+l^{\prime}-3k^{\prime}, k-k^{\prime})}
\oplus\mathbf{6}_{(l-2l^{\prime}+3k^{\prime}, k+2k^{\prime}-l^{\prime})}
\oplus\mathbf{6}_{(l+l^{\prime},
k-k^{\prime}+l^{\prime})}$\\
\begin{equation}
\begin{array}{lll}
\mathbf{6}_{(l+l^{\prime}, k+k^{\prime})}\sim\begin{pmatrix}
\alpha_{1}\beta_{1}\\
\alpha_{2}\beta_{2} \\
\alpha_{3}\beta_{3} \\
\alpha_{4}\beta_{4}\\
\alpha_{5}\beta_{5} \\
\alpha_{6}\beta_{6}
\end{pmatrix},
& \mathbf{6}_{(l+l^{\prime}-3k^{\prime},
k-2k^{\prime}+l^{\prime})}\sim
\begin{pmatrix}
\alpha_{1}\beta_{2}\\
\alpha_{2}\beta_{3} \\
\alpha_{3}\beta_{1} \\
\alpha_{4}\beta_{5}\\
\alpha_{5}\beta_{6} \\
\alpha_{6}\beta_{4}
\end{pmatrix},
& \mathbf{6}_{(l-2l^{\prime}+3k^{\prime},
k+k^{\prime}-l^{\prime})}\sim\begin{pmatrix}
\alpha_{1}\beta_{3}\\
\alpha_{2}\beta_{1} \\
\alpha_{3}\beta_{2} \\
\alpha_{4}\beta_{6}\\
\alpha_{5}\beta_{4} \\
\alpha_{6}\beta_{5}
\end{pmatrix},\\[-7pt]\\[4pt]
\mathbf{6}_{(l+l^{\prime}-3k^{\prime},
k-k^{\prime})}\sim\begin{pmatrix}
\alpha_{1}\beta_{4}\\
\alpha_{2}\beta_{6} \\
\alpha_{3}\beta_{5} \\
\alpha_{4}\beta_{1}\\
\alpha_{5}\beta_{3} \\
\alpha_{6}\beta_{2}
\end{pmatrix}, &
\mathbf{6}_{(l-2l^{\prime}+3k^{\prime},
k+2k^{\prime}-l^{\prime})}\sim
\begin{pmatrix}
\alpha_{1}\beta_{5}\\
\alpha_{2}\beta_{4} \\
\alpha_{3}\beta_{6} \\
\alpha_{4}\beta_{2}\\
\alpha_{5}\beta_{1} \\
\alpha_{6}\beta_{3}
\end{pmatrix},
& \mathbf{6}_{(l+l^{\prime},
k-k^{\prime}+l^{\prime})}\sim
\begin{pmatrix}
\alpha_{1}\beta_{6}\\
\alpha_{2}\beta_{5} \\
\alpha_{3}\beta_{4} \\
\alpha_{4}\beta_{3}\\
\alpha_{5}\beta_{2} \\
\alpha_{6}\beta_{1}
\end{pmatrix}\,.
\end{array}
\end{equation}
We would like to point out that certain three-dimensional and six-dimensional representations in the above tensor product decompositions may be reducible, and accordingly it should be reduced into smaller irreducible representations of the $D^{(1)}_{9n, 3n}$ group. The reducible conditions and corresponding reduction formulae are summarized in table~\ref{tab:reducible}.

\end{itemize}

\begin{table}[hptb]
\begin{center}
\begin{tabular}{|c|c|c|c|}\hline\hline

 & \multicolumn{3}{c|}{$l~(\text{mod}~9n)=3n\lambda$, $\lambda=0, 1, 2$}\\ \hline
  & \multicolumn{3}{c|}{$\mathbf{3}_{l, i}\cong\mathbf{1}_{i, \lambda} \oplus\mathbf{2_{\lambda}}$} \\ \hline

$\mathbf{3}_{l, i}~\sim\begin{pmatrix}
\gamma_{1}\\
\gamma_{2} \\
\gamma_{3}\\
\end{pmatrix}$ & \multicolumn{3}{|c|}{$
\begin{array}{l}
\mathbf{1}_{i,\lambda}~\sim \gamma_{1}+\gamma_{2}+\gamma_{3}\\
 \\[-0.1in]
\mathbf{2_{\lambda}}~\sim\begin{pmatrix}
-2\gamma_{1}+\gamma_{2}+\gamma_{3}\\
\sqrt{3}\gamma_{2}-\sqrt{3}\gamma_{3}
\end{pmatrix},~~\text{for}~~i=0 \\
 \\[-0.1in]
\mathbf{2_{\lambda}}~\sim\begin{pmatrix}
\sqrt{3}\gamma_{2}-\sqrt{3}\gamma_{3} \\
2\gamma_{1}-\gamma_{2}-\gamma_{3}
\end{pmatrix},~~\text{for}~~i=1
\end{array}$ }  \\\hline\hline

 &  $\left\{\begin{array}{c}
 3k=0~(\text{mod}~9n)\\
 l\neq0~(\text{mod}~3n)
 \end{array}\right.$  & $\left\{\begin{array}{c}
 3l-3k=0~(\text{mod}~9n)\\
 l\neq0~(\text{mod}~3n)
 \end{array}\right.$ & $\left\{\begin{array}{c}
 3l-6k=0~(\text{mod}~9n)\\
 l, k \neq0~(\text{mod}~3n)
 \end{array}\right.$ \\
 & &  & \\[-0.18in] \hline
& &  & \\[-0.18in]
 & $\mathbf{6}_{(l , k)}\cong\mathbf{3}_{l, 0}\oplus\mathbf{3}_{l, 1}$ & $\mathbf{6}_{(l , k)}\cong\mathbf{3}_{l, 0}\oplus\mathbf{3}_{l, 1}$ & $\mathbf{6}_{(l , k)}\cong\mathbf{3}_{l-3k, 0}\oplus\mathbf{3}_{l-3k, 1}$\\ \hline
 & &  & \\[-0.18in]
$\mathbf{6}_{(l, k)}\sim \begin{pmatrix}
\gamma_{1}\\
\gamma_{2} \\
\gamma_{3} \\
\gamma_{4}\\
\gamma_{5} \\
\gamma_{6}
\end{pmatrix}$   &~ $\begin{array}{c}
\mathbf{3}_{l, 0}\sim \begin{pmatrix}
\gamma_{1}+\gamma_{4}\\
\gamma_{2}+\gamma_{6} \\
\gamma_{3}+\gamma_{5}
\end{pmatrix}\\
\\[-0.1in]
\mathbf{3}_{l, 1}\sim \begin{pmatrix}
\gamma_{1}-\gamma_{4}\\
\gamma_{2}-\gamma_{6} \\
\gamma_{3}-\gamma_{5}
\end{pmatrix}
\end{array}$ &~
$\begin{array}{c}
\mathbf{3}_{l, 0}\sim \begin{pmatrix}
\gamma_{3}+\gamma_{6}\\
\gamma_{1}+\gamma_{5} \\
\gamma_{2}+\gamma_{4}
\end{pmatrix}\\
\\[-0.1in]
\mathbf{3}_{l, 1}\sim \begin{pmatrix}
\gamma_{3}-\gamma_{6}\\
\gamma_{1}-\gamma_{5} \\
\gamma_{2}-\gamma_{4}
\end{pmatrix}
\end{array}$ &~
$\begin{array}{c}
\mathbf{3}_{l-3k, 0}\sim \begin{pmatrix}
\gamma_{2}+\gamma_{5}\\
\gamma_{3}+\gamma_{4} \\
\gamma_{1}+\gamma_{6}
\end{pmatrix}\\
\\[-0.1in]
\mathbf{3}_{l-3k, 1}\sim \begin{pmatrix}
\gamma_{2}-\gamma_{5}\\
\gamma_{3}-\gamma_{4} \\
\gamma_{1}-\gamma_{6}
\end{pmatrix}
\end{array}$ \\
& &  & \\[-0.18in] \hline\hline
 &  \multicolumn{3}{c|}{$k~(\text{mod}~3n)=0$,~~ $l~(\text{mod}~9n)=3n\lambda$, $\lambda=0, 1, 2$} \\ \hline

 & \multicolumn{3}{c|}{$\mathbf{6}_{(l , k)}\cong\mathbf{1}_{0, \lambda}\oplus\mathbf{1}_{1, \lambda}\oplus\mathbf{2}_{\lambda}\oplus\mathbf{2}_{\lambda}$}\\ \hline
$\mathbf{6}_{(l, k)}\sim \begin{pmatrix}
\gamma_{1}\\
\gamma_{2} \\
\gamma_{3} \\
\gamma_{4}\\
\gamma_{5} \\
\gamma_{6}
\end{pmatrix}$  &  \multicolumn{3}{|c|}{$\begin{array}{l}
\mathbf{1}_{0, \lambda}\sim\gamma_1+\gamma_2+\gamma_3+\gamma_4+\gamma_5+\gamma_6,\\
\\[-0.1in]
\mathbf{1}_{1, \lambda}\sim\gamma_1+\gamma_2+\gamma_3-\gamma_4-\gamma_5-\gamma_6,\\
\\[-0.1in]
\mathbf{2}_{\lambda}\sim\begin{pmatrix}
-2\gamma_1+\gamma_2+\gamma_3-2\gamma_4+\gamma_5+\gamma_6 \\
\sqrt{3}\left(\gamma_2-\gamma_3-\gamma_5+\gamma_6\right)
\end{pmatrix},~\\
\\[-0.1in]
\mathbf{2}_{\lambda}\sim\begin{pmatrix}
\sqrt{3}\left(\gamma_2-\gamma_3+\gamma_5-\gamma_6\right) \\
2\gamma_1-\gamma_2-\gamma_3-2\gamma_4+\gamma_5+\gamma_6
\end{pmatrix}
\end{array}
$}\\
& &  & \\[-0.18in]\hline\hline

\end{tabular}
\caption{\label{tab:reducible}The reducible conditions for $\mathbf{3}_{l, 0}$, $\mathbf{3}_{l, 1}$ and $\mathbf{6}_{(l, k)}$, and the decomposition of reducible three-dimensional and six-dimensional representations into smaller irreducible representations of $D^{(1)}_{9n, 3n}$. Notice that the expression for the doublet vector $\mathbf{2}_{\lambda}$ is not unique in the decomposition of $\mathbf{6}_{(3n\lambda, 0)}\cong\mathbf{1}_{0, \lambda}\oplus\mathbf{1}_{1, \lambda}\oplus\mathbf{2}_{\lambda}\oplus\mathbf{2}_{\lambda}$.}
\end{center}
\end{table}

\end{appendix}

\end{document}